\documentclass[letterpaper]{article}
\usepackage{emulateapj}
\usepackage{epsfig}
\usepackage{times}
\submitted{The Astrophysical Journal, 537: ...--..., 2000 July 10}

\newcommand{\Berat}{$^{10}$Be/$^9$Be} 
\newcommand{\gray}{$\gamma$-ray\ } 
\newcommand{\grays}{$\gamma$-rays\ }
\def\fwb{78mm}
\def\fwc{57mm}
\def\fwd{75mm}
\newcommand{\hs}{15mm}
\newcommand{\fns}{\footnotesize}
\lefthead{STRONG, MOSKALENKO, AND REIMER} 
\righthead{DIFFUSE CONTINUUM GAMMA RAYS FROM THE GALAXY}

\begin{document}

\title{Diffuse continuum gamma rays from the Galaxy}

\author{Andrew W.~Strong\altaffilmark{1},  
Igor V.~Moskalenko\altaffilmark{1,2,3,4},  and 
Olaf Reimer\altaffilmark{1}}

\affil{\altaffilmark{1}Max-Planck-Institut f\"ur extraterrestrische
   Physik, Postfach 1603, D-85740 Garching, Germany}
\affil{\altaffilmark{2}Institute for Nuclear Physics,
   M.V.Lomonosov Moscow State University, 119 899 Moscow, Russia}
\affil{\altaffilmark{3}Present address: Laboratory for High Energy
   Astrophysics  NASA/GSFC, Code 660, Greenbelt, MD 20771, U.S.A.}
\affil{\altaffilmark{4}NRC Senior Research Associate}

\authoremail{aws@mpe.mpg.de; imos@milkyway.gsfc.nasa.gov;
olr@mpe.mpg.de}

\begin{abstract}

A new study of the diffuse Galactic \gray continuum radiation is
presented, using a cosmic-ray propagation model which includes
nucleons, antiprotons, electrons, positrons, and synchrotron
radiation.  Our treatment of the inverse Compton scattering
includes the effect of anisotropic scattering in the Galactic
interstellar radiation field (ISRF) and a new evaluation of the ISRF
itself.  Models based on locally
measured electron and nucleon spectra and synchrotron constraints are
consistent with \gray measurements in the 30--500 MeV range, but
outside this range excesses are apparent.  A harder nucleon spectrum
is considered but fitting to \grays causes it to violate limits from
positrons and antiprotons.  A harder interstellar electron spectrum
allows the \gray spectrum to be fitted above 1 GeV as well, and this
can be further improved when combined with a modified nucleon spectrum
which still respects the limits imposed by antiprotons and positrons.

A large electron/inverse Compton halo is proposed which reproduces well the
high-latitude variation of \gray emission; this is taken as support
for the halo size for nucleons deduced from studies of cosmic-ray
composition.  Halo sizes in the range 4--10 kpc are favoured
by both analyses. The halo contribution of Galactic emission to the
high-latitude \gray intensity is large, with implications for the
study of the diffuse extragalactic component and signatures of dark
matter.  The constraints provided by the radio synchrotron spectral
index do not allow all of the $<$30 MeV \gray emission to be explained
in terms of a steep electron spectrum unless this takes the form of a
sharp upturn below 200 MeV.  This leads us to prefer a source
population as the origin of the excess low-energy $\gamma$-rays, which
can then be seen as a continuation of the hard X-ray continuum
measured by OSSE, GINGA and RXTE.

\end{abstract}

\section{Introduction}
%######################################################################

Despite much effort the origin of the diffuse Galactic continuum \gray
emission is still subject to considerable uncertainties.  While the
main \gray production mechanisms are agreed to be inverse Compton (IC) 
scattering, $\pi^0$-production, and bremsstrahlung, their individual 
contributions depend on many details such as interstellar electron and nucleon
spectra, interstellar radiation and magnetic fields, gas distribution
etc.  At energies above $\sim1$ GeV and below $\sim30$ MeV the
dominant physical mechanisms are yet to be established (see, e.g.,
\cite{Hunter97}, \cite{Skibo97}, \cite{PohlEsposito98}, 
Moskalenko, Strong, \& Reimer 1998, hereafter \cite{MSR98}, 
Moskalenko \& Strong 1999a, hereafter \cite{MS99a}).

The spectrum of Galactic \grays as measured by EGRET shows enhanced
emission above 1 GeV in comparison with calculations based on locally
measured proton and electron spectra (\cite{Hunter97}).  Mori (1997)
and Gralewicz et al.\ (1997) proposed a harder interstellar proton
spectrum as a solution. This possibility has been tested
using cosmic-ray antiprotons and positrons (\cite{MSR98},
\cite{MS99a}).  Another explanation has been proposed by Porter \&
Protheroe (1997) and Pohl \& Esposito (1998), who suggested that the
average interstellar electron spectrum can be harder than that locally
observed due to the spatially inhomogeneous source distribution and
energy losses.  Pohl \& Esposito (1998) made detailed Monte Carlo
simulations of the high-energy electron spectrum in the Galaxy taking
into account the spatially inhomogeneous source distribution, and
showed that the \gray excess could indeed be explained in terms of
inverse Compton emission from a hard electron spectrum.

The situation below several MeV is also unclear; Skibo et al.\ (1997)
showed that the diffuse flux measured by OSSE below 1 MeV
(\cite{Purcell96}) can be explained by bremsstrahlung only if there is
a steep upturn in the electron spectrum at low energies, but that this
requires very large energy input into the interstellar medium.  A
model for the acceleration of low-energy electrons has been proposed
by Schlickeiser (1997).  An analysis of the emission in the 1--30 MeV
range, based on the latest COMPTEL data, has been made by \cite{MS99a},
who found that the predicted intensities are  significantly below the
observations, and that a point-source component is probably necessary.
Solving these puzzles requires a systematic study including all
relevant astrophysical data and a corresponding self-consistent
approach to be adopted.

With this motivation a numerical method and corresponding computer
code (`GALPROP') for the calculation of Galactic cosmic-ray
propagation has been developed (Strong \& Moskalenko 1998, hereafter 
\cite{SM98}).  Primary and secondary
nucleons, primary and secondary electrons, secondary positrons and
antiprotons, as well as \grays and synchrotron radiation are
included. The basic spatial propagation mechanisms are diffusion and
convection, while in momentum space energy loss and diffusive
reacceleration are treated.  Fragmentation and energy losses are
computed using realistic distributions for the interstellar gas and
radiation fields.  Our preliminary results were presented in
Strong \& Moskalenko (1997, hereafter \cite{SM97})
and full results for protons, Helium, positrons, and
electrons in Moskalenko \& Strong (1998a, hereafter \cite{MS98a}).  
The evaluation of the B/C and \Berat\
ratios, evaluation of diffusion/convection and reacceleration models,
and full details of the numerical method are given in \cite{SM98}.
Antiprotons have been evaluated in the context of  the `hard
interstellar nucleon spectrum' hypothesis in \cite{MSR98}.  The effect
of anisotropy on the inverse Compton scattering of cosmic-ray
electrons in the Galactic radiation field is described in 
Moskalenko \& Strong (2000, hereafter \cite{MS00}).
As an application of our model, the Green's functions for the 
propagation of positrons from dark-matter particle annihilations in the
Galactic halo have been evaluated in \cite{MS99b}.

The rationale for our approach was given previously
(\cite{SM98}, \cite{MS98a}, \cite{MSR98}, \cite{SMR00}).  
Briefly, the idea is to
develop a model which simultaneously reproduces observational data of
many kinds related to cosmic-ray origin and propagation: directly via
measurements of nuclei, electrons, and positrons, indirectly via
\grays and synchrotron radiation.  These data provide many independent
constraints on any model and our approach is able to take advantage of
this since it aims to be consistent with many types of observation.
We emphasize also the use of realistic astrophysical input (e.g.\ for
the gas distribution) as well as theoretical developments (e.g.\
reacceleration).  The code is sufficiently general that new physical
effects can be introduced as required. We aim for a `standard model'
which can be improved with new astrophysical input and additional
observational constraints.

Comparing our approach with the model for EGRET data by Hunter et al.\
(1997), which used a spiral-arm model with cosmic-ray/gas coupling, we
concentrate less on obtaining an exact fit to the angular distribution
of \grays and more on the relation to cosmic-ray propagation theory
and data.

With this paper we complete the description of our model by describing
the \gray calculation, and make a new derivation of the
ISRF\footnote{Since the IC scattering is one of the central points in our
analysis, we feel that the derivation of the ISRF deserves a short
description which we place in Sect.~\ref{isrf}, while more details
will be given in a forthcoming  paper. }.   The
\grays allow us to test some aspects of the model, such as halo size,
which come from the previous work based on nucleon propagation
(\cite{SM98}).  We then use the complete model to try to answer the
question: what changes to the `conventional' approach are required to
fit the \gray data, and which are consistent with other constraints
imposed by synchrotron, positrons, antiprotons, etc.\ ?  Although no
final answer is provided, we hope to have made a contribution to the
solution.

For interested users our model including software and result datasets
is available in the public domain on the World Wide
Web\footnote{http://www.gamma.mpe--garching.mpg.de/$\sim$aws/aws.html}.

\section{Basic features of the GALPROP models} \label{Description}
%######################################################################
The GALPROP models have been described in full detail elsewhere
(\cite{SM98}); here we just summarize briefly their basic features.

The models are three dimensional with cylindrical symmetry in the
Galaxy, and the basic coordinates are $(R,z,p)$ where $R$ is
Galactocentric radius, $z$ is the distance from the Galactic plane and
$p$ is the total particle momentum. In the models the  propagation
region is bounded by $R=R_h$, $z=\pm z_h$ beyond which free escape is
assumed. We take $R_h=30$ kpc. For a given $z_h$ the diffusion
coefficient as a function of momentum  and the reacceleration
parameters are determined by B/C.  Reacceleration provides a natural
mechanism to reproduce the B/C ratio without an ad-hoc form for the
diffusion coefficient.  The spatial diffusion coefficient is taken as
$\beta D_0(\rho/\rho_0)^\delta$.  Our reacceleration treatment assumes
a Kolmogorov spectrum   with $\delta=1/3$.  For the case of
reacceleration the momentum-space diffusion coefficient $D_{pp}$ is
related to the spatial coefficient (\cite{SeoPtuskin94},
\cite{Berezinskii90}).  The injection spectrum of nucleons is assumed
to be a power law in momentum,  $dq(p)/dp \propto p^{-\gamma}$ for the
injected particle density, if necessary with a break.

The total magnetic field is assumed to have the form
\begin{equation}%=====================================================+
\label{eq.1}
B_{tot}=B_0\,e^{ - (R-R_\odot)/R_B-|z|/z_B}.
\end{equation}%=====================================================+
The values of the parameters ($B_0, R_B, z_B$) are adjusted to match
the 408 MHz synchrotron longitude and latitude distributions.  The
interstellar hydrogen distribution uses HI and CO surveys and
information on the ionized component; the Helium fraction of the gas
is taken as 0.11 by number.  Energy losses for electrons by
ionization, Coulomb interactions, bremsstrahlung, inverse Compton, and
synchrotron are included, and for nucleons by ionization and Coulomb
interactions following Mannheim \& Schlickeiser (1994).  The
distribution of cosmic-ray sources is chosen to reproduce the
cosmic-ray distribution determined by analysis of EGRET \gray data
(\cite{StrongMattox96}). The source distribution adopted  was
described in \cite{SM98}.  It adequately reproduces the  observed
\gray based gradient, while being significantly flatter than the
observed distribution of supernova remnants.

The ISRF, which is used for calculation of the IC emission and
electron energy losses, is based on stellar population models and COBE
results, plus the cosmic microwave background (CMB), more details are
given in Sect.~\ref{isrf}.  IC scattering is treated using the
formalism for an anisotropic radiation field described in \cite{MS00}.

Gas related \gray intensities are computed from the emissivities as a
function of $(R,z,E_\gamma)$ using the column densities of HI and
H$_2$ for Galactocentric annuli based on 21-cm and CO surveys
(\cite{StrongMattox96})\footnote{ While the propagation model uses
cylindrically symmetric gas distributions this influences only the
ionization energy losses, which affect protons below 1 GeV, so the
lack of a full 3D treatment here has a negligible effect on the
$\pi^0$-decay emission.  For the line-of-sight integral on the other
hand, our use of the HI and H$_2$ annuli is important since it traces
Galactic structure as seen from the solar position.  }.   Our
$\pi^0$-decay calculation is given in \cite{MS98a}. In addition our
bremsstrahlung and synchrotron calculations are described in the
present paper in Appendices~\ref{bremss}, \ref{synchrotron}; together
with previous papers in this series this completes the full
presentation of the details of our model.

In our analysis we distinguish the following main cases: the
`conventional' model which after propagation matches the observed
electron and nucleon spectra, the `hard nucleon spectrum' model, and
the `hard electron spectrum' model.  The `hard spectrum' models are
chosen so that the calculated \gray spectrum matches the \gray EGRET
data.

\begin{figure*}[tbh]%***************************************************** 1
\centerline{ 
\psfig{file=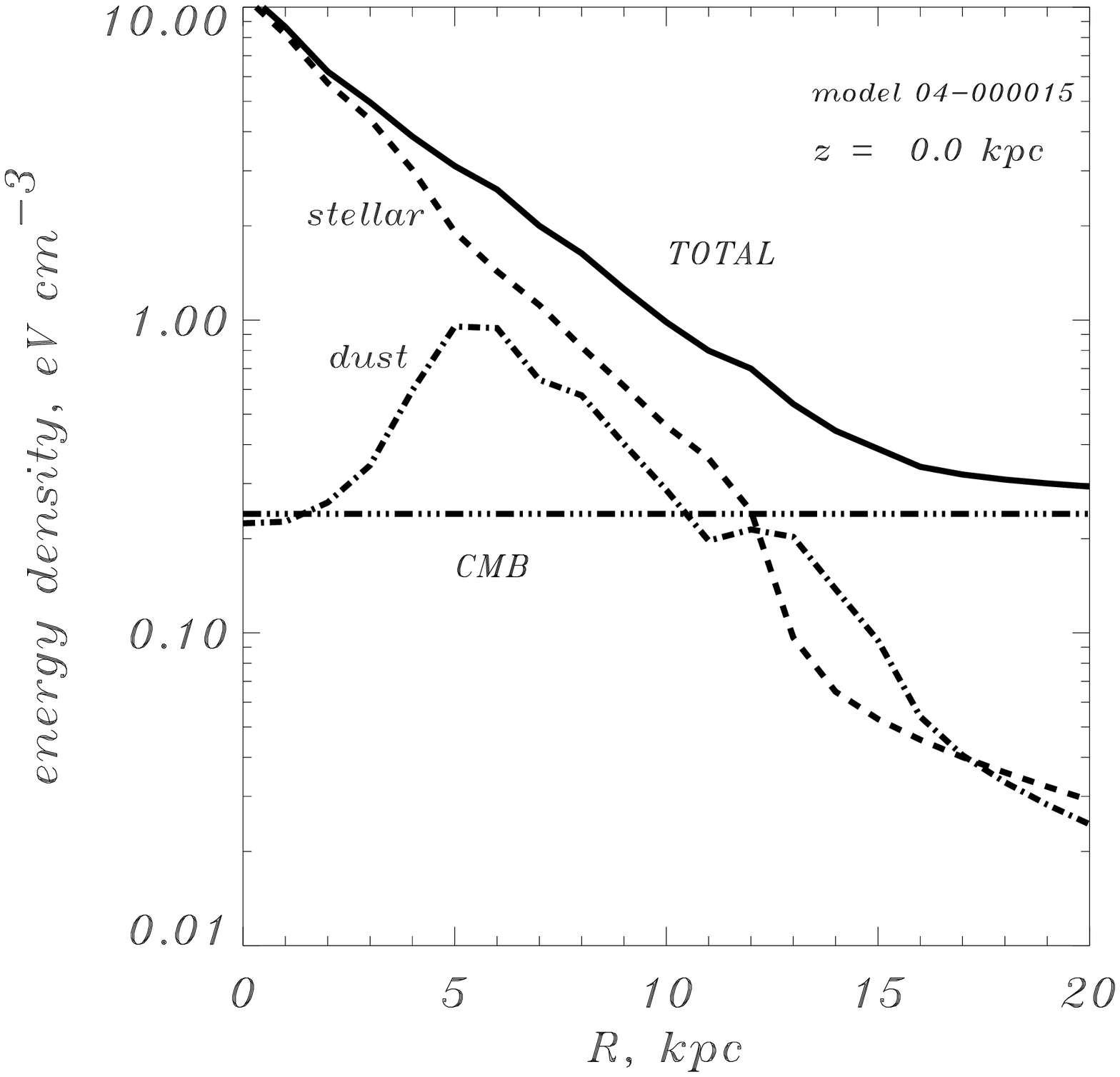,width=\fwb,clip=}
\psfig{file=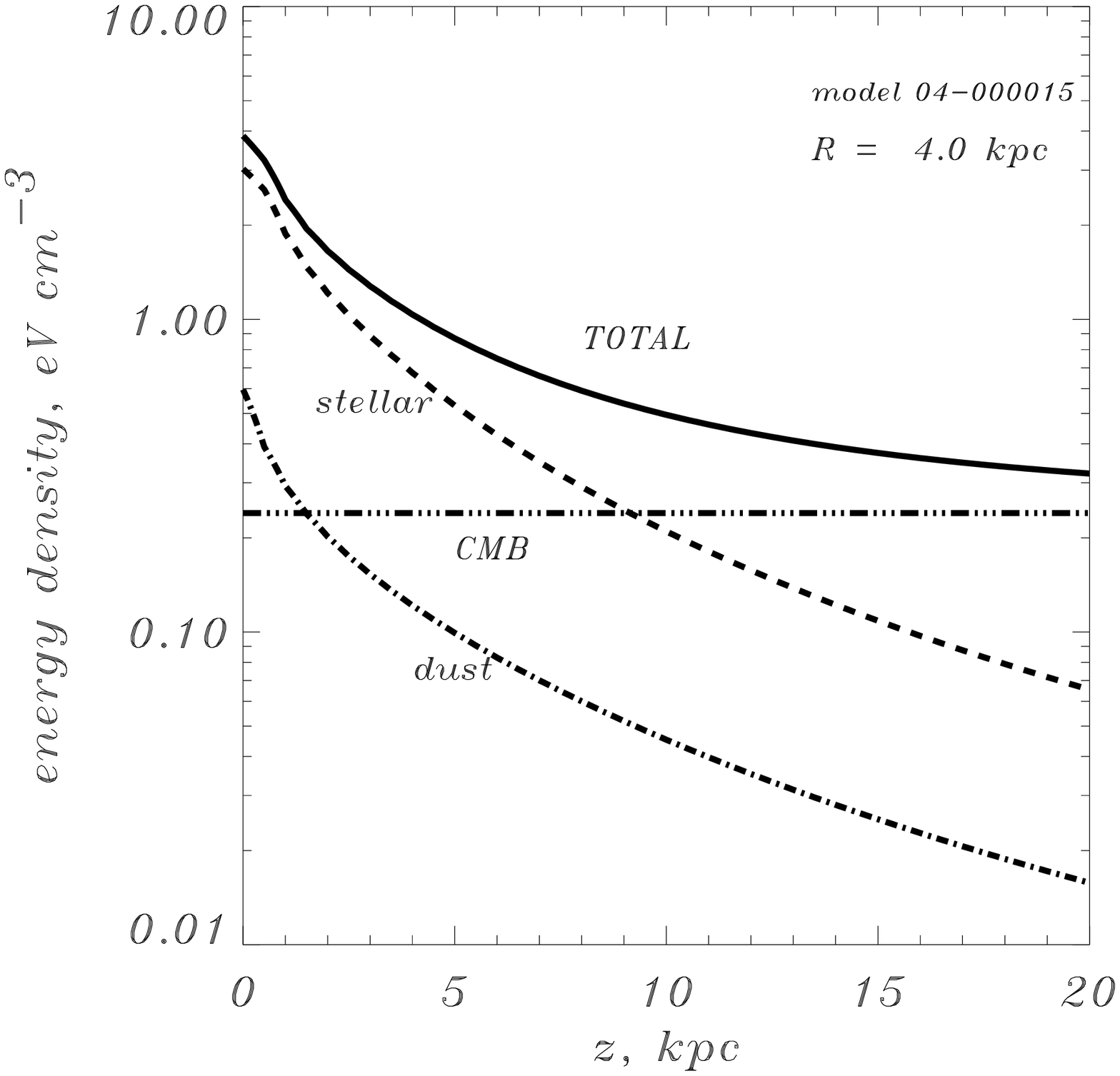,width=\fwb,clip=} }
\figcaption[fig1a.ps,fig1b.ps]{ 
ISRF energy density as function
of $R$ at $z = 0$ (left), and of $z$ at $R = 4$ kpc
(right). Shown are the contributions of stars (dashed), dust
(dash-dot), CMB (dash-3-dots), and total (full line).
\label{Fig_isrf}}
\end{figure*}

\placefigure{Fig_isrf}

\subsection{Interstellar radiation field} \label{isrf}
%######################################################################

Since Mathis, Mezger, \& Panagia (1983), Bloemen (1985), Cox \& Mezger
(1989), and Chi \& Wolfendale (1991) no calculations of the
large-scale Galactic ISRF have appeared in the literature despite the
considerable amount of new information now available especially from
IRAS and COBE.  These results reduce significantly the uncertainties
in the calculation, especially regarding the distribution of stars and
the emission from dust.  In view of the importance of the ISRF for
\gray models, a new calculation is justified.  Moreover, we require
the full ISRF as a function of $(R,z,\nu)$, which is not available in
the literature.  Our ISRF calculation uses emissivities based on
stellar populations and dust emission; as in the rest of the model,
cylindrical symmetry is assumed. The dust and stellar
components are stored separately in order to allow for
their different source distributions in the anisotropic IC scattering
calculation (\cite{MS00}).   Here we give only a brief summary of our ISRF
calculation; a fuller presentation will be given in a separate paper
(in preparation).   The resulting datasets are available at the
address given in the Introduction.

%\subsection{Dust component.}
%######################################################################
The infrared emissivities per atom of HI and H$_2$ are based on
COBE/DIRBE data from Sodrowski et al.\ (1997), combined with the
distribution of HI and H$_2$ described in \cite{SM98}.  The spectral
shape is based on the silicate, graphite and PAH synthetic spectrum
using COBE data from Dwek et al.\  (1997).

%\subsection{Stellar component}
%######################################################################
For the distribution of the old stellar disk component we use the
model of Freudenreich (1998) based on the COBE/DIRBE few micron
survey.  This has an exponential disk with radial scale length of 2.6
kpc, a vertical $\cosh^2(z)$ form with scale height of 0.346 kpc, and
a central bar.  We also use the Freudenreich single-temperature
($T=3800$ K) spectrum to compute the ISRF for 1--10 $\mu$m to
calibrate the more extensive stellar population treatment. Since the
Freudenreich model is based directly on COBE/DIRBE maps it should give
an accurate ISRF at wavelengths of a few $\mu$m and serves as a
reference datum for the more model-dependent shorter wavelength range.

The stellar luminosity function is taken from Wainscoat et al.\
(1992).  For each stellar class the local density and absolute
magnitude in standard optical and near-infrared bands is given, and
these are used to compute the local stellar emissivity by
interpolation in wavelength.  The $z$-scaleheight for each class and
the spatial functions (disk, halo, rings, arms) given by Wainscoat et
al.\ (1992) then give the volume emissivity as a function of position
and wavelength.  All their main-sequence and AGB types were explicitly 
included.

%\subsection{Absorption}
%######################################################################
Absorption is based on the specific extinction per H atom given by
Cardelli, Clayton, \& Mathis (1989) and Mathis (1990).  The albedo of
dust particles is taken as 0.63 (\cite{Mathis83}) and scattering is
assumed to be sufficiently in the forward direction as not to affect
the ISRF calculation too much.  Again the gas model described in
\cite{SM98} is used.

%\subsection{ISRF results}
%######################################################################
%The ISRF spectrum at $R=0, 4$, and 8 kpc is shown in Fig.~\ref{fig1}.

The calculated $R$- and $z$-distributions
of the total energy density are shown in Fig.~\ref{Fig_isrf} in order to
illustrate the ISRF distribution in 3D.

\section{Summary of models}
\begin{figure*}[tbh]%***************************************************** 2
\centerline{ 
\psfig{file=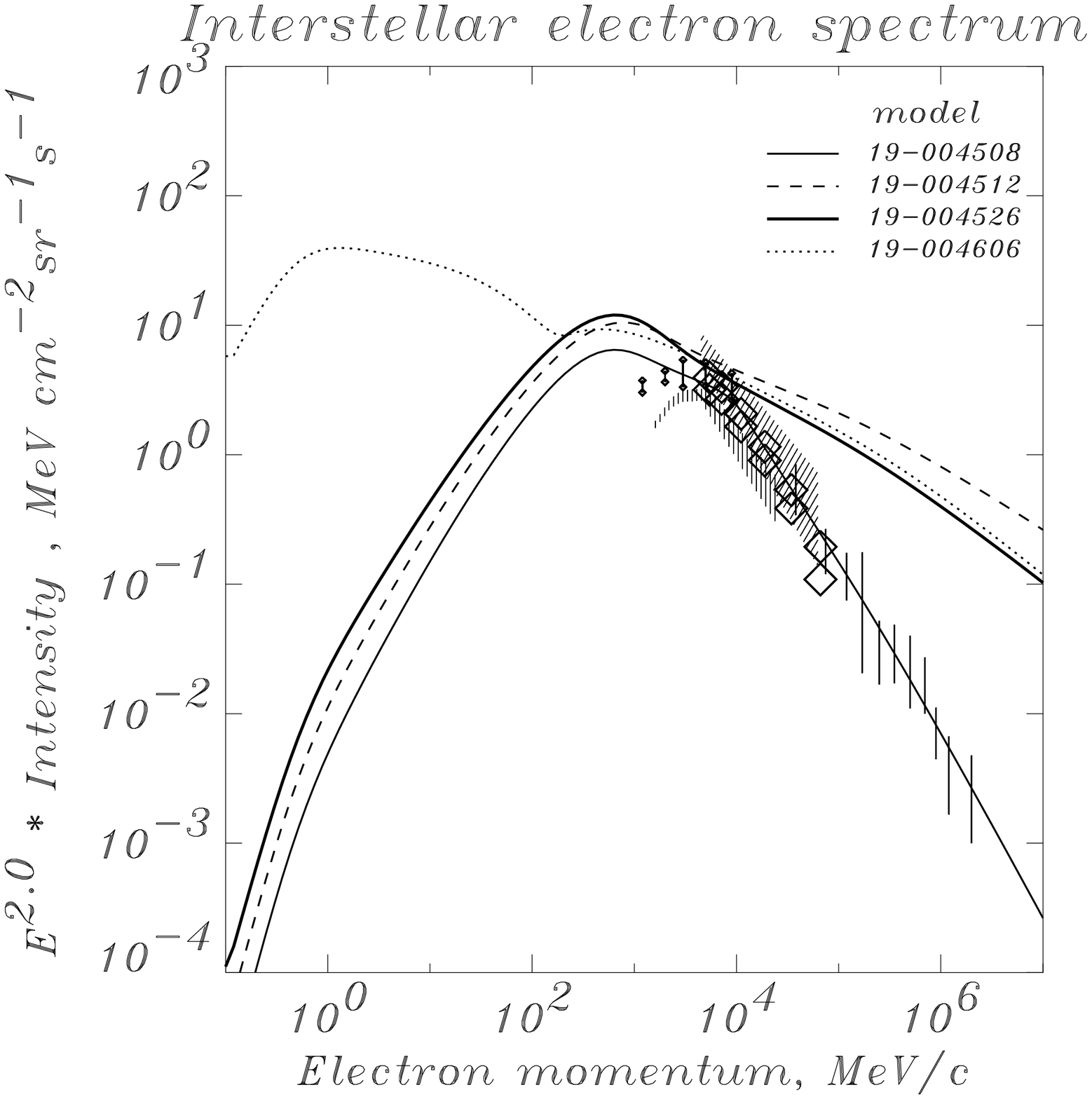,width=\fwb,clip=}
\psfig{file=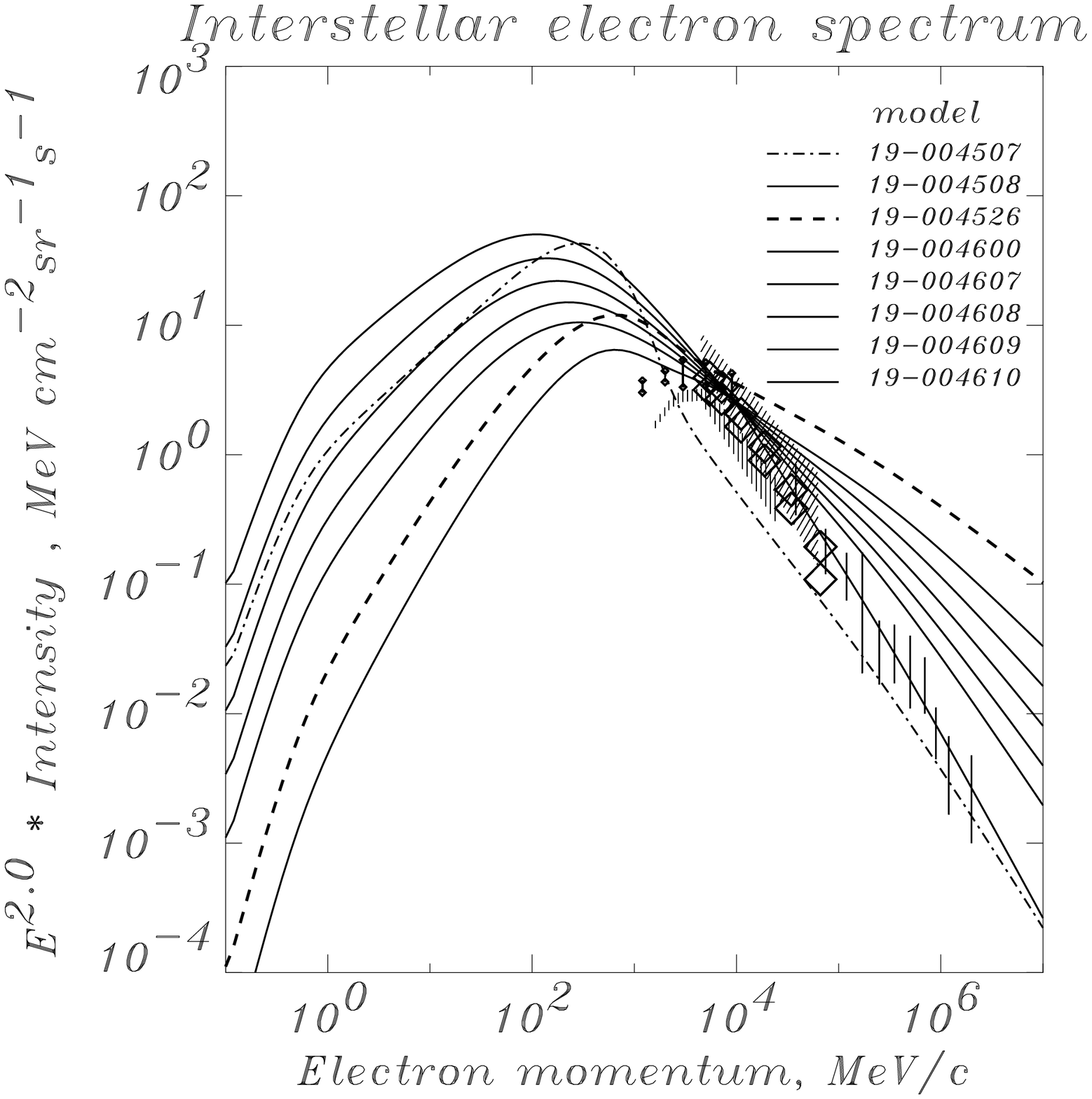,width=\fwb,clip=} }
\figcaption[fig2a.ps,fig2b.ps]{ 
Electron
spectra as obtained after propagation in our models compared
with direct measurements.   Data:  Taira et al.\ (1993)
(vertical lines), Golden et al.\ (1984, 1994) (shaded areas),
Ferrando et al.\ (1996) (small diamonds), Barwick et al.\ (1998)
(large diamonds).  Left: Thin solid line: C model, dashes: HE
model, thick solid line: HEMN model, dots: SE model. Right:
Electron injection spectral indices 2.0--2.4, no reacceleration.
\label{Fig_electrons}}
\end{figure*}

\begin{figure*}[thb]%***************************************************** 3
\centerline{ 
\psfig{file=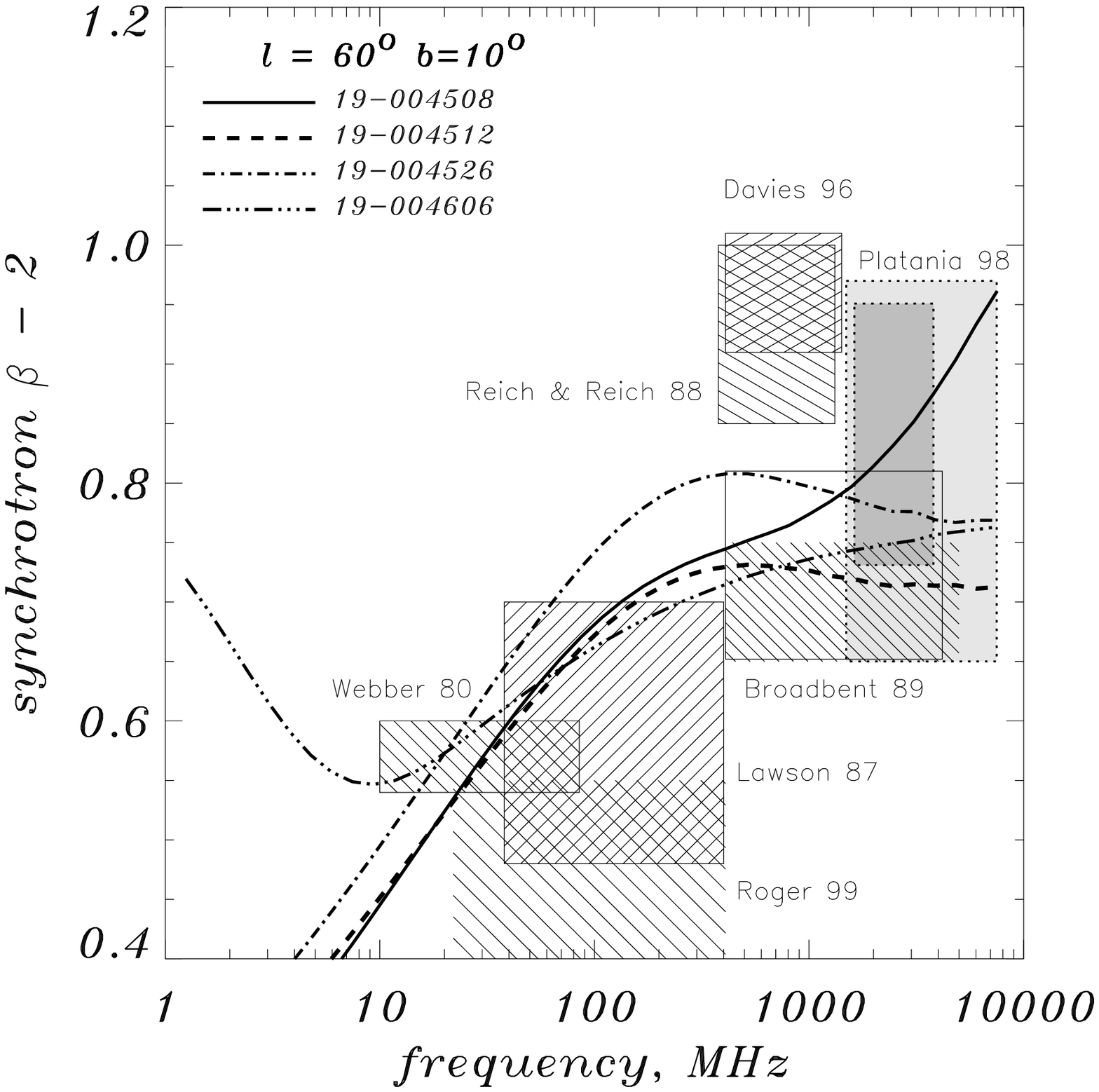,width=\fwb,clip=}
\psfig{file=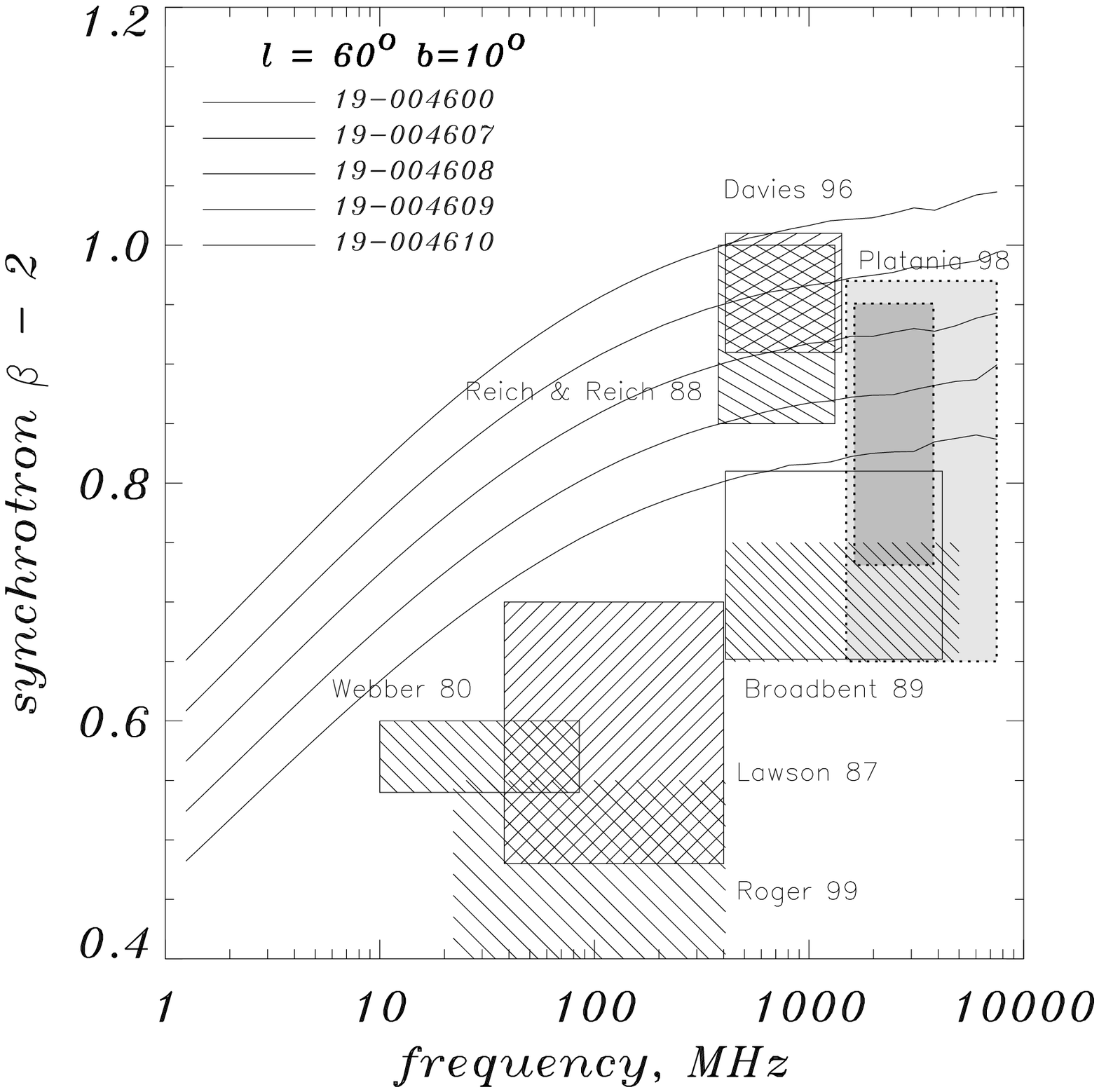,width=\fwb,clip=} }
\figcaption[fig3a.ps,fig3b.ps]{
Synchrotron spectral index for selected models.  Measurements by
different authors are shown by boxes.  Data references:  Webber,
Simpson, \& Cane (1980), Lawson et al.\  (1987),  Roger et al.\
(1999),  Broadbent, Haslam, \& Osborne (1989), Platania et al.\
(1998) (grey boxes and open box), Reich \& Reich (1988), Davies,
Watson, \& Guti\'errez (1996).  Note that the error bar given by
Webber, Simpson, \& Cane (1980) is probably too small due to the
difficulties of low-frequency radio measurements.  Left: Solid
line: model C, dashes: model HE, dash-dot: model HEMN,
dash-3-dots: model SE.  Right: Electron injection spectral
indices 2.0--2.4 (from bottom to top), no reacceleration;
electron spectra as in Fig.~\ref{Fig_electrons} (right).
\label{Fig_sync_index}}
\end{figure*}

\begin{figure*}[tbh]%***************************************************** 4 
\centerline{ 
\psfig{file=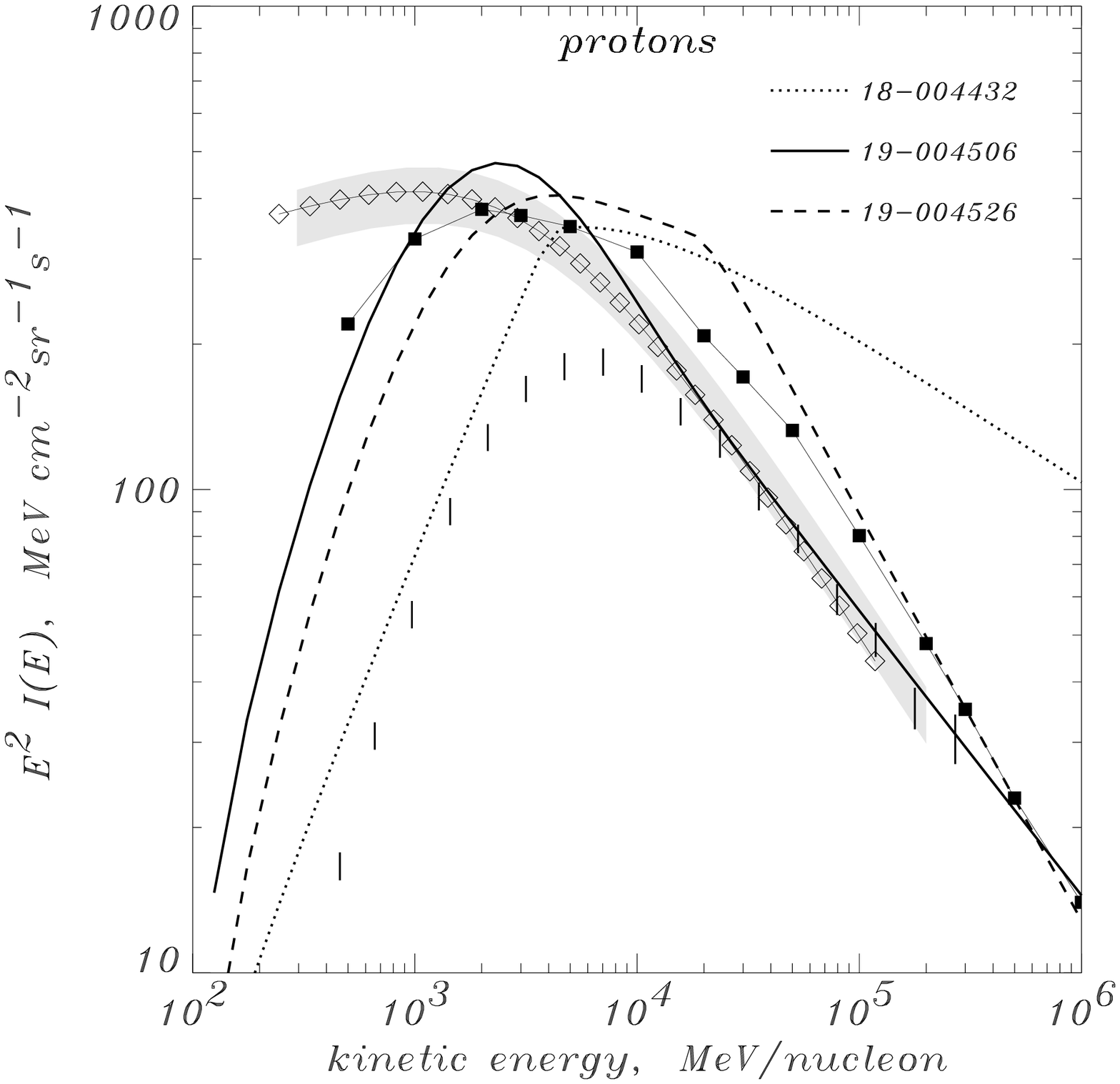,width=\fwb,clip=} }
\figcaption[fig4.ps]{ 
Proton spectra as obtained after
propagation in our models compared with IMAX data and published
estimates of the interstellar spectrum.  Solid line: using
power-law injection spectrum (models C, HE), dashed line: with
break in injection spectrum at 20 GeV (model HEMN), dotted line:
hard nucleon spectrum (model HN).  Vertical bars: IMAX direct
measured values (\protect\cite{Menn97}).  Evaluations of the
interstellar spectrum:  shaded area: based on IMAX data
(\protect\cite{Menn00}), connected filled squares: Webber \&
Potgieter (1989) and Webber (1998), connected open diamonds:
based on LEAP and IMP-8 (\protect\cite{Seo91}).
\label{Fig_protons}}
\end{figure*}

%######################################################################
We consider 6 different models to illustrate the possible options
available.  They differ mainly in their assumptions about  the
electron and nucleon spectra.  The parameters of the models and the
main motivation for considering each one are summarized in Table 1.
The electron and proton spectra and the synchrotron spectral index for
all these models are shown in Figs.~\ref{Fig_electrons},
\ref{Fig_sync_index}, and \ref{Fig_protons}.

\placetable{table1}

\begin{deluxetable}{cccccccl}% NB must have correct number of columns
\tablefontsize{\footnotesize}
\tablecolumns{7} \tablewidth{0mm} \footnotesize \tablecaption{
Parameters and objectives of models. \label{table1}} \tablehead{%
\colhead{} & \colhead{} & \colhead{} & \colhead{} &
\multicolumn{3}{c}{Injection index} & \colhead{} \\ 
\cline{5-7}
\colhead{Model\tablenotemark{a}} & \colhead{GALPROP} & \colhead{$z_h$}
& \colhead{$D_0$} & \colhead{} & \colhead{} & \colhead{} &
\colhead{Motivation/Comments}  \\ 
\colhead{} & \colhead{code} &
\colhead{kpc} & \colhead{cm$^2$ s$^{-1}$} & \colhead{electrons} &
\colhead{protons} & \colhead{He} & \colhead{} } 
\startdata
%%%%
C       & 19-004508&  4 & $6\times10^{28}$ & 1.6/2.6\tablenotemark{b}
& 2.25     & 2.45 &
\parbox[t]{55mm}{matches local electron, nucleon data and synchrotron;
consistent with $\bar{p}$ and $e^+$ constraints}
\medskip\nl
%%%%
HN\tablenotemark{f} & 18-004432&  4 & $3.5\times10^{28}$ 
& 2.0/2.4\tablenotemark{b}  & 1.7 & 1.7     &
\parbox[t]{55mm}{matches high-energy \grays using hard nucleon spectrum;
inconsistent with $\bar{p}$ and $e^+$ constraints}
\medskip\nl
%%%%
HE\tablenotemark{c} & 19-004512&  4 & $6\times10^{28}$ & 1.7    & 2.25
& 2.45    &
\parbox[t]{55mm}{matches high-energy \grays using hard electron spectrum}
\medskip\nl
%%%%
HEMN    & 19-004526&  4 & $6\times10^{28}$ & 1.8    &
1.8/2.5\tablenotemark{d} & 1.8/2.5\tablenotemark{d} &
\parbox[t]{55mm}{optimized to match high-energy \grays using hard
electron spectrum and broken nucleon spectrum;  consistent with
$\bar{p}$ and $e^+$ constraints}
\medskip\nl
%%%%
HELH    & 19-010526 & 10 & $12\times10^{28}$ & 1.8   &
1.8/2.5\tablenotemark{d} & 1.8/2.5\tablenotemark{d} &
\parbox[t]{55mm}{HEMN with large halo}
\medskip\nl
%%%%
SE      & 19-004606&  4 & $6\times10^{28}$ & 3.2/1.8\tablenotemark{e}
& 2.25     & 2.45    &
\parbox[t]{55mm}{matches low energy \grays using upturn in electron
spectrum} \nl  \enddata
%%%
\parbox{6in}{\tablenotetext{a}{\fns Propagation parameters are given in 
\cite{SM98} (C, HE, HEMN models: 15-004500; HELH: 15-010500; HN: 15-004100).
All models except SE and HN are with
reacceleration (Alfv\'en speed $v_A=20$ km s$^{-1}$). $D_0$ is
the diffusion coefficient at 3 GV (5 GV for HN model).  SE:
$\delta=1/3$, no reacceleration.}}
\tablenotetext{b}{\fns Electron injection index shown is below/above 10 GeV.}
\tablenotetext{c}{\fns Nucleon spectrum normalization is 0.8 relative 
to model C.} 
\tablenotetext{d}{\fns Injection index shown is below/above 20 GeV/nucleon.} 
\tablenotetext{e}{\fns Electron injection index shown is below/above 200 MeV.}
\tablenotetext{f}{\fns $\delta=-0.60/0.60$ below/above 5 GV, no convection.}
\end{deluxetable}

\placefigure{Fig_electrons} 
\placefigure{Fig_sync_index}
\placefigure{Fig_protons}

In model C (`conventional') the electron spectrum is adjusted to agree
with the locally measured one from 10 GeV to 1 TeV and to satisfy the
stringent synchrotron spectral index constraints.  We show that the
simple C model is inadequate for $\gamma$-rays; the remaining models
represent various possibilities for improvement.  Model HN (`hard
nucleon spectrum') uses  the same electron spectrum  as in model C,
while the nucleon spectrum is adjusted to fit the \gray emission above
1 GeV.  This model is tested against antiproton and positron data.  In
model HE (`hard electron spectrum') the electron spectrum is adjusted
to match the \gray emission above 1 GeV via IC emission, relaxing the
requirement of fitting the locally measured electrons above 10 GeV.
Model HEMN has the same electron spectrum as the HE model but has a
modified nucleon spectrum to obtain an improved fit to the \gray data.
Model HELH (`large halo') is like the HEMN model but with 10 kpc halo
height, to illustrate the possible influence on extragalactic
background estimates. Finally, in model SE (`soft electron spectrum')
a spectral upturn in the electron spectrum below 200 MeV is invoked to
reproduce the low-energy ($<$30 MeV) \gray emission without violating
synchrotron constraints.
 
Even given the particle injection spectra we still have the choice of
halo size and whether to include reacceleration. We have used
reacceleration models here except for the more exploratory cases HN
and SE.  The propagation is obviously  also subject to many
uncertainties.  The modelling of propagation can however simply be
seen as a way to obtain a physically motivated set of particle spectra
to be tested against \gray and other observations; in the end we test
just the ambient electron and nucleon spectra against the data,
independent of the physical nature of their origin.  In this sense our
investigation does not depend on the details of the propagation models
but still retains the constraints imposed by antiproton and positron
data.

\section{Synchrotron emission} \label{synch_emis}
%######################################################################
Observations of synchrotron intensity and spectral index provide
essential and stringent constraints on the interstellar electron
spectrum and on our magnetic field model. For this reason we discuss
it first, before considering the more complex subject of $\gamma$-rays.

The synchrotron emission in 10 MHz -- 10 GHz band constrains the
electron spectrum in the $\sim$1--10 GeV range (see e.g.\
\cite{Webber80}).  Out of the plane, free-free absorption is only
important below 10 MHz (e.g. \cite{Strong78}) and so can be neglected
here.  In particular the synchrotron spectral index
($T\propto\nu^{-\beta}$) provides information on the ambient electron
spectral index $\gamma$ in this range (approximately given by $\beta =
2 + {\gamma-1\over 2}$ but note that we perform the correct
integration over our electron spectra after propagation).

While there is considerable variation on the sky and scatter in the
observations, and local variations due to loops and spurs, it is
agreed that a general steepening with increasing frequency from
$\beta=2.5$ to $\beta=2.8-3$ is present.  Webber, Simpson, \& Cane
(1980) found $\beta= 2.57\pm0.03$ for 10--100 MHz.  Lawson et al.\
(1987) give values for 38--408 MHz between $\beta= 2.5$ and 2.6 using
drift-scan simulations which lead to more reliable results than the
original analyses (e.g.  \cite{Sironi74}: $\beta\sim 2.4$).   A recent
reanalysis of a DRAO 22 MHz survey (\cite{Roger99}) finds a rather
uniform 22 -- 408 MHz spectral index, with most of the emission
falling in the range $\beta = 2.40-2.55$.  Reich \& Reich (1988)
consider $\beta$(408--1420 MHz)$ = 3.1$ after taking into account
thermal emission.  Broadbent, Haslam, \& Osborne (1989) find
$\beta$(408-5000 MHz) $\sim 2.7$ in the Galactic plane, using far IR
data to model the thermal emission, but remark that 3.0  may be more
appropriate for a full sky average (cf. the Reich \& Reich value).
Davies, Watson, \& Guti\'errez (1996) find an index range for
408--1420 MHz of $\beta= 2.6 - 3.3$ for a high-latitude band, and
state that 3.0 is a typical value.  Recent new experiments give
reliable spectral indices up to several GHz (\cite{Platania98}); they
used a catalogue of HII regions to account for thermal emission.

Fig.~\ref{Fig_sync_index} summarizes these estimates of the Galactic
nonthermal spectral index as a function of frequency.  Since the
electron spectrum around 1 GeV is steepened both by energy losses and
energy-dependent diffusion, we can conclude from the low-frequency
$\beta\sim2.5$ that the {\it injection} spectrum must have $\gamma\le
2.0$.  In fact our models require an injection $\gamma = 1.6-1.8$ to
compensate the steepening and give reasonable agreement with the
observed $\beta(\nu)$. Since all the models we will describe are
chosen to have this injection index {\it in the energy range producing
radio synchrotron}, they are all consistent with the synchrotron index
constraints.

The comparison with models also depends on the $z$-distribution of the
magnetic field, since this affects how the spectral index is weighted
with $z$, and will give larger indices for larger extents of $B$ due
the spectral steepening with $z$.  Since the $z$-variation of $B$ is
unknown and otherwise plays a rather secondary r\^ole in our model we
use our predicted $\beta$ just for a representative intermediate
Galactic direction ($l=60^\circ, b=10^\circ$), which is taken as
typical of the data with which we compare.  The analysis is quite
insensitive to the choice of direction.

\begin{figure*}[thb]%***************************************************** 5
\centerline{ 
\psfig{file=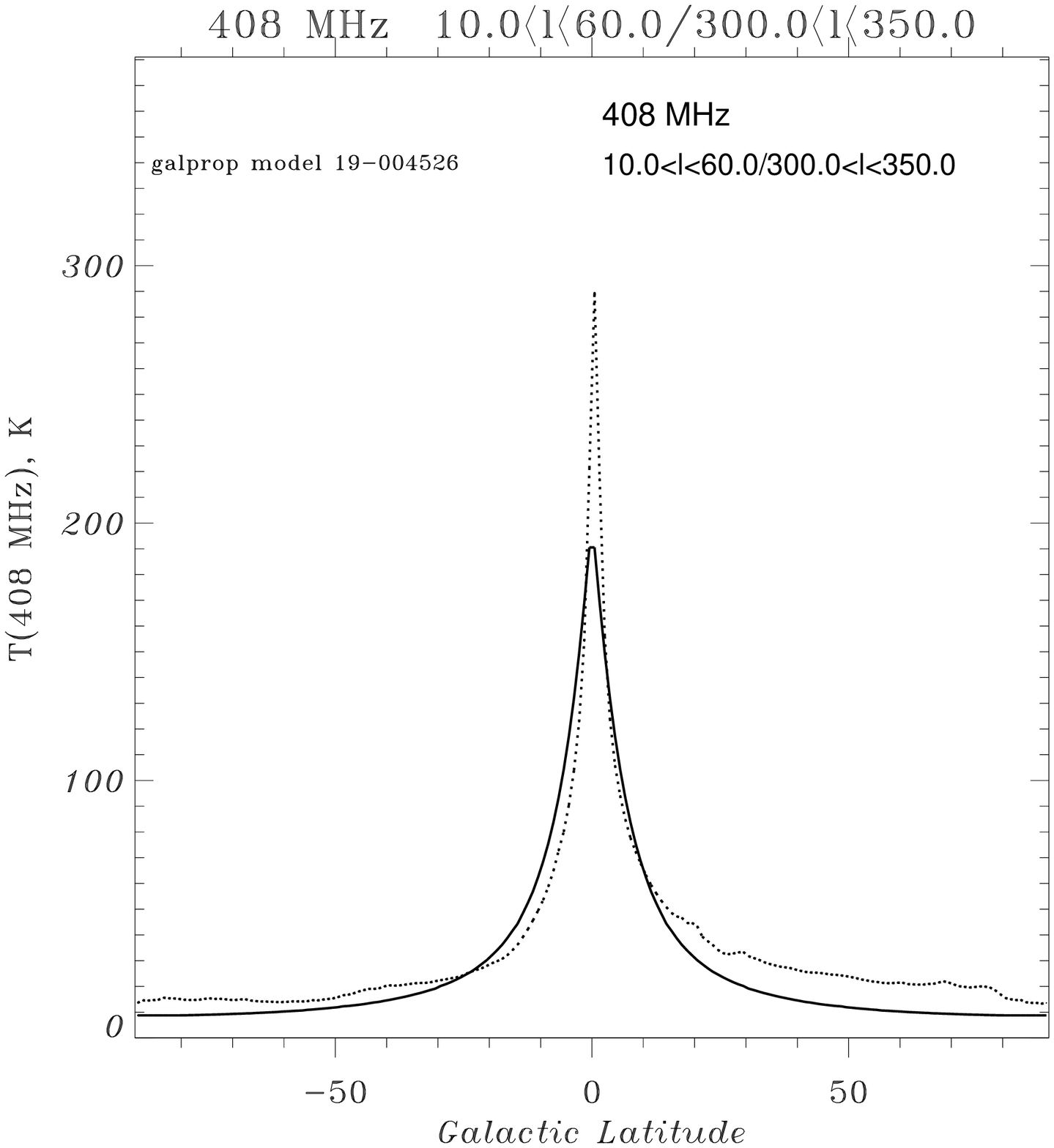,width=\fwb,clip=}
\psfig{file=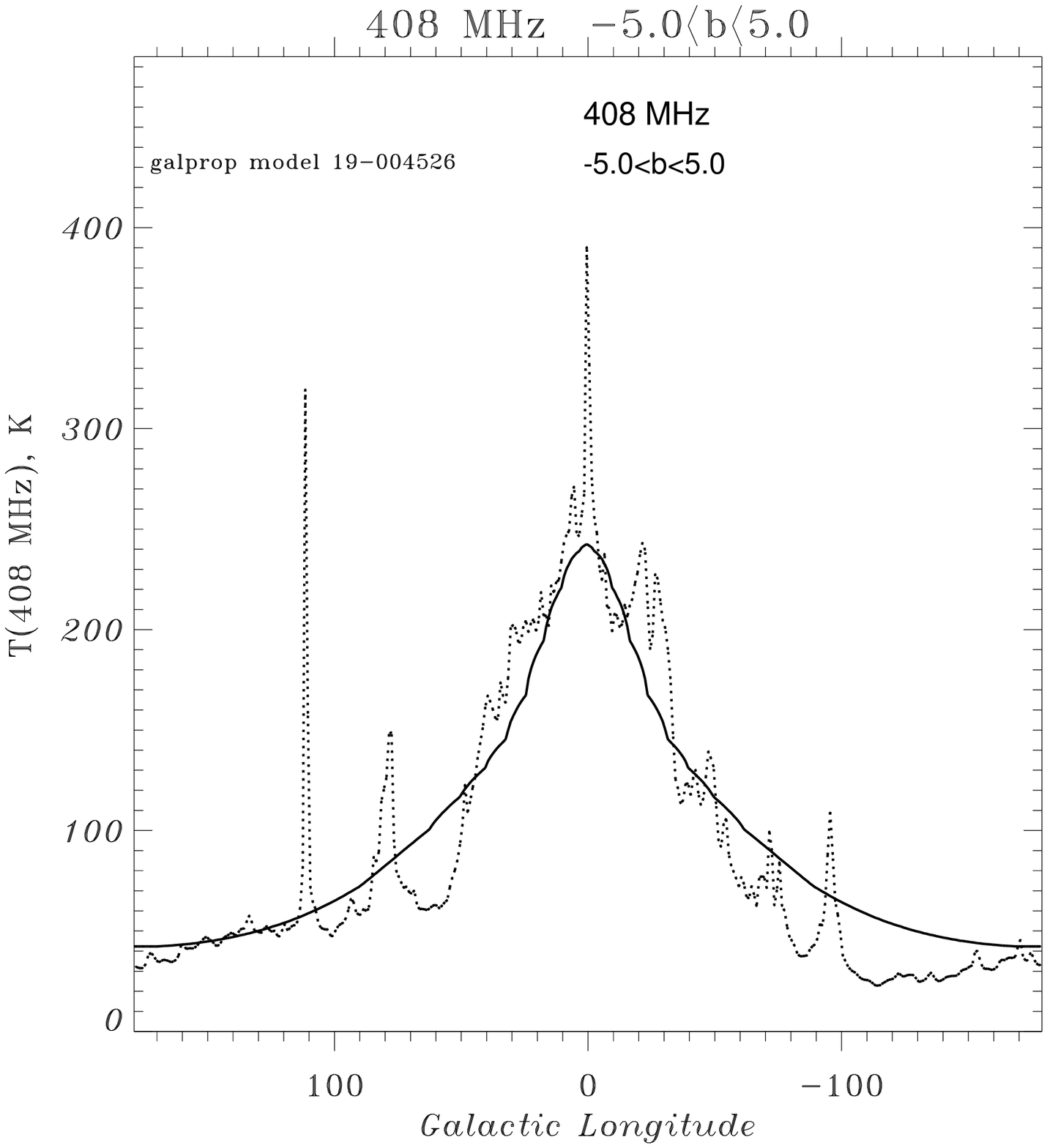,width=\fwb,clip=} }
\figcaption[fig5a.ps,fig5b.ps]{
Intensity profiles of synchrotron emission at 408 MHz in
latitude $(10^\circ \le l\le 60^\circ, 300^\circ \le l\le
350^\circ)$ and longitude ($|b|\le 5^\circ$) for the HEMN model.
Data:  Haslam et al.\ (1982).
\label{Fig_sync_profile}}
\end{figure*}

\begin{figure*}[thb]%***************************************************** 6
\centerline{ 
\psfig{file=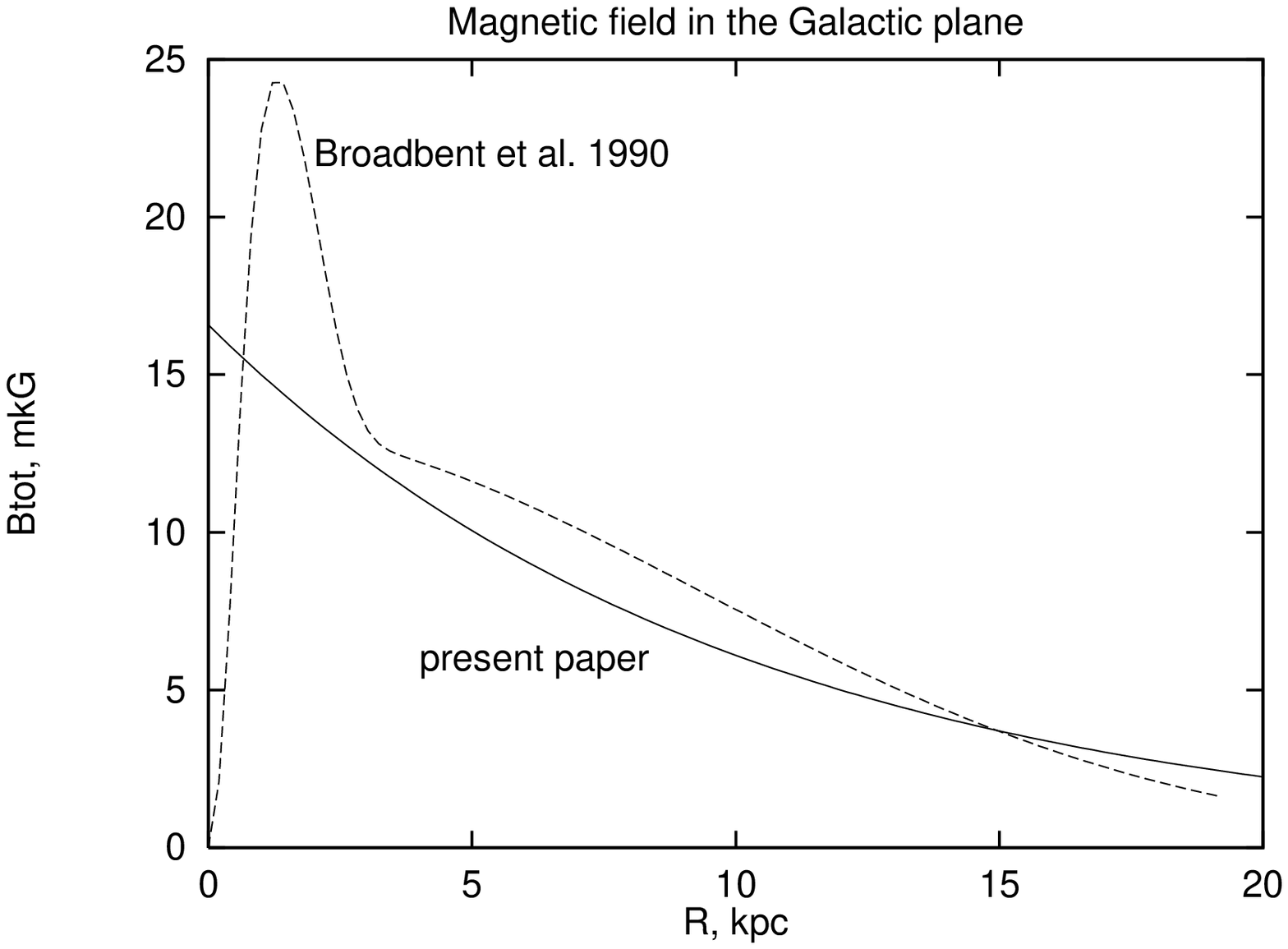,width=\fwb,clip=}
      \hspace{\hs}
\psfig{file=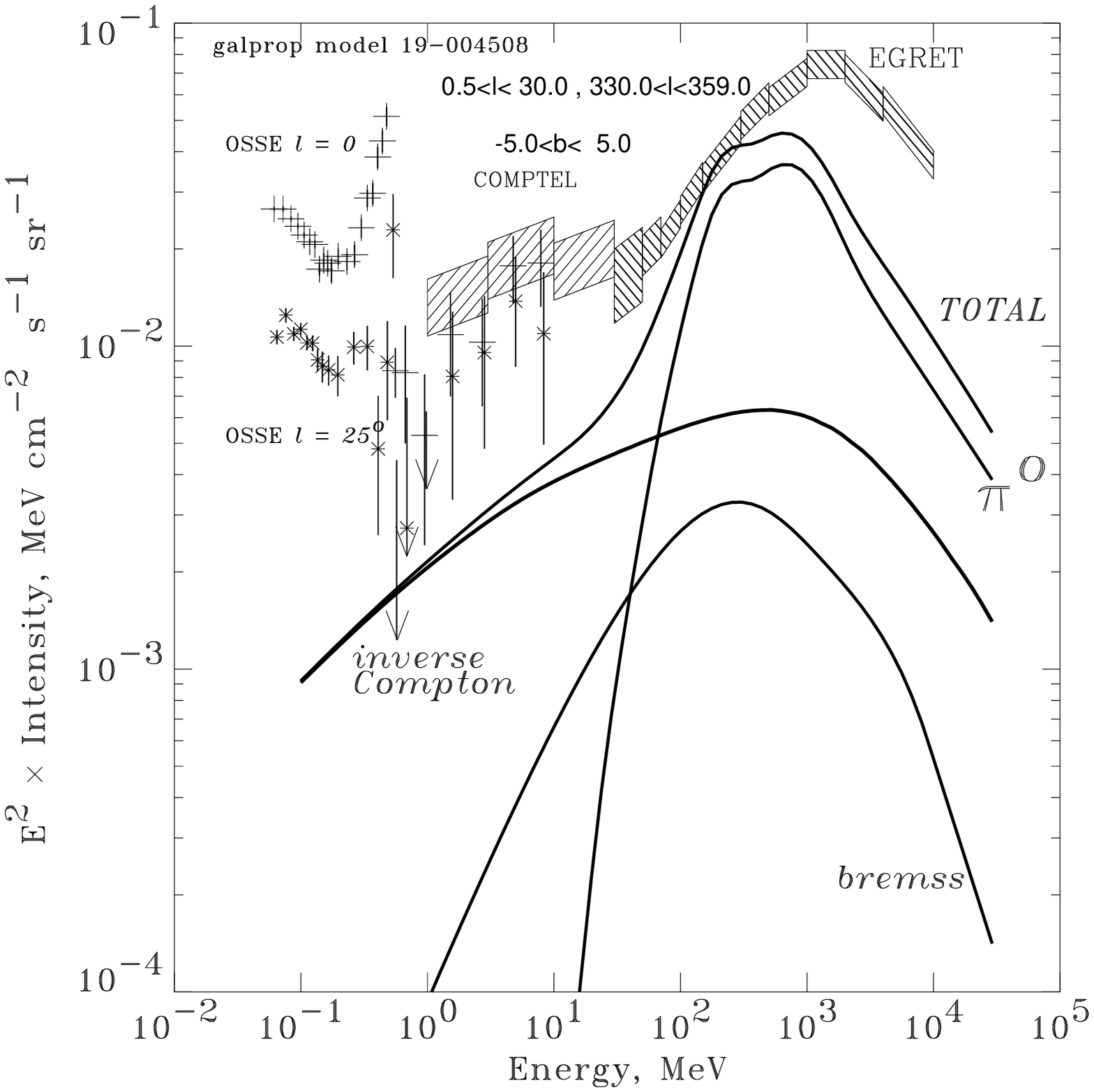,width=\fwb,clip=} } 
\parbox{89mm}{%
\figcaption[fig6.ps]{  
Magnetic field ($B_{tot}$)
distribution at $z=0$ as used in our models (solid line)
compared to parametrization by Broadbent et al.\ (1990) (dashes).
\vspace{4\baselineskip}\vspace{1pt}
\label{Fig_magnetic_field}}
}\hspace{7mm}
\parbox{89mm}{%
\figcaption[fig7.ps]{ 
Gamma-ray energy spectrum of
the inner Galaxy ($300^\circ \le l\le 30^\circ$, $|b|\le 5^\circ$)
compared with calculations in C model (the injection spectrum chosen
to fit local electron spectrum and synchrotron index). Curves show the
contribution of IC, bremsstrahlung, and $\pi^0$-decay, and the total.
Data: EGRET (\protect\cite{StrongMattox96}), COMPTEL
(\protect\cite{Strongetal99}), OSSE ($l=0, 25^\circ$:
\protect\cite{Kinzer99}).
\label{Fig_gamma_spectrum_C}}
}
\end{figure*}

\placefigure{Fig_sync_profile} 
\placefigure{Fig_magnetic_field}

\placefigure{Fig_gamma_spectrum_C} 

We evaluate  synchrotron emission in more detail only for our model
with  hard electron and modified nucleon injection spectra (HEMN)
since this is preferred from our \gray analysis in the following
sections.  The values of the parameters adopted in equation
(\ref{eq.1}), $B_0$ = 6.1$\mu$G,  $R_B$ = 10 kpc, $z_B$ = 2 kpc, were
found to reproduce sufficiently well the synchrotron index
(Fig.~\ref{Fig_sync_index}), and the absolute magnitude and profiles
of the 408 MHz emission (\cite{Haslam82}) as shown in
Fig.~\ref{Fig_sync_profile}.  The thermal contribution in the plane at
this frequency is only about $\sim$15\% (\cite{Broadbent89}).   A
significantly smaller field would give too low synchrotron intensites
as well as a spectral index distribution which disagrees with the
data, shifting the curve in the $\beta(\nu)$ plot to the left.  $R_B$
is constrained by the longitude profile, and $z_B$ by the latitude
profile of synchrotron emission.

For comparison,  Heiles (1996) gives $B_0\sim$ 5$\mu$G for the volume
and azimuthally averaged (uniform + random) field at the solar
position based on pulsar rotation measures and synchrotron
data. Vall\'ee (1996) gives similar values.  Our $B$ value follows
from the attempt to include \gray information on the electron spectrum
throughout the Galaxy and is consistent with these other estimates.
The radial distribution and magnitude of the magnetic field is also
consistent with that used by Broadbent et al.\ (1990), as shown in
Fig.~\ref{Fig_magnetic_field}.

Our model cannot reproduce the asymmetries in latitude or fine details
of the longitude distribution of synchrotron emission and this is not
our goal.  An exact fit to the profiles, involving spiral structure as
well as explicit modelling of random and non-random field components,
as in Phillipps et al.\ (1981), Broadbent et al.\ (1990), Beuermann,
Kanbach, \& Berkhuijsen (1985), is not attempted here.

\section{Gamma-ray data} \label{gray_data}
%######################################################################
For comparison of longitude and latitude profiles we use EGRET data
from Cycle 1--4 in the form of standard counts and exposure maps in 10
energy ranges (bounded by 30, 50, 70, 100, 150, 300, 500, 1000, 2000,
4000, and 10000 MeV).   The systematic errors in the EGRET profiles
due to calibration uncertainty and time-dependent sensitivity
variations are up to 13\% (\cite{Sreekumar98}).  These dominate over
statistical errors in the EGRET profiles.

The contribution from point sources was removed using the following
procedure. Point-like \gray excesses were determined in four energy
regimes (30--100, 100--300, 300--1000, and $\ge$1000 MeV) using a
likelihood method (\cite{Mattox96}). A detection threshold similar to
EGRET source catalogs was applied, and $\sim280$ sources were selected
by comparing with sources from the 3rd EGRET catalog
({\cite{Hartman99}).  Using either the catalog spectral indices, or a
--2.0 spectral index when no spectrum could be obtained, the flux of
each source was subsequently integrated for the standard energy
intervals.  For the three brightest sources on the sky (Vela, Crab,
Geminga pulsar)  the flux determined in each energy interval was used
directly.  The simulated count distributions of the selected sources
were subtracted from the summed count maps of the EGRET Cycle 1--4
data.

The \gray sky maps computed in our models are convolved with the EGRET
point-spread function generated for an $E^{-2}$ input spectrum (the
convolution is insensitive to the exact form of the spectrum).  For
spectral comparison at low latitudes it is better to use results based
on multicomponent fitting, which accounts for the angular resolution
of the instrument; here we use the results of Strong \& Mattox (1996),
synthesizing the skymaps of Galactic emission from their model
components and parameters.  Cross-checks between this approach and the
direct method show excellent agreement.  At high latitudes, where the
convolution has negligible effect, we generate spectra directly from
the  EGRET Cycle 1--4 data described above.

For energies below 30 MeV only spectral data for the inner Galaxy
($330^\circ\le l\le 30^\circ, |b|\le 5^\circ$) are considered,
COMPTEL: Strong et al.\ (1999), OSSE: Kinzer, Purcell, \& Kurfess (1999).  
The COMPTEL low-latitude spectrum is a recent improved analysis which is
about a factor 2 above that given in Strong et al.\ (1997) and there
remains some uncertainty as discussed in Strong et al.\ (1999);
however the difference has negligible effect on our conclusions.  The
COMPTEL high-latitude spectra are from Bloemen et al.\ (1999),
Kappadath (1998), and Weidenspointner et al.\ (1999).

\section{Model C (conventional model)}
%######################################################################
We start with a `conventional' model which reproduces the local
directly measured electron, proton, and Helium spectra above 10 GeV
(where solar modulation is small) and which also satisfies the
synchrotron constraints.  The propagation parameters are taken from
\cite{SM98}.  This model has $z_h = 4$ kpc, reacceleration with $v_A =
20$ km s$^{-1}$ and a normalization chosen to best fit the local
electron spectrum above 10 GeV.  A break in the injection spectrum  is
required to fit both the synchrotron spectrum and the directly
measured electron spectrum;   we adopted a steepening from  $-1.6$ to
$-2.6$ at 10 GeV.  The local electron spectrum, the synchrotron
spectral index, the local proton spectrum, and the \gray spectrum of
the inner Galaxy, are shown in Figs.~\ref{Fig_electrons},
\ref{Fig_sync_index}, \ref{Fig_protons}, and
\ref{Fig_gamma_spectrum_C}, respectively.

\placefigure{Fig_gamma_spectrum_GE} 
\placefigure{Fig_antiprotons}
\placefigure{Fig_positrons}

\begin{figure*}[p]%***************************************************** 8
\centerline{
\psfig{file=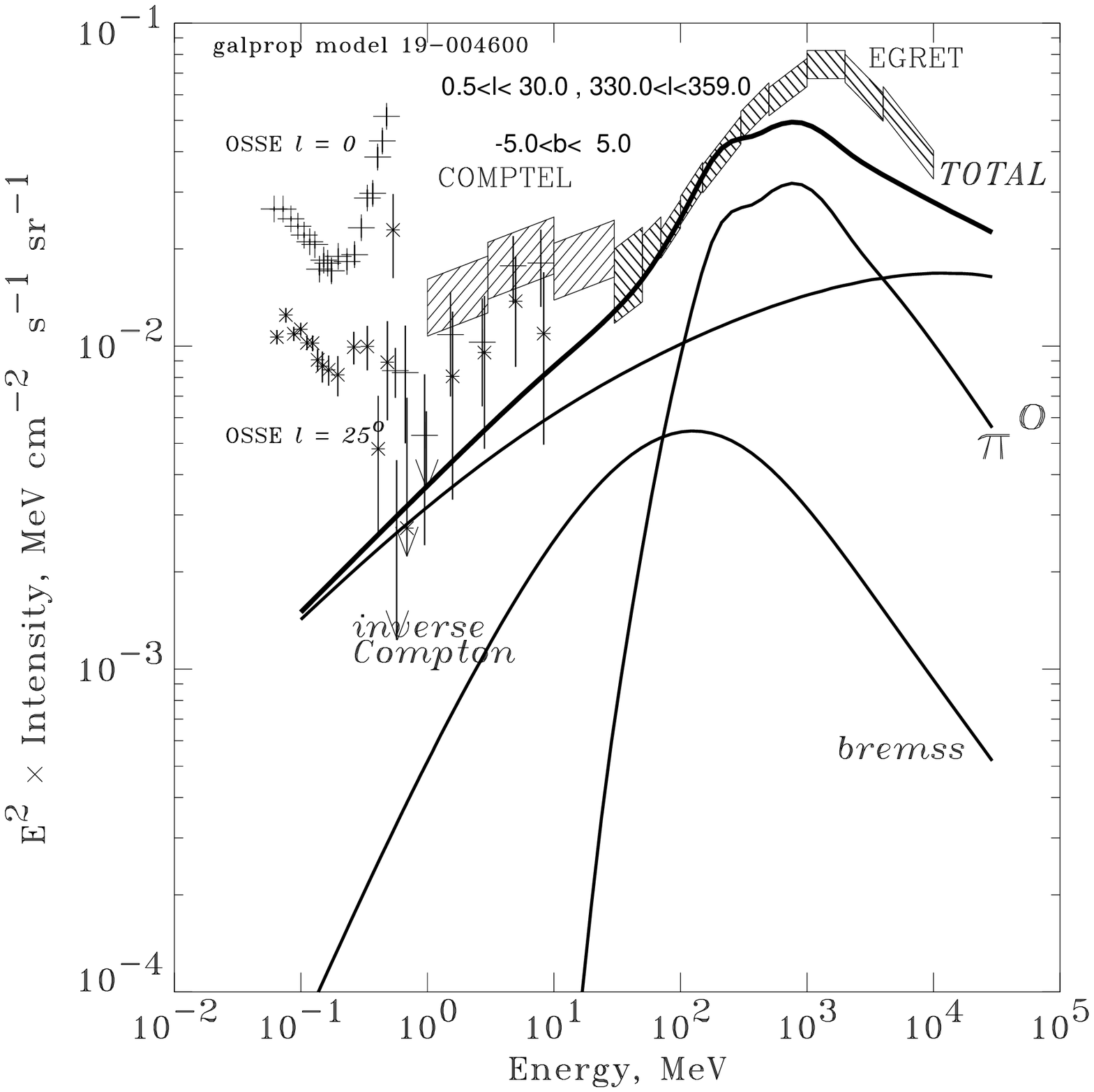,width=\fwd,clip=}
\psfig{file=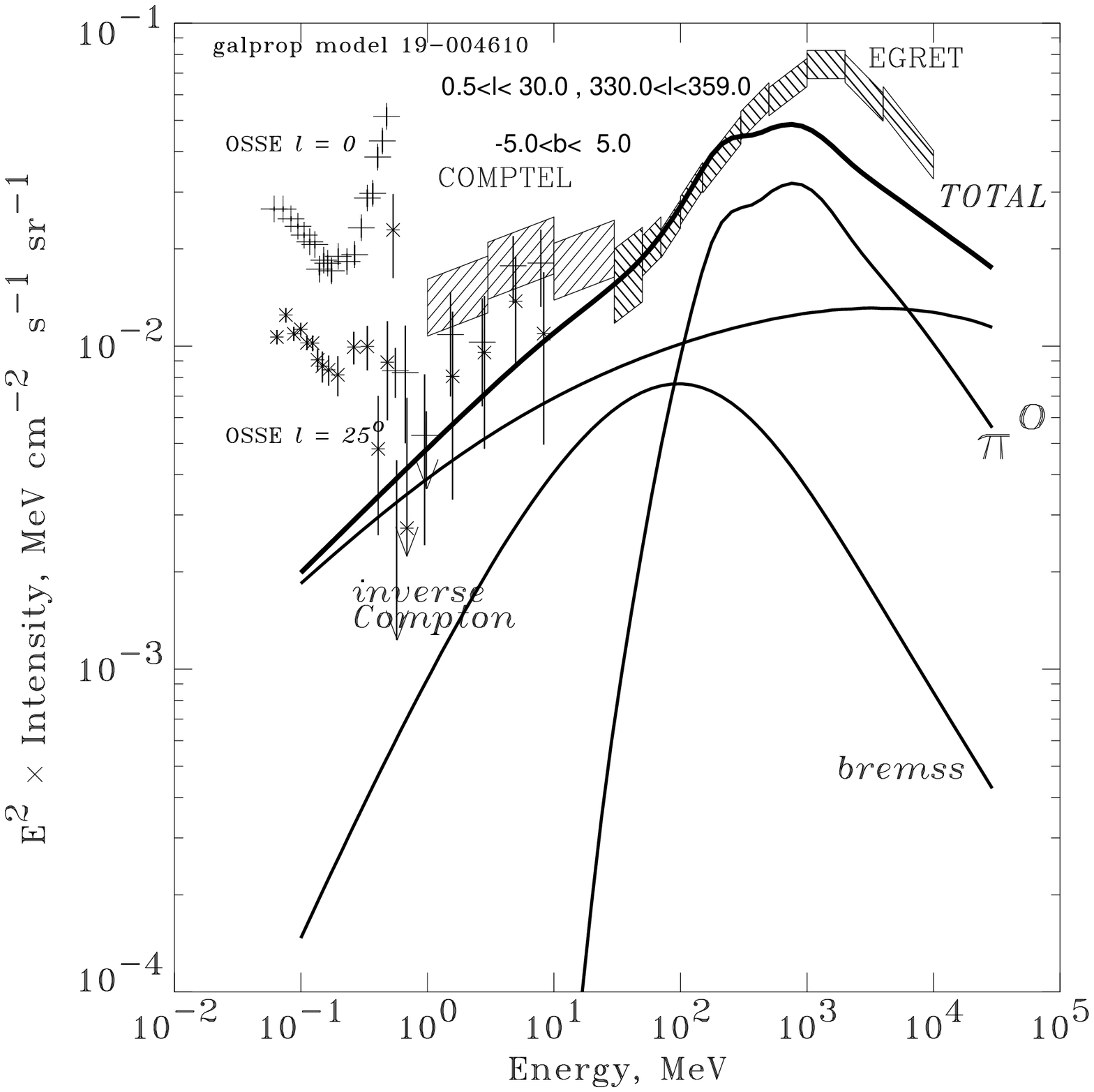,width=\fwd,clip=}}
\centerline{
\psfig{file=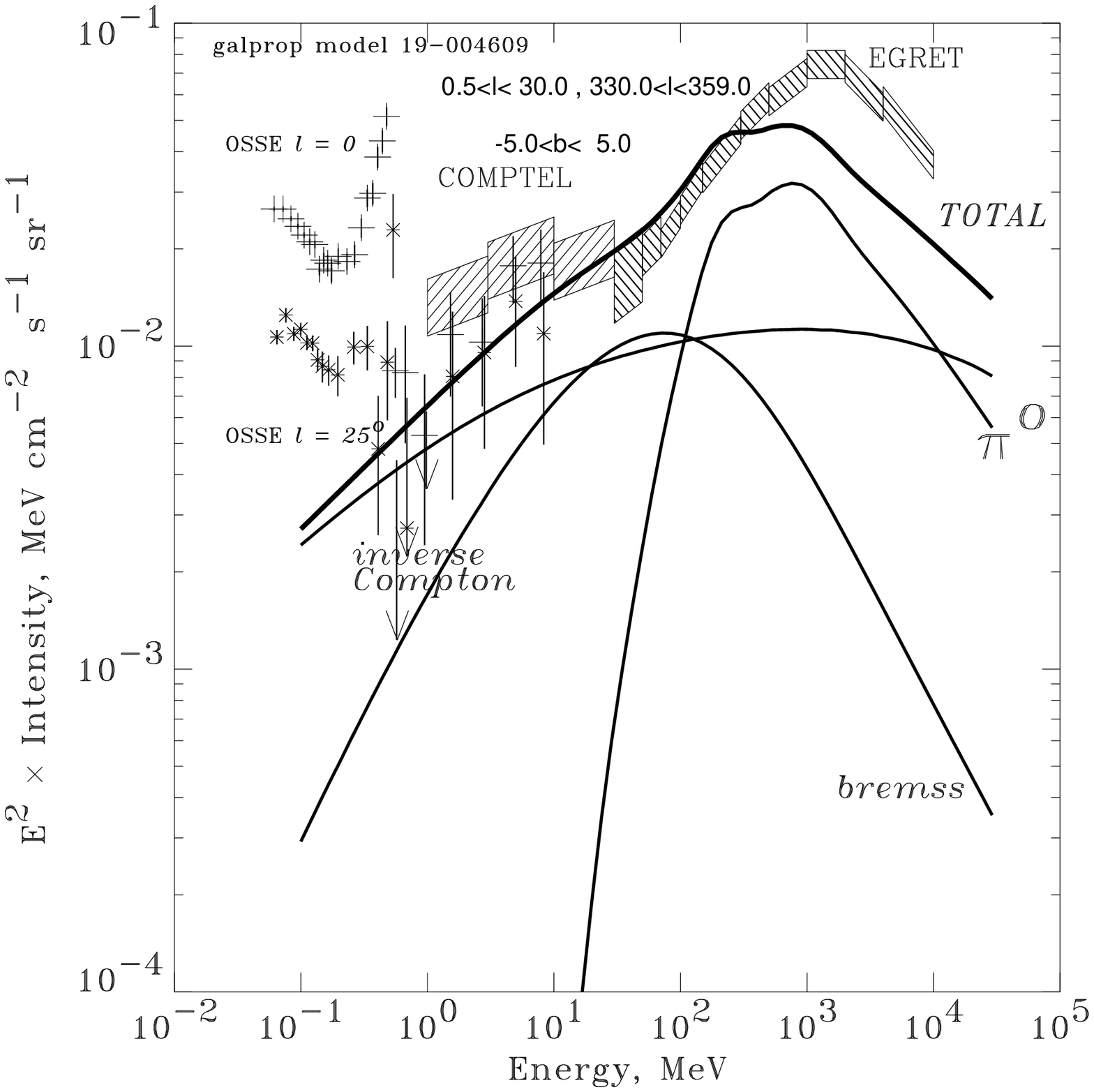,width=\fwd,clip=}
\psfig{file=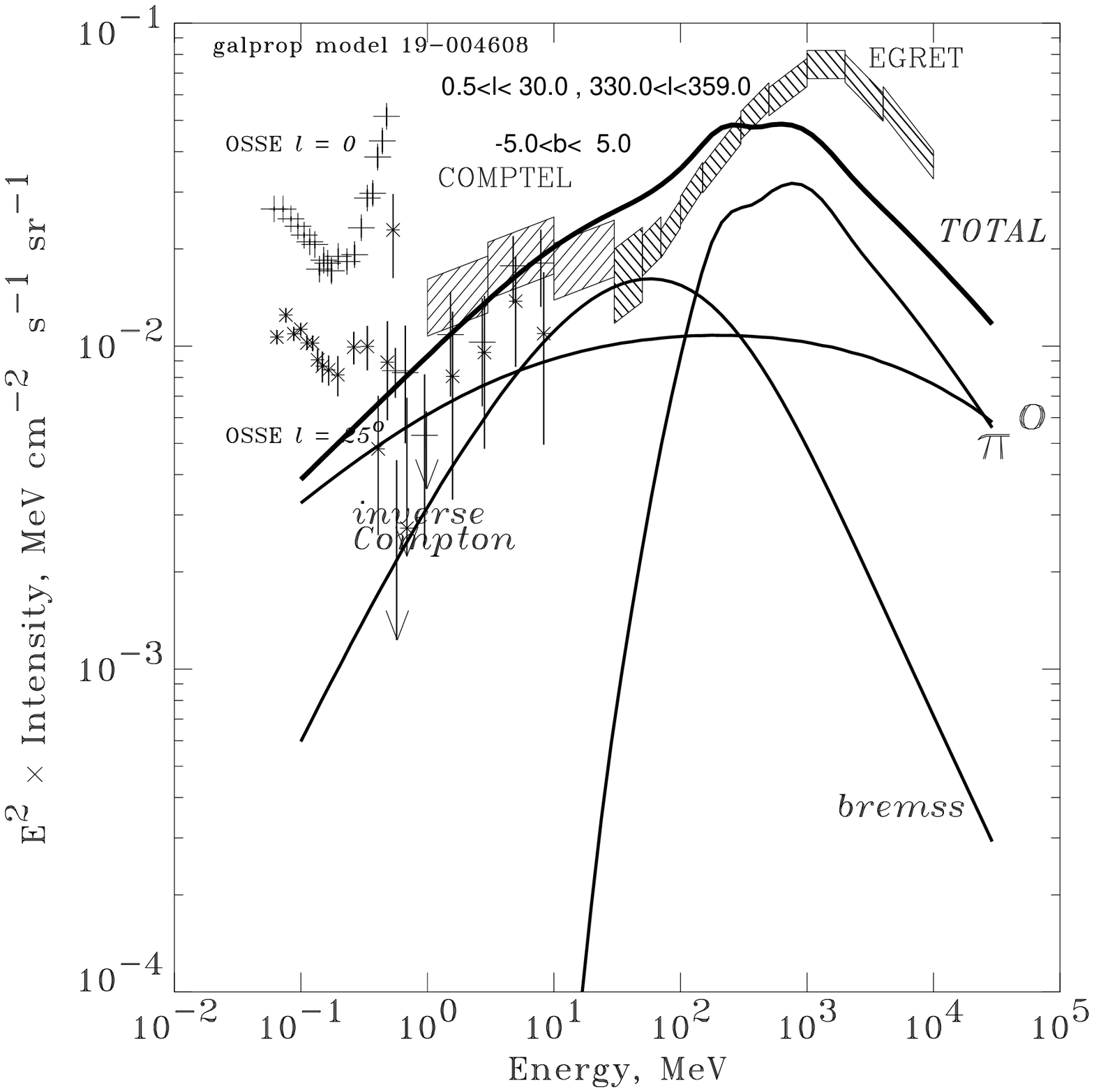,width=\fwd,clip=}}
\centerline{
\psfig{file=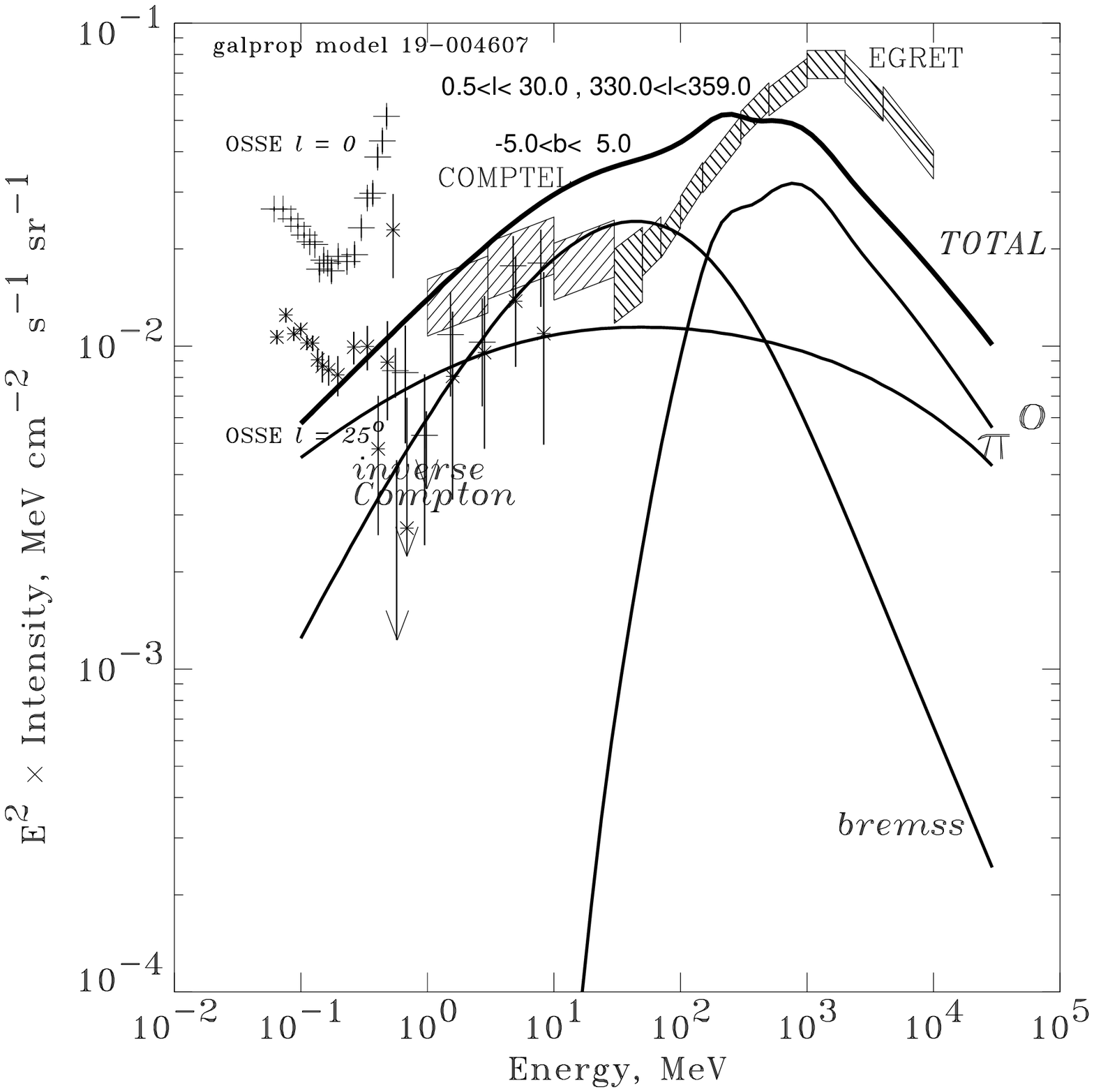,width=\fwd,clip=}}
\figcaption[fig8a.ps,fig8b.ps,fig8c.ps,fig8d.ps,fig8e.ps]{ 
Gamma-ray energy spectrum of the
inner Galaxy ($300^\circ \le l\le 30^\circ$, $|b|\le 5^\circ$)
compared with calculations for electron injection spectral
indices 2.0--2.4 (from left to right, top to bottom) as shown in
Fig.~\ref{Fig_sync_index}.  Curves show the contribution of IC,
bremsstrahlung, and $\pi^0$-decay, and the total.  Data: as in
Fig.~\ref{Fig_gamma_spectrum_C}.  
\label{Fig_gamma_spectrum_GE}}
\end{figure*} 

\begin{figure*}[tbh]%***************************************************** 9
\centerline{ 
\psfig{file=fig9.ps,width=\fwb,clip=}
      \hspace{\hs}
\psfig{file=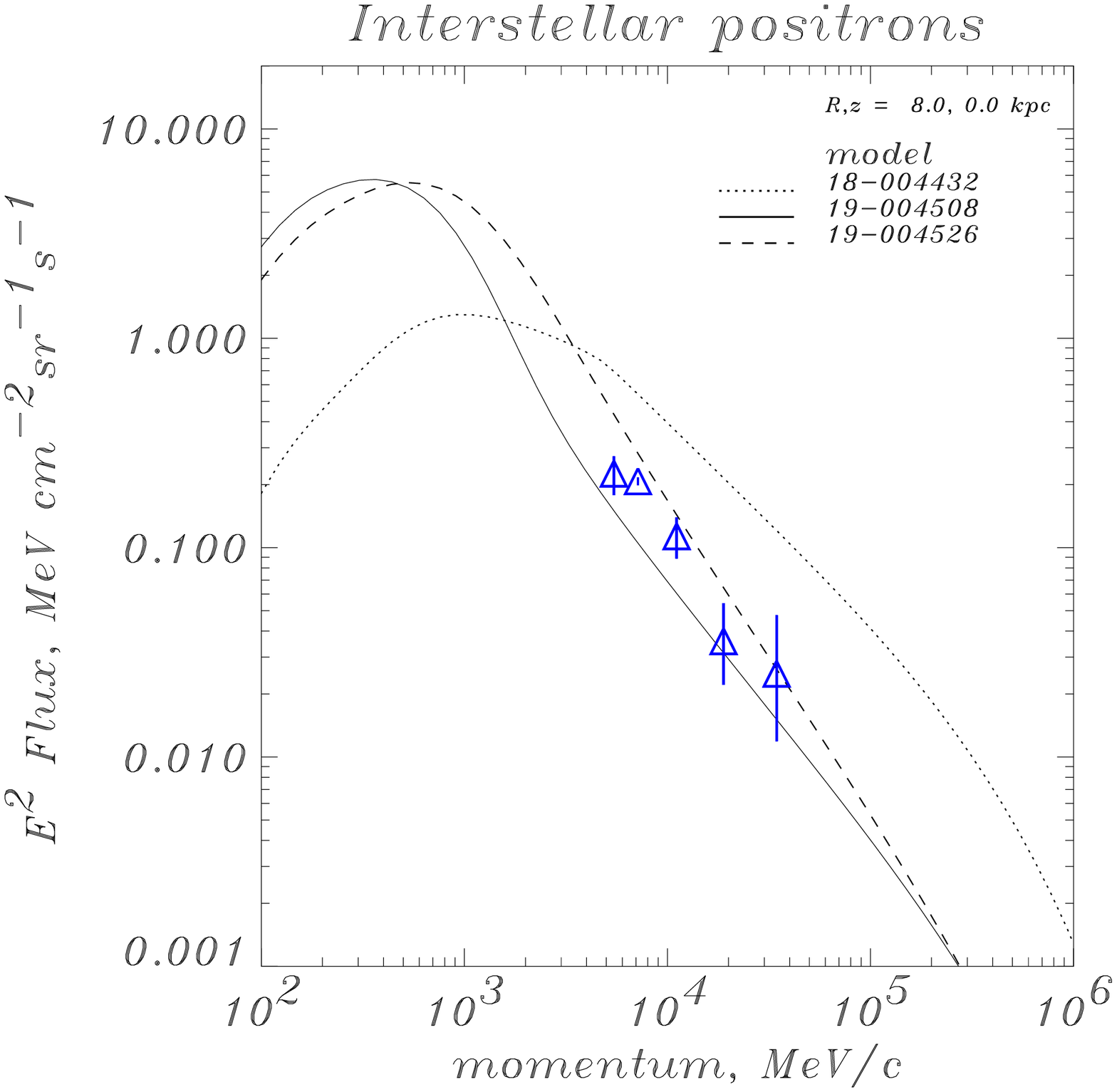,width=\fwb,clip=} }
\parbox{89mm}{%
\figcaption[fig9.ps]{ 
Interstellar antiproton flux for models 
(\protect\cite{MSR98}) compared with local measurements.
Solid line: C, dotted line: HN, dashed line: HEMN.  Data (direct
measurements): triangles: MASS91 experiment (\protect\cite{Basini99}),
other points: see references in
\protect\cite{MSR98}.
\vspace{1pt}
\label{Fig_antiprotons}}
}\hspace{7mm}
\parbox{89mm}{%
\figcaption[fig10.ps]{
Interstellar positron spectra for
models compared with data.  Solid line: model C, dotted line:
HN, dashed line: HEMN.  Data (direct measurements): HEAT
experiment (\protect\cite{Barwick98}).
\vspace{1\baselineskip}
\label{Fig_positrons}}
}
\end{figure*}

Model C is based entirely on non-\gray data but still approximates the
\gray data within a factor 3 over three decades of energy (10 MeV to
10 GeV).  It also satisfies the limits imposed by antiprotons and
positrons (\cite{MSR98}, Moskalenko \& Strong 1998b, hereafter \cite{MS98b}), 
though our new
calculation\footnote{Small differences from \cite{MS98b} are due to
use of our new ISRF and magnetic field model which modifies the energy
losses, and use of $z_h$ = 4 kpc instead of 3 kpc, which changes the
propagation slightly.} shows some deficit of positrons below $\approx
10$ GeV (Figs.~\ref{Fig_antiprotons} and \ref{Fig_positrons}, see also
discussion in Section~\ref{hn}).   It still remains a useful first
approximation to serve as the basis for the developments which follow.

The fit to the inner Galaxy \gray spectra is satisfactory from 30 to
500 MeV but a large excess in the EGRET spectrum relative to the
predictions above 1 GeV is evident, as first pointed out by Hunter et
al.\ (1997).  Simple rescaling of either electron or nucleon spectra
does not allow the agreement to be signficantly improved.  Harder
nucleon or electron spectra are therefore investigated below.

The model also fails to account for the \gray intensities below 30 MeV
as observed by COMPTEL and OSSE; attempting to account for this with a
steeper electron spectrum immediately violates the synchrotron
constraints, unless the steepening occurs at electron energies below a
few hundred MeV, as discussed in Section~\ref{se} (model SE). 
To prove this important point, we consider a series of electron
injection indices 2.0--2.4 in a model without reacceleration and
diffusion coefficient index $\delta=1/3$; this conveniently
spans the range of reasonably simple electron
spectra. Fig.~\ref{Fig_electrons} (right) shows these electron
spectra, and Fig.~\ref{Fig_sync_index} (right) the corresponding
synchrotron indices. The corresponding gamma-ray spectra are shown in
Fig.~\ref{Fig_gamma_spectrum_GE}. It is clear that an index 2.2--2.3 is
required to fit the $\gamma$-rays, while this produces a synchrotron
index about 0.8 which is substantially above that allowed by the
data. Although this is for a particular propagation model, note that
any combination of injection and propagation which fits the \grays
will have the same problem, except for extreme models (e.g.\ our SE
model).

We emphasize that it is the synchrotron constraint on the electron
index which forces us to this conclusion; in the absence of this we
would be free to adopt a uniformly steep electron injection spectrum
to obtain a fit to the low-energy $\gamma$-rays. This is the essential
difference between present and earlier work (e.g.\ \cite{Strong96},
\cite{SM97}) where the consequences of the synchrotron constraints were
not fully appreciated.

\placefigure{Fig_gamma_spectrum_HN}

\begin{figure*}[tbh]%***************************************************** 11
\centerline{ 
\psfig{file=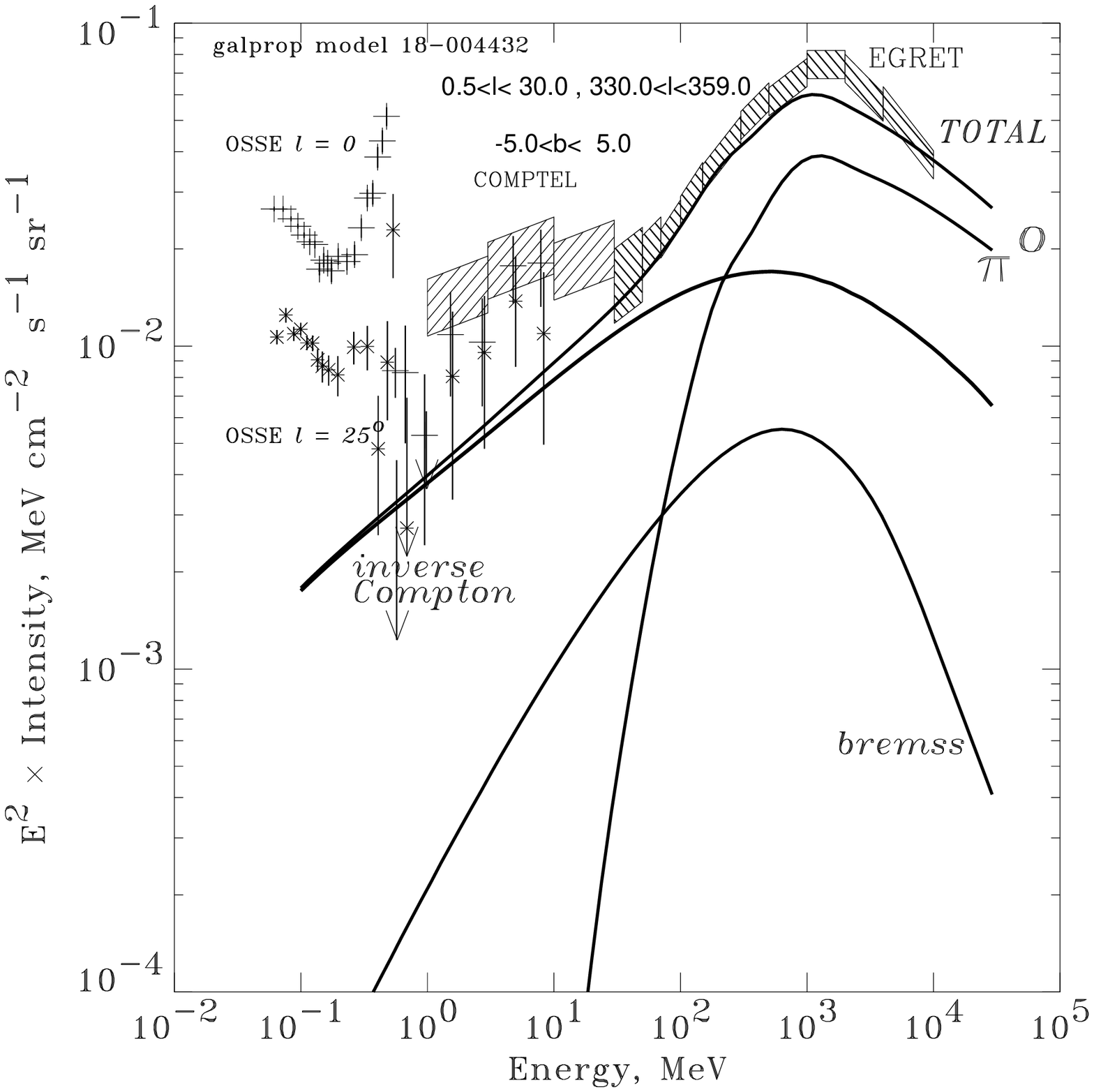,width=\fwb,clip=} }
\figcaption[fig11.ps]{ 
Gamma-ray data as in
Fig.~\ref{Fig_gamma_spectrum_C} compared with HN model (hard
nucleon spectrum chosen to fit EGRET data above 1 GeV).
\label{Fig_gamma_spectrum_HN}}
\end{figure*}

\section{Beyond the conventional models}
%######################################################################
\subsection{HN model (hard nucleon spectrum)} \label{hn}
%######################################################################
One possibility to reproduce the \gray excess above 1 GeV is to invoke
interstellar proton and Helium spectra which are harder than those
directly observed in the heliosphere (\cite{Gralewicz97},
\cite{Mori97}).  Spatial variations in the nucleon spectrum are quite
possible over the Galaxy so that such an option is worth serious
consideration.  This model has been studied in detail in \cite{MSR98}
in relation to antiprotrons, so that here we just summarize the
results and also extend the calculation to secondary positrons.  The
\gray spectrum of the inner Galaxy for a model with a hard nucleon
injection spectrum chosen to match the \grays (no reacceleration, proton
and He injection index = 1.7) is shown in
Fig.~\ref{Fig_gamma_spectrum_HN}.  The corresponding propagated
interstellar proton spectrum is shown in Fig.~\ref{Fig_protons}.

As pointed out in \cite{MSR98}, the same nucleons which contribute to
the GeV \gray emission through the decay of $\pi^0$-mesons produce
also secondary antiprotons and positrons (on the same interstellar
matter).  The harder nucleon spectrum hypothesis, therefore, can be
tested with measurements of CR $\bar{p}$'s and $e^+$'s (\cite{MSR98},
\cite{MS98b}).  Above $T_p\sim$ few 10 GeV (for a power-law proton
spectrum) the mean kinetic energy of parent protons is about 10 times
larger than that of produced secondary $\bar{p}$'s, and roughly the
same holds for $\gamma$-rays, so 10 GeV $\bar{p}$'s and $\gamma$'s
both are produced by $\sim100$ GeV nucleons. This relation is also
valid for secondary positrons.  Such tests are therefore well tuned,
and sample the Galactic-scale properties of CR $p$ and He rather than
just the local region, independent of fluctuations due to local
primary CR sources.  

The conclusion of \cite{MSR98}, based on the $\bar{p}/p$ ratio as
measured by Hof et al.\ (1996), was that antiprotons provide a sensitive 
test of the interstellar nucleon spectra, and that a hard nucleon
spectrum overproduces $\bar{p}$'s at GeV energies.
In Fig.~\ref{Fig_antiprotons} the predicted flux of antiprotons is 
compared with new {\it absolute} $\bar{p}$ fluxes above 3 GeV
from the MASS91 experiment (\cite{Basini99}). While the agreement is good
for the normal nucleon spectrum, the hard nucleon spectrum produces too
many antiprotons by a factor of $\sim5$, well outside the error bars
of the three data points. We conclude that such a hard nucleon spectrum
is inconsistent with the antiproton data.  

Fig.~\ref{Fig_positrons} shows the interstellar positron spectrum for
the conventional and hard nucleon spectra, where we used the formalism
given in \cite{MS98a}. The flux for the conventional case agrees with
recent data (\cite{Barwick98}) at high energies, where solar
modulation is not so important. For the hard nucleon spectrum the flux
is higher than observed by factor $\sim$4; this provides more evidence
against a hard nucleon spectrum.  However this test is less direct
than $\bar{p}$ due to the difference in particle type, the large
effect of energy losses, and the effect of solar modulation at lower
positron energies.

Taken together, the antiproton and positron data provide rather
substantial evidence against the idea of explaining the $>$1 GeV \gray
excess with a hard nucleon spectrum.

\placefigure{Fig_gamma_spectrum_HE}
\placefigure{Fig_gamma_spectrum_HEMN}

\begin{figure*}[tbh]%***************************************************** 12
\centerline{ 
\psfig{file=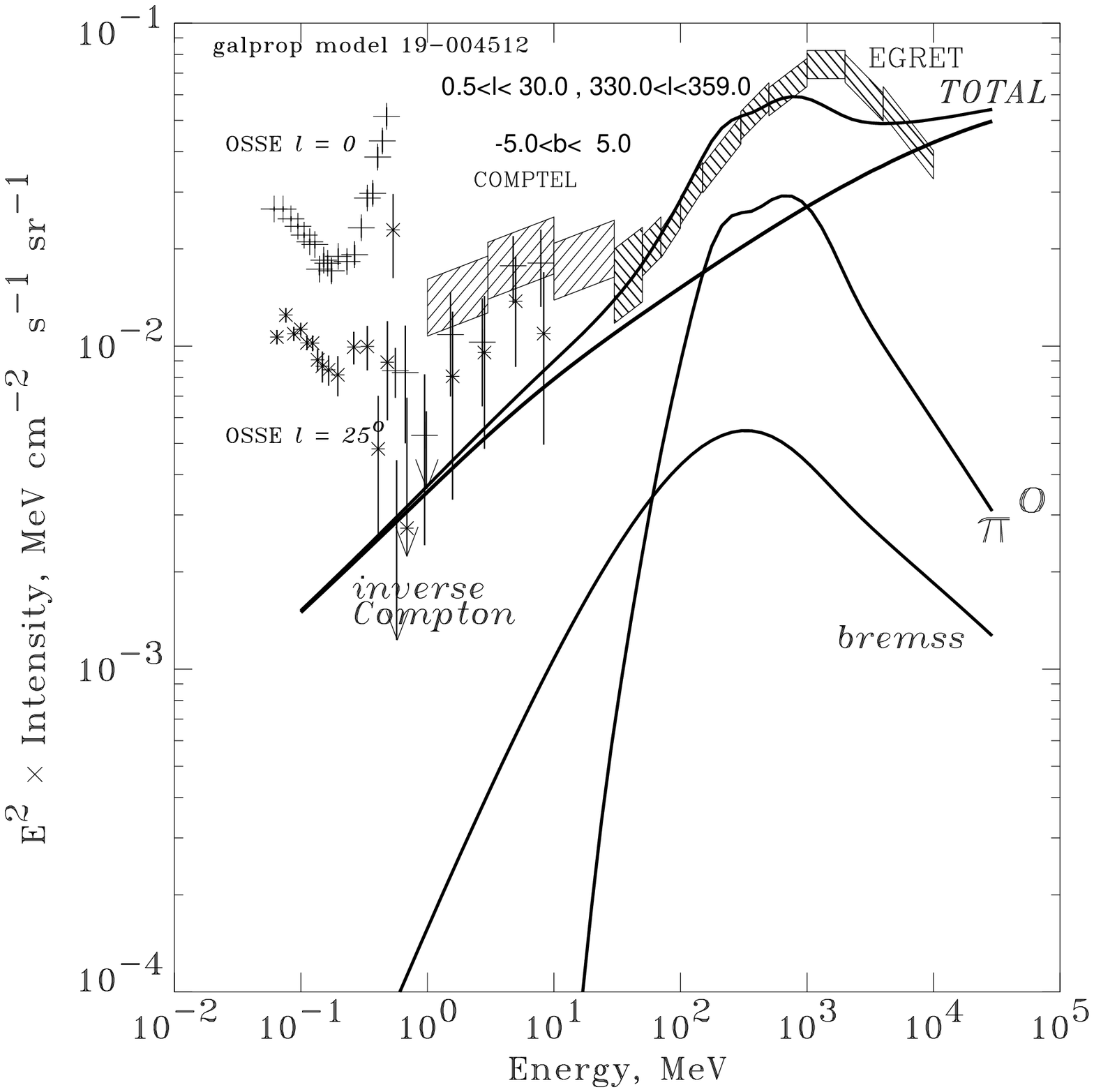,width=\fwb,clip=}
      \hspace{\hs}
\psfig{file=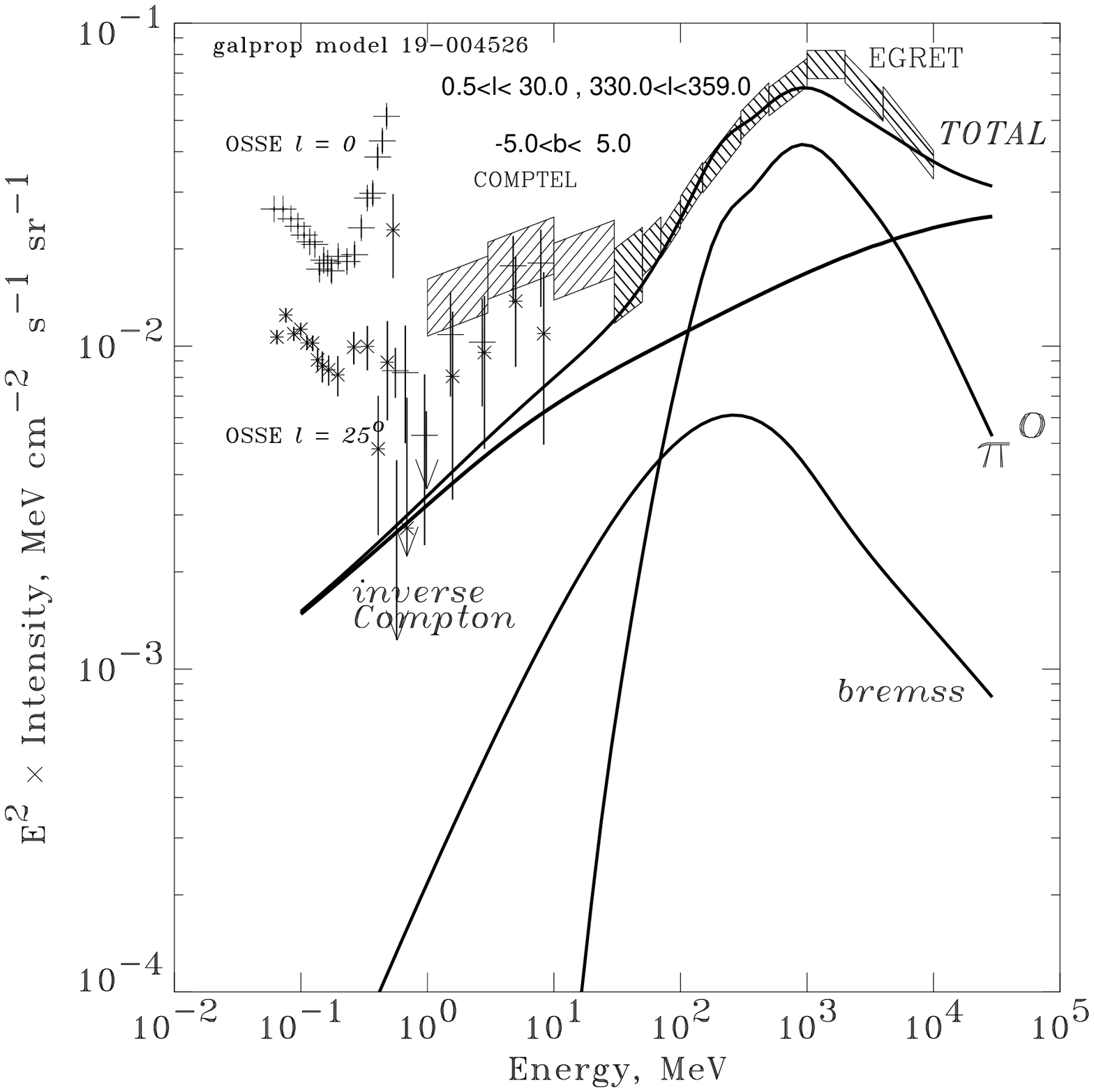,width=\fwb,clip=} }
\parbox{89mm}{%
\figcaption[fig12.ps]{ 
Gamma-ray data as in
Fig.~\ref{Fig_gamma_spectrum_C} compared with HE model (electron
injection index $-1.7$).
\label{Fig_gamma_spectrum_HE}}
}\hspace{7mm}
\parbox{89mm}{%
\figcaption[fig13.ps]{  
Gamma-ray data as in
Fig.~\ref{Fig_gamma_spectrum_C} compared with HEMN model
(electron injection index $-1.8$, and modified nucleon spectrum).
\label{Fig_gamma_spectrum_HEMN}}
}
\end{figure*}

Note that in these tests we assume that only the protons have a local
spectrum which is different from that on Galactic scales. We assume
that the propagation parameters can still be derived from B/C, so that
implicitly the heavier nuclei C, O are not affected.  We could
alternatively adopt a picture in which the C, O etc.\ also have a
local spectrum different from Galactic scales, but then fitting the
B/C ratio would imply $\delta > 0.6$ for the index in the diffusion
coefficient (nonreacceleration case) which is certainly problematic
for high-energy anisotropy (see \cite{SM98} and references therein)
and larger than predicted by standard diffusion theory.  Therefore we
consider the only case worth testing at this stage is the one where
only the protons are affected.  In any case, pursuit of more complex
options is beyond the scope of this paper.

\subsection{A harder interstellar electron spectrum (HE \& HEMN models)}
\label{hard_e}
%######################################################################
An obvious way to improve the fit to the EGRET data above 1 GeV is to
adopt a harder interstellar electron injection spectrum.  Such  models
will {\it not} match the directly-observed electron spectrum above 10
GeV, but this is not critical since the large energy losses in this
region mean that large spatial fluctuations are expected
(\cite{PohlEsposito98}).  Hence we relax the constraint of consistency
with the locally measured electron spectrum.

For the HE model we adjust the electron injection index and the
absolute electron flux to optimize the fit to the inner-Galaxy \gray
spectrum; also the absolute nucleon intensity ($\pi^0$-component) was
reduced slightly (factor 0.8) within the limits allowed by the proton
and Helium data.  The inner-Galaxy \gray spectrum is shown in
Fig.~\ref{Fig_gamma_spectrum_HE}.  This model with its harder electron
injection index ($-1.7$) satisfies the synchrotron constraints and also
leads to a better fit to the \grays above 1 GeV,  but produces too few
\grays at energies below 30 MeV by a factor 2--4.   In this case the
additional low energy \grays must be attributed to another component
as discussed in Section~\ref{se} in the context of SE model.

Pohl \& Esposito (1998) use an electron injection index 2.0 with a
Gaussian distribution of 0.2 for their \gray model. This has the
effect of an upwards curvature which is equivalent to a harder
spectrum similar to ours.  Quantitatively, from their Fig.~4, the
difference in effective index for their spiral arm model going from a
single index to the index with dispersion gives in fact a difference
0.2 at high energies. So the dispersed index is equivalent to a single
index, 2.0--0.2 = 1.8, which is similar to our 1.7 (HE model).

Since the fit of HE model to the EGRET detailed spectral shape is
still not very good above 1 GeV we can ask whether it can be improved
by allowing more freedom in the nucleon spectrum also (model HEMN).
Some freedom is allowed since solar modulation affects direct
measurements of nucleons below 20 GeV, and the locally measured
nucleon spectrum may not  necessarily be representative of the average
on Galactic scales either in spectrum or intensity due to details of
Galactic structure (e.g.\ spiral arms).  Because of the hard electron
spectrum the required modification to the nucleon spectrum is much
less drastic than in model HN.  By introducing an {\it ad hoc}
flattening of the nucleon spectrum below 20 GeV, a small steepening
above 20 GeV, and a suitable normalization,  an improved match to the
inner Galaxy EGRET spectrum is indeed possible
(Fig.~\ref{Fig_gamma_spectrum_HEMN}). The spectral parameters are
given in Table 1.  For the modified nucleon spectrum
(Fig.~\ref{Fig_protons})  we must invoke departures from cylindrical
symmetry so that the local value still agrees with direct measurements
at the solar position.  However this modification of the nucleon
spectrum must be checked against the stringent constraints on the {\it
interstellar} spectrum provided by antiprotons and positrons (as in
model HN).  The predictions of this model are shown in
Figs.~\ref{Fig_antiprotons} and \ref{Fig_positrons}.  As expected the
predictions are larger than the conventional model but still within
the antiproton and positron limits.

\placefigure{Fig_gamma_latitude_profile} 
\placefigure{Fig_gamma_longitude_profile}
\placefigure{Fig_gamma_high_latitude_profile}

\begin{figure*}[!p]%***************************************************** 14
\centerline{
\psfig{file=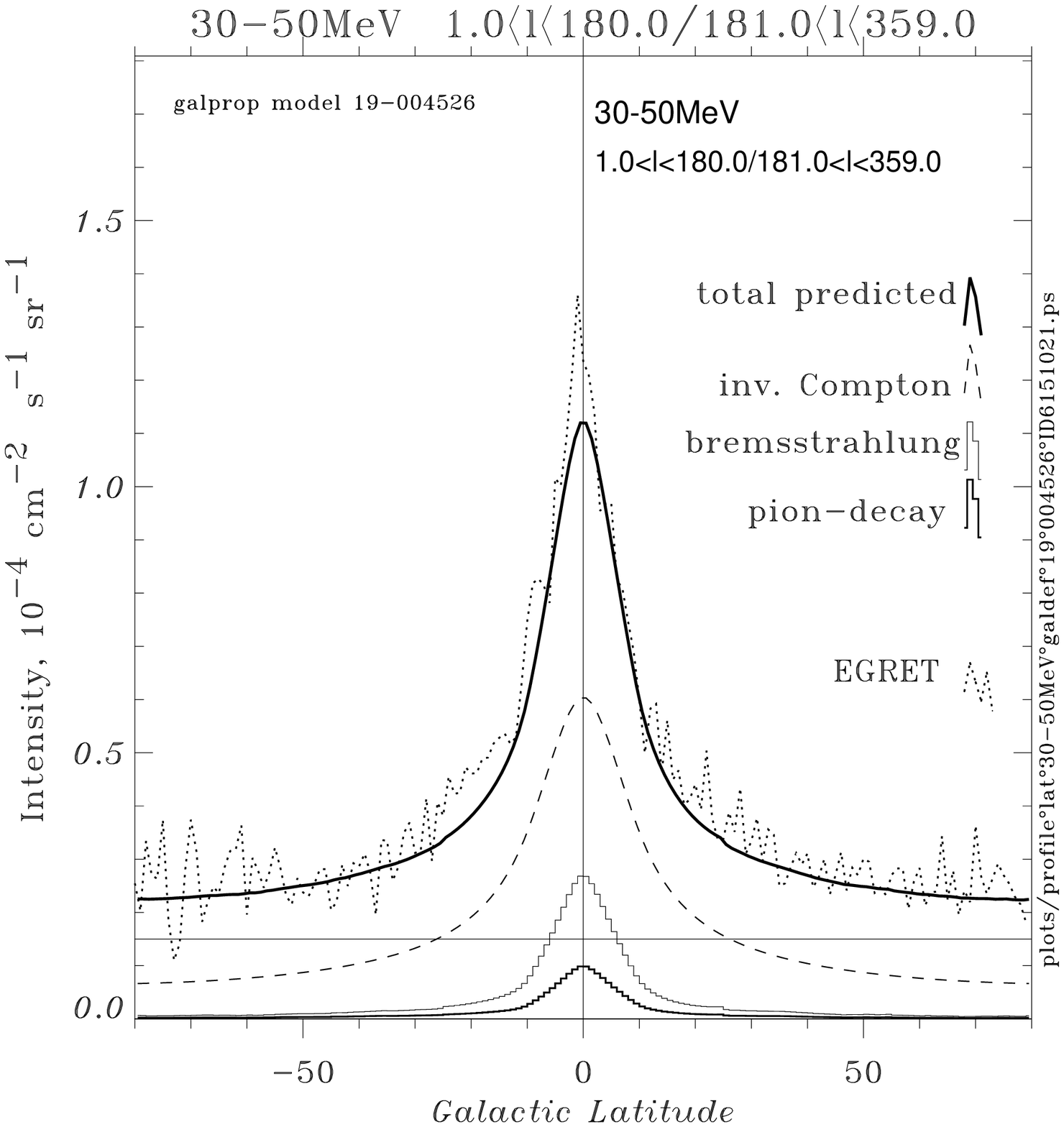,width=\fwc,clip=}
\psfig{file=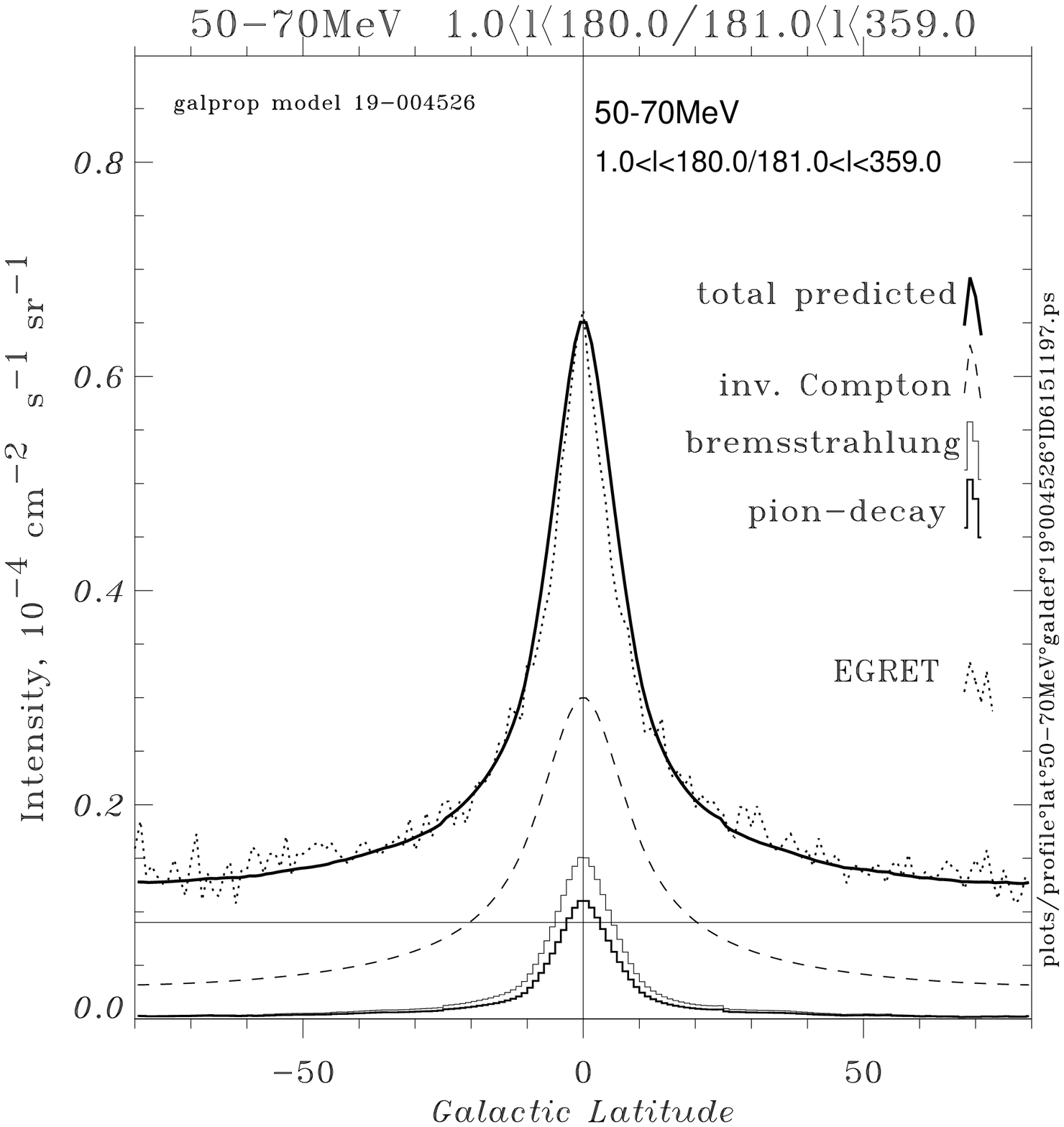,width=\fwc,clip=}
\psfig{file=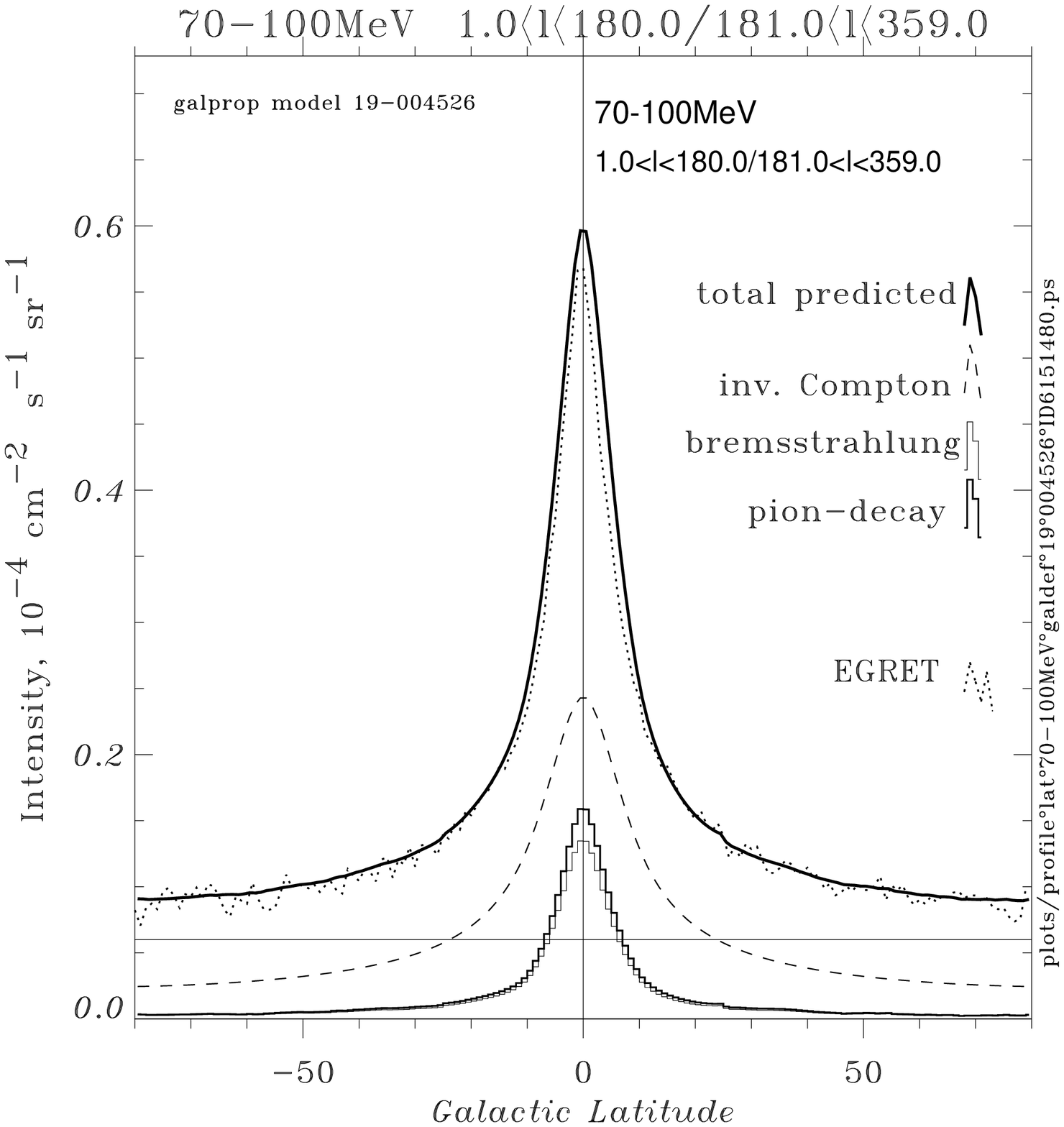,width=\fwc,clip=} }
\centerline{
\psfig{file=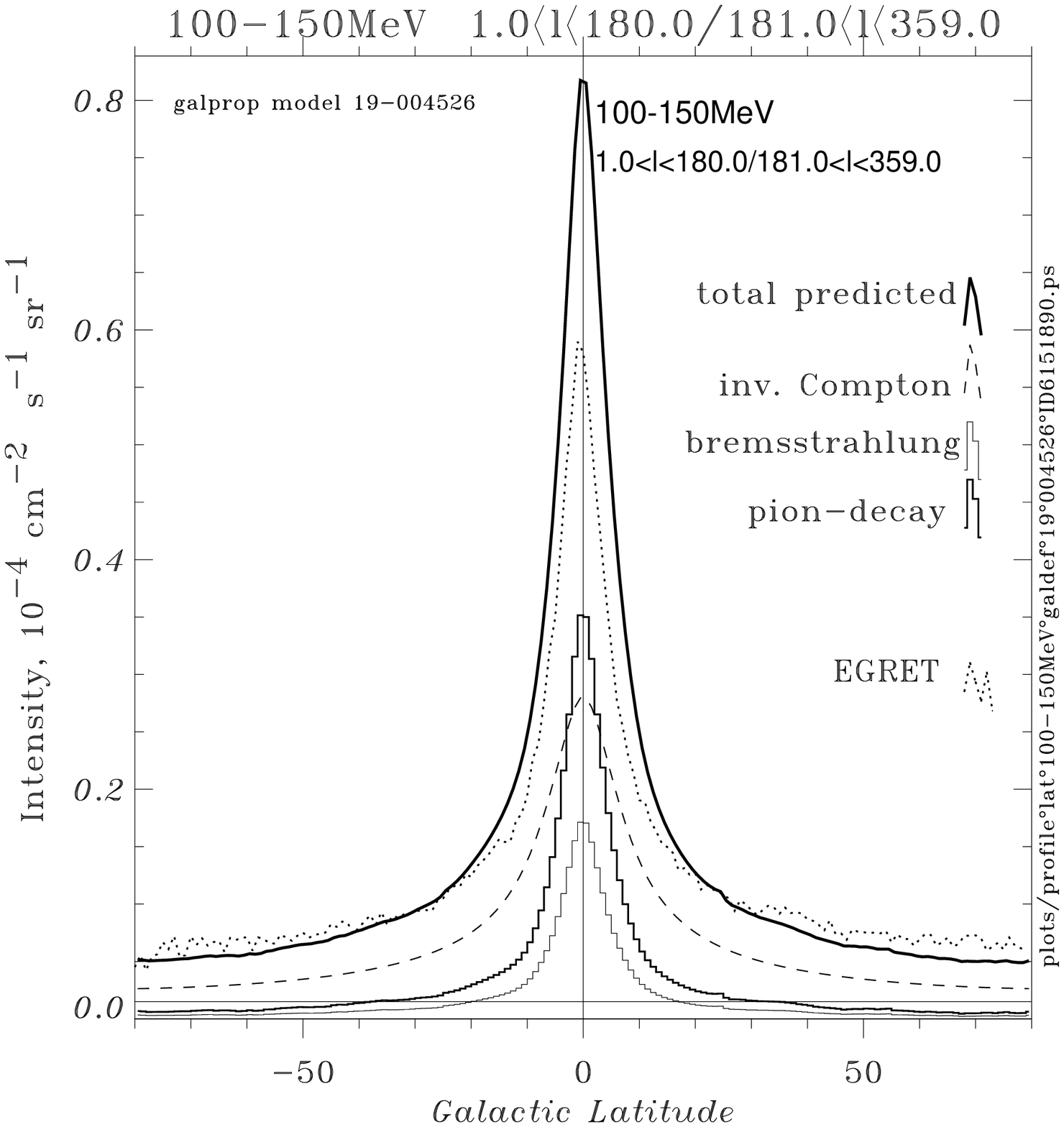,width=\fwc,clip=}
\psfig{file=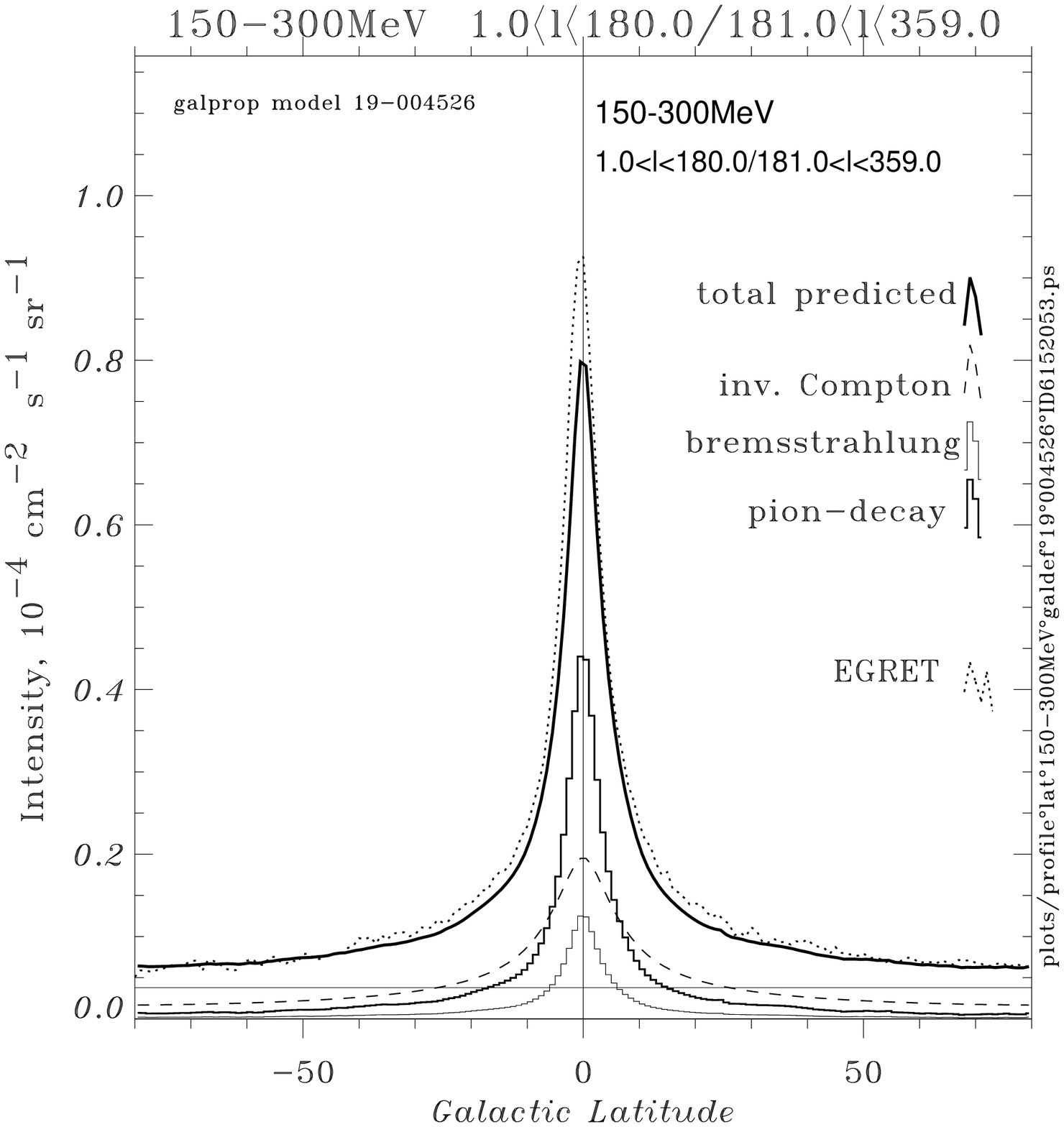,width=\fwc,clip=}
\psfig{file=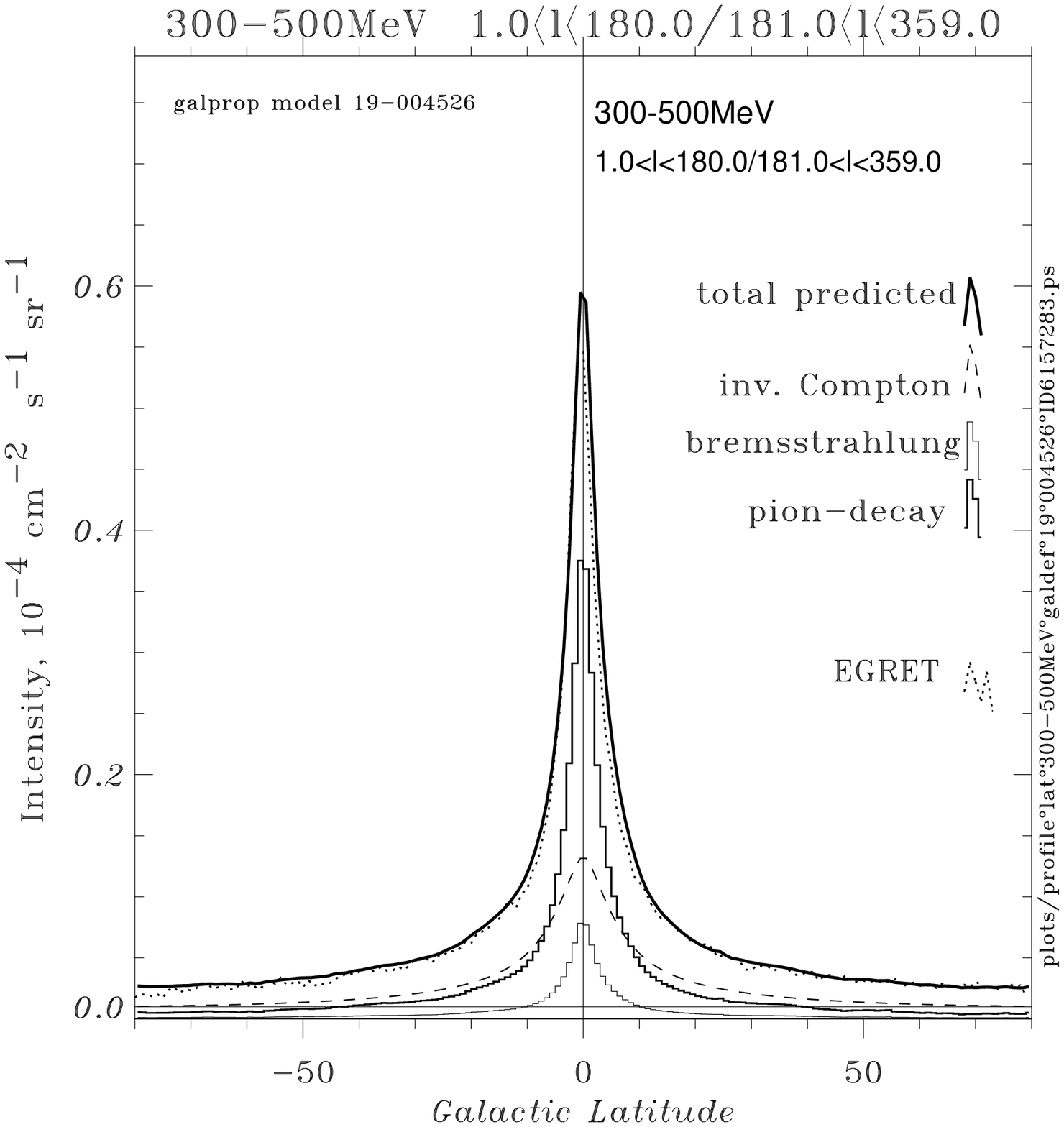,width=\fwc,clip=} }
\centerline{
\psfig{file=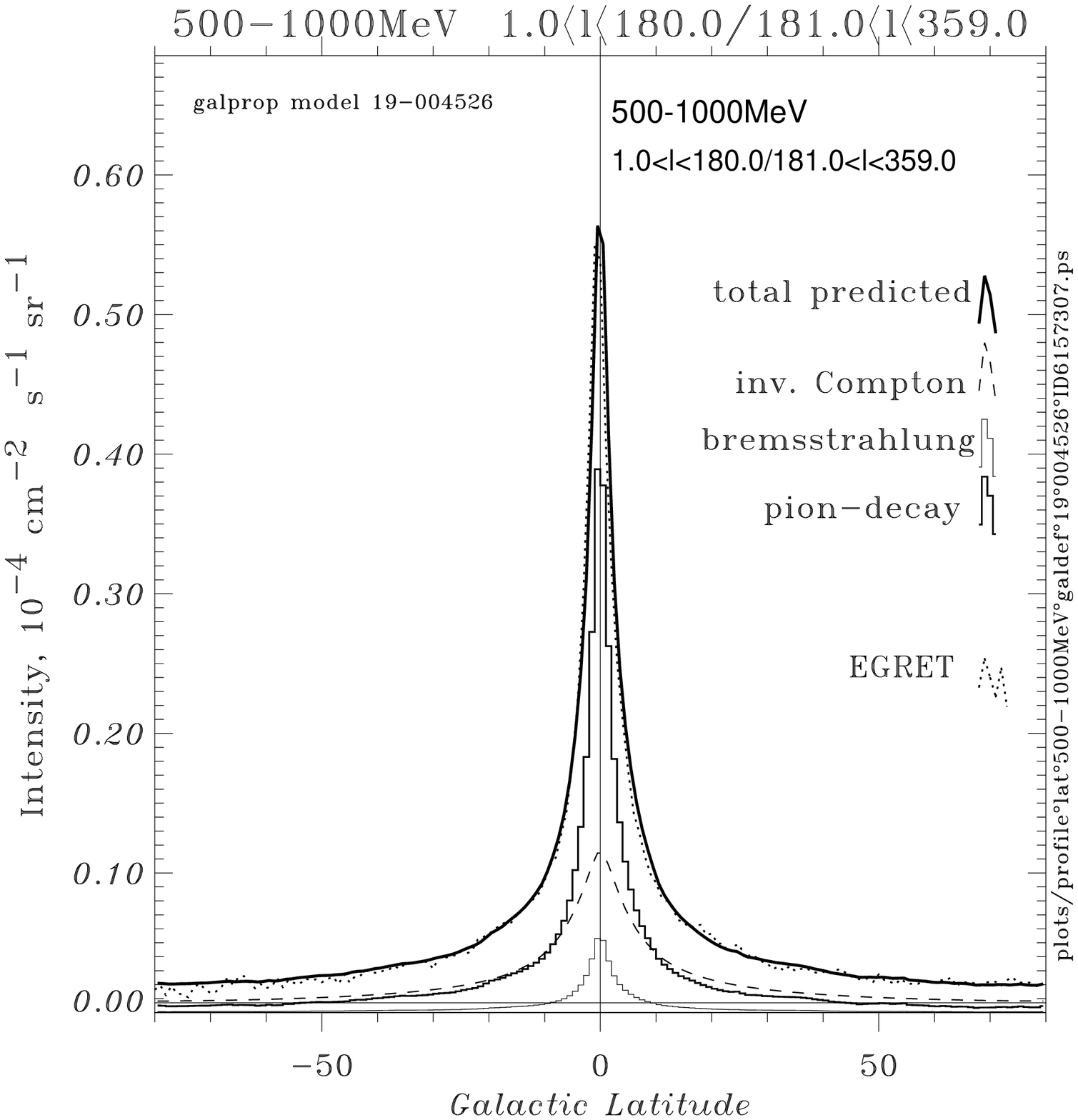,width=\fwc,clip=}
\psfig{file=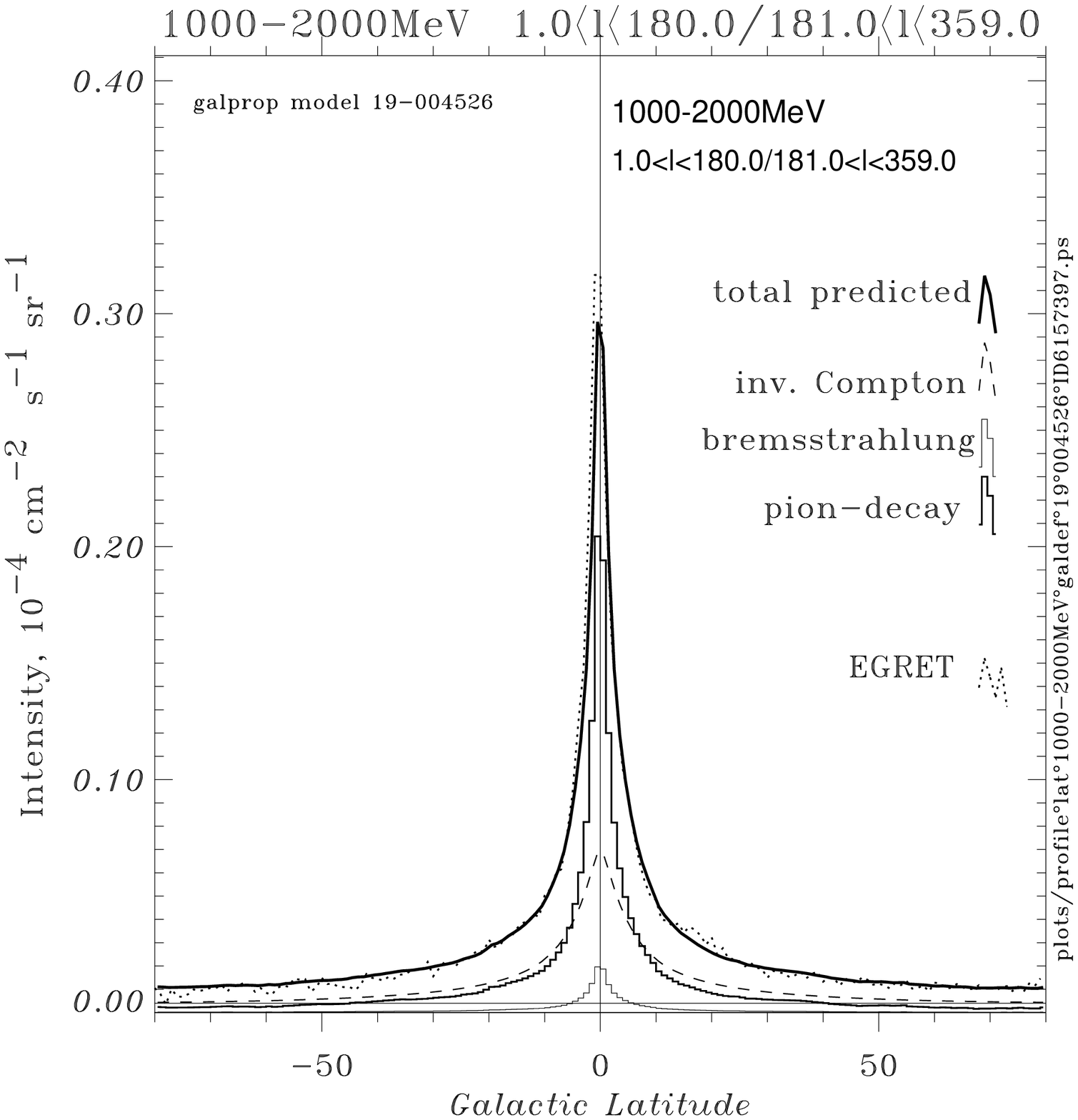,width=\fwc,clip=}
\psfig{file=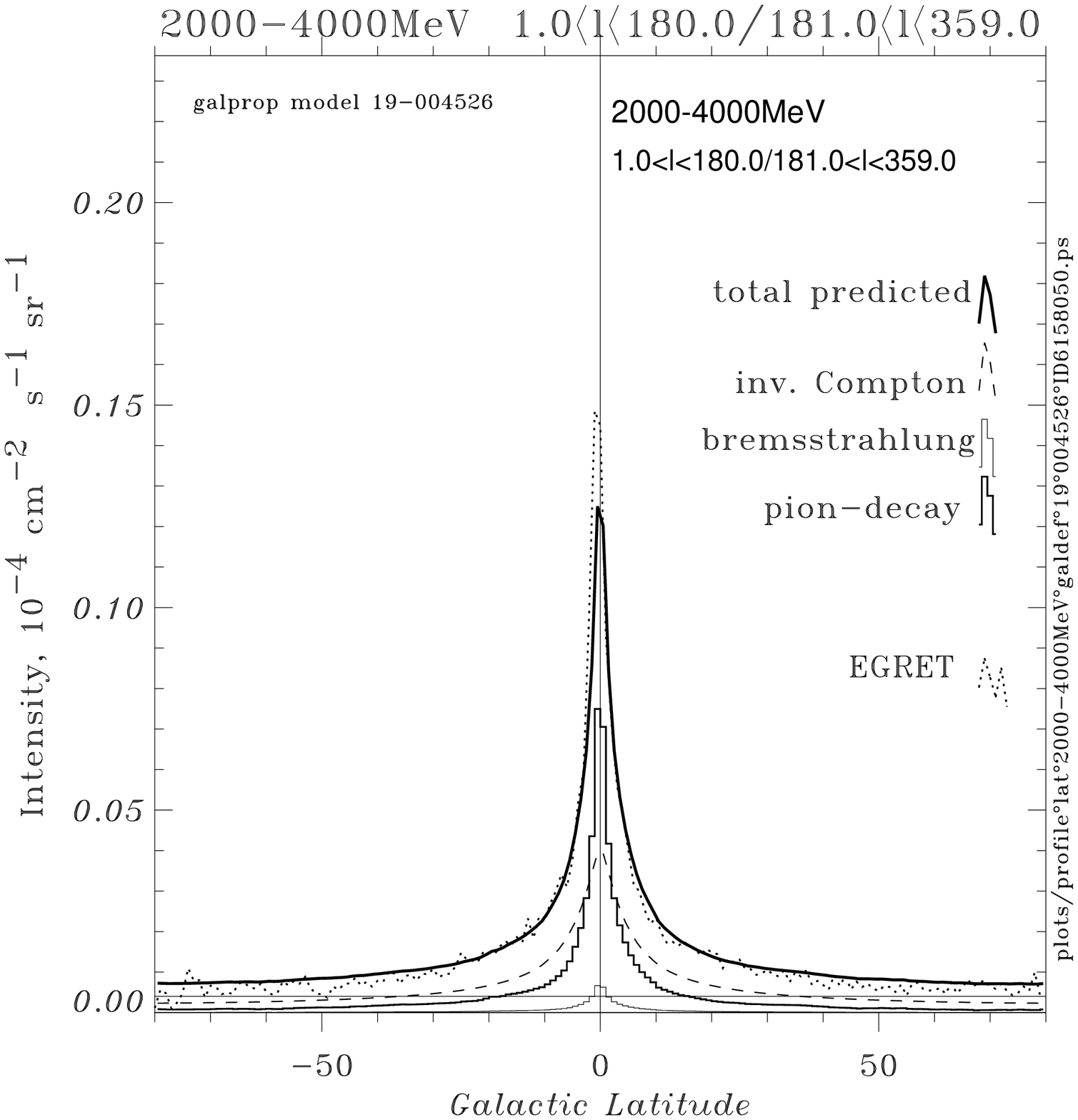,width=\fwc,clip=} }
\centerline{
\psfig{file=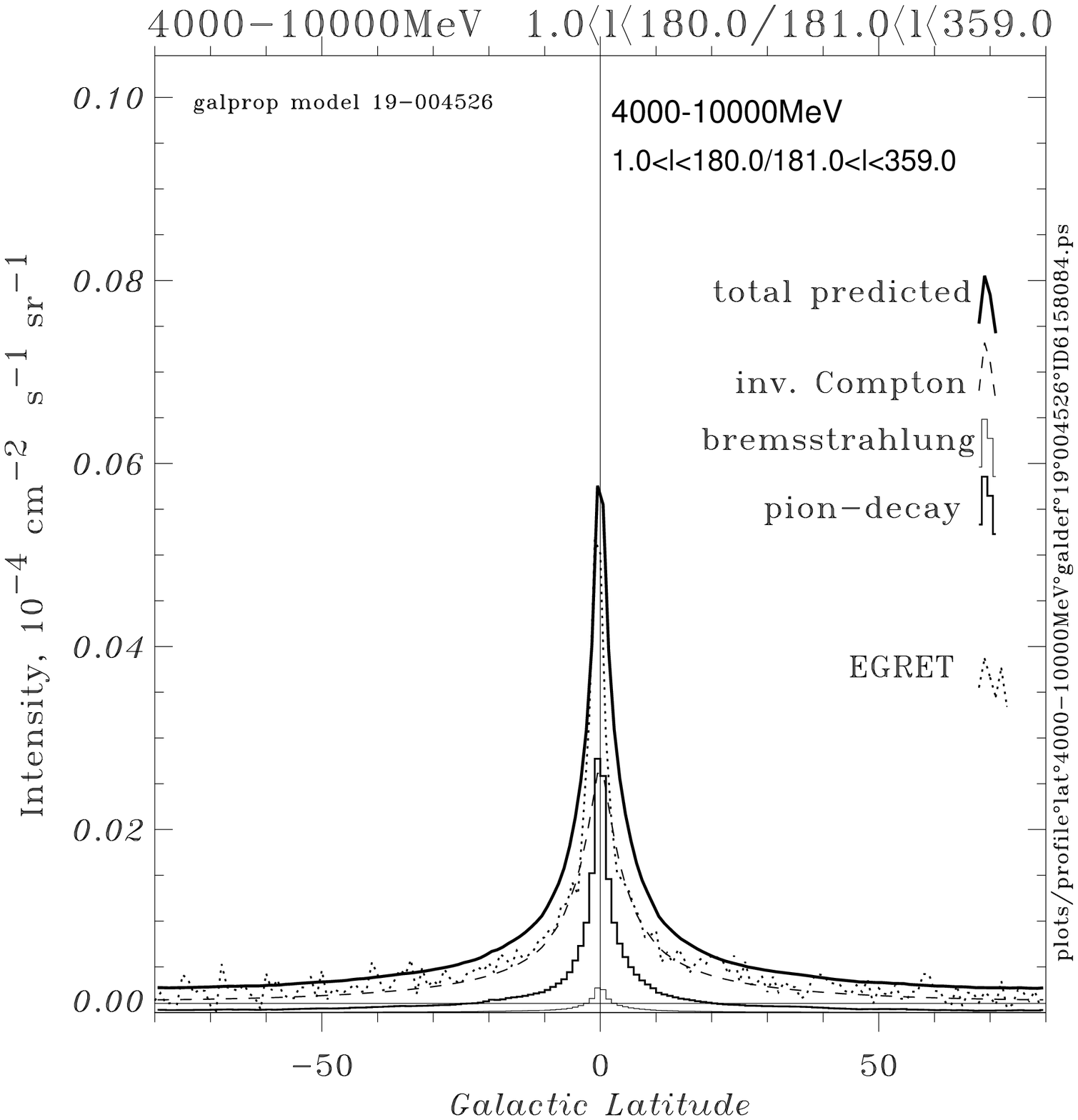,width=\fwc,clip=} }
\figcaption[fig14a.ps,fig14b.ps,fig14c.ps,fig14d.ps,fig14e.ps,fig14f.ps,%
fig14g.ps,fig14h.ps,fig14i.ps,fig14j.ps]{
Latitude distribution of $\gamma$-rays ($0^\circ \le l\le
360^\circ$) for model HEMN (thick solid line).  Separate
components show the contribution of IC (dashes), bremsstrahlung
(thin histogram), $\pi^0$-decay (thick histogram), horizonal
line: isotropic background.  EGRET data: dotted line.
\label{Fig_gamma_latitude_profile}}
\end{figure*}

\begin{figure*}[!p]%***************************************************** 15
\centerline{
\psfig{file=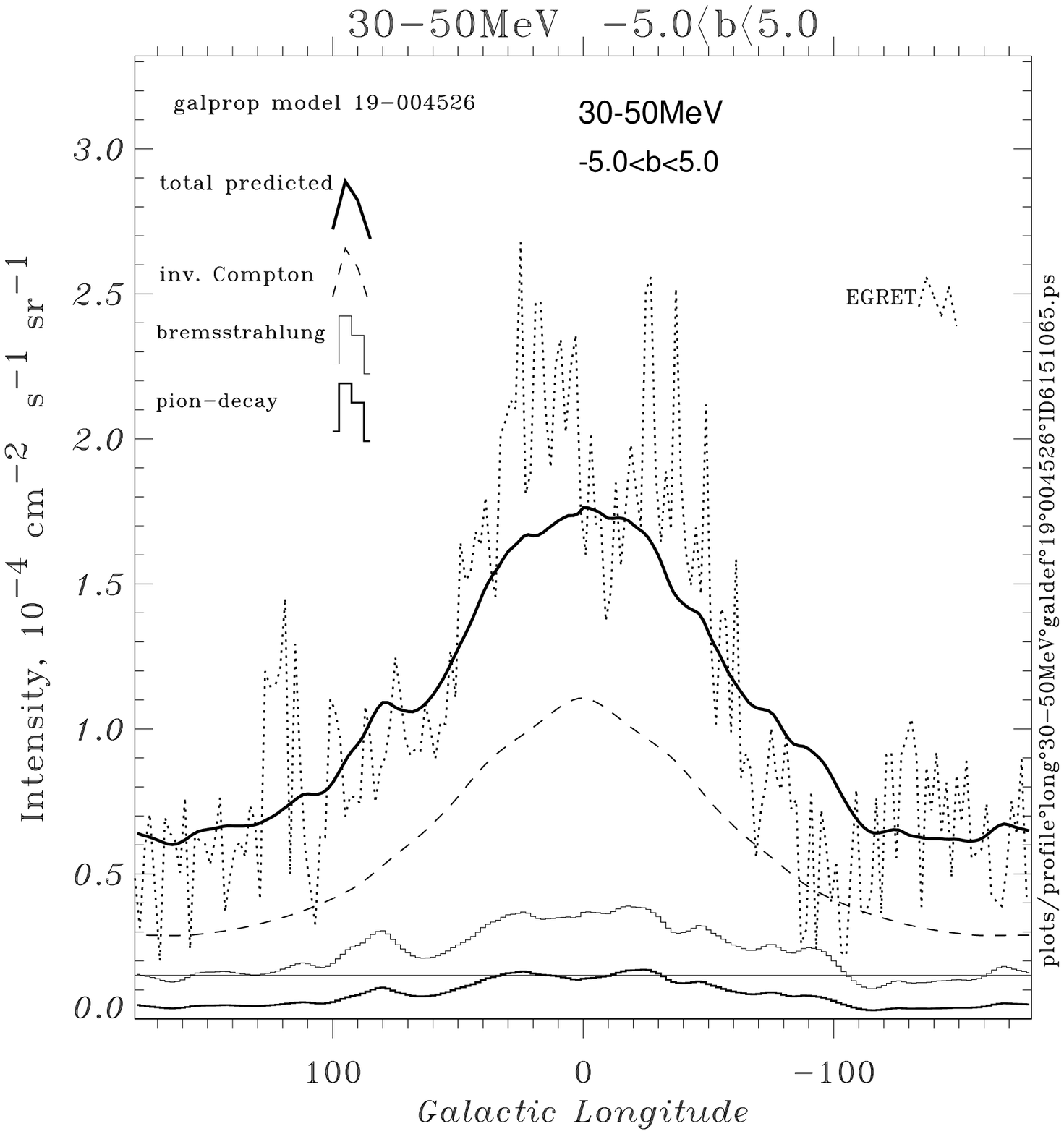,width=\fwc,clip=}
\psfig{file=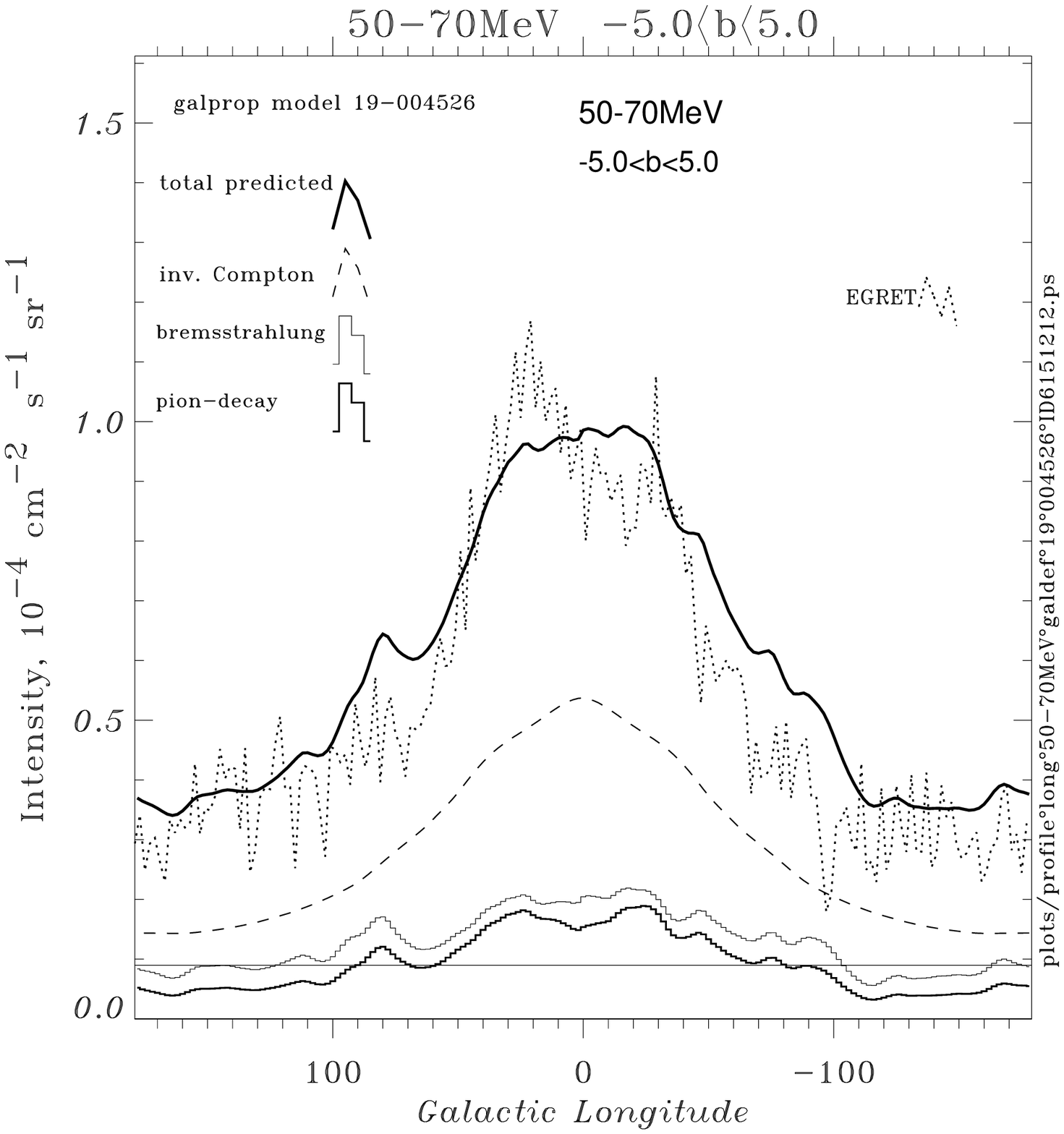,width=\fwc,clip=}
\psfig{file=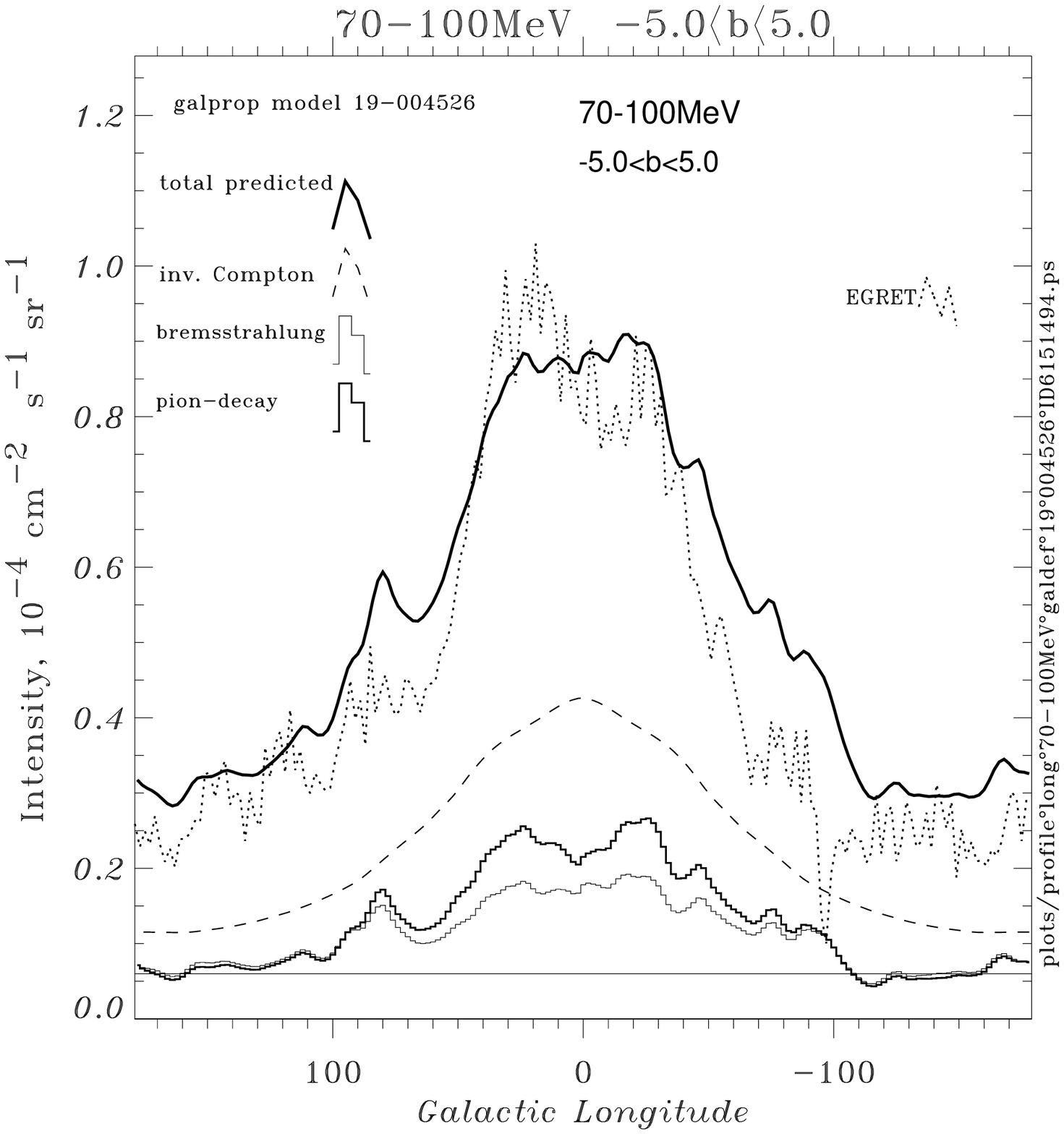,width=\fwc,clip=} }
\centerline{
\psfig{file=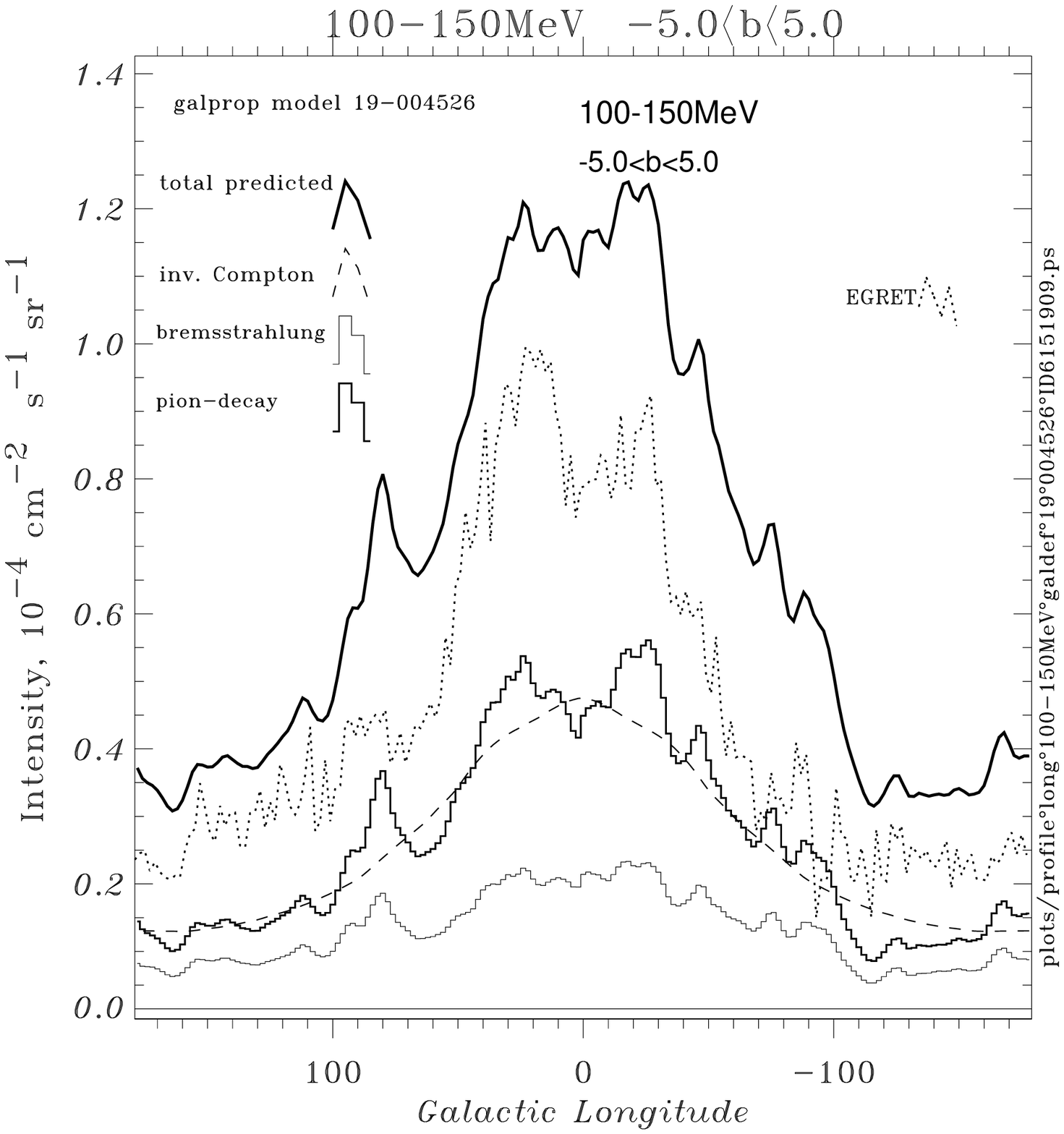,width=\fwc,clip=}
\psfig{file=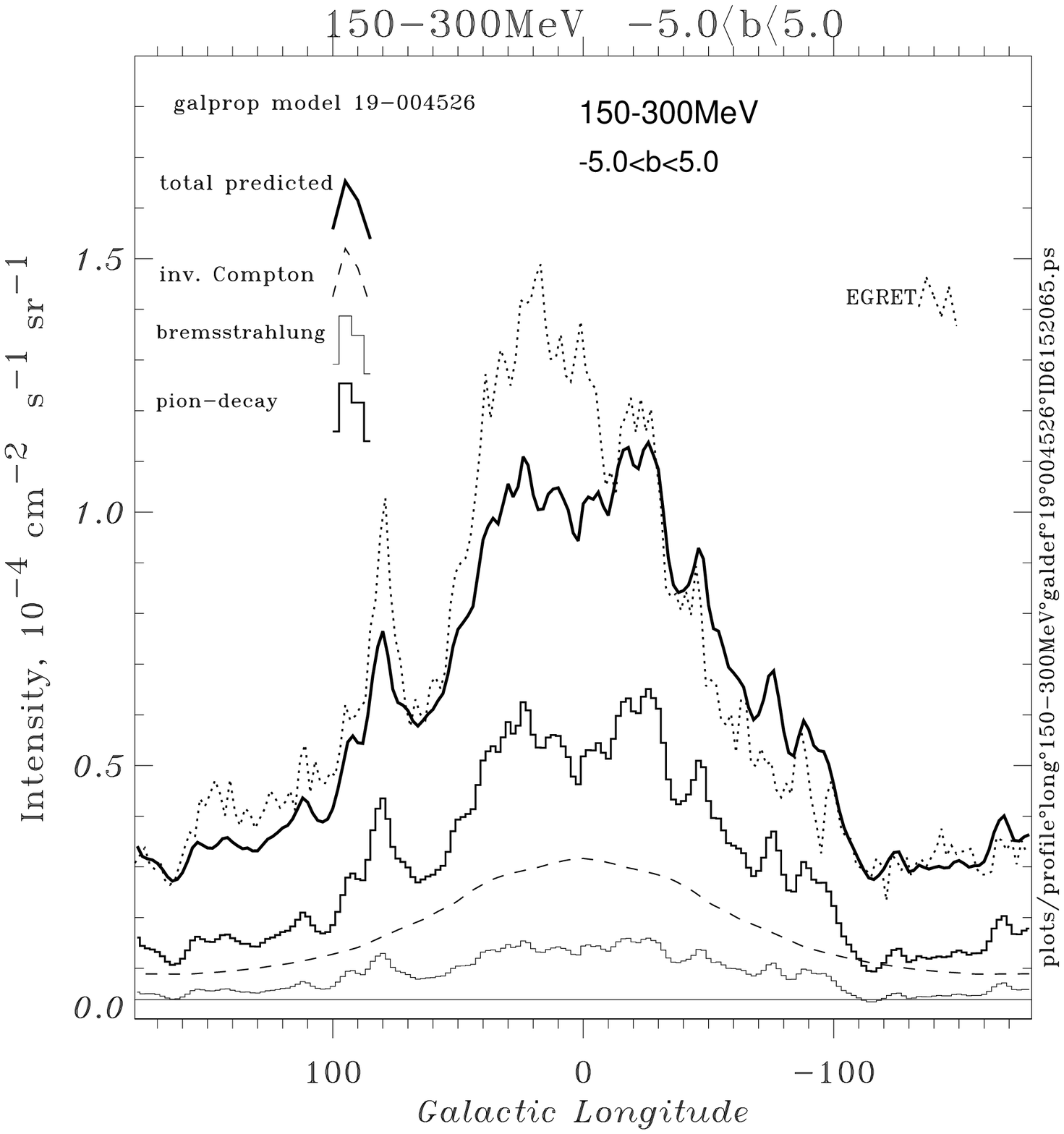,width=\fwc,clip=}
\psfig{file=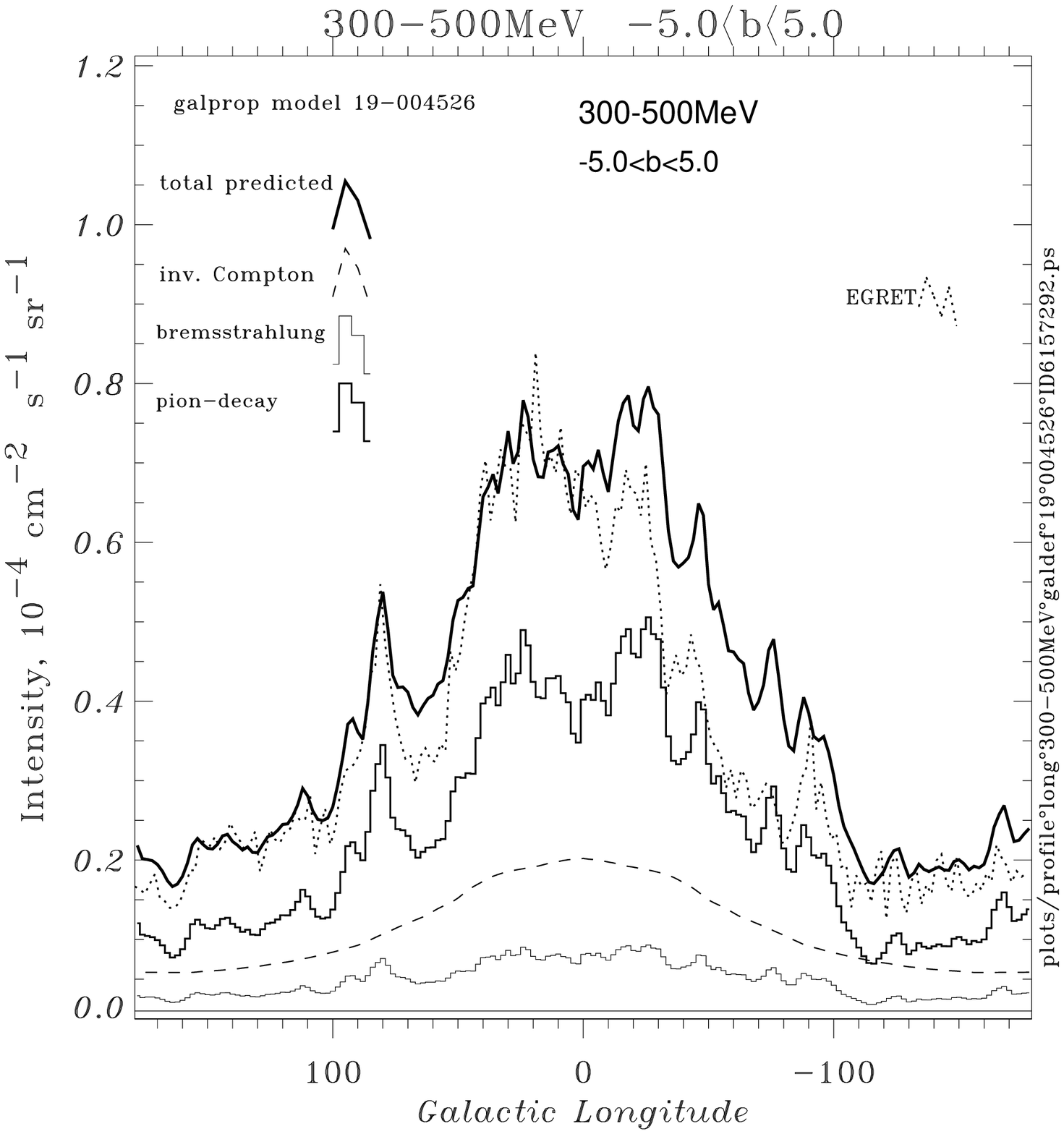,width=\fwc,clip=} }
\centerline{
\psfig{file=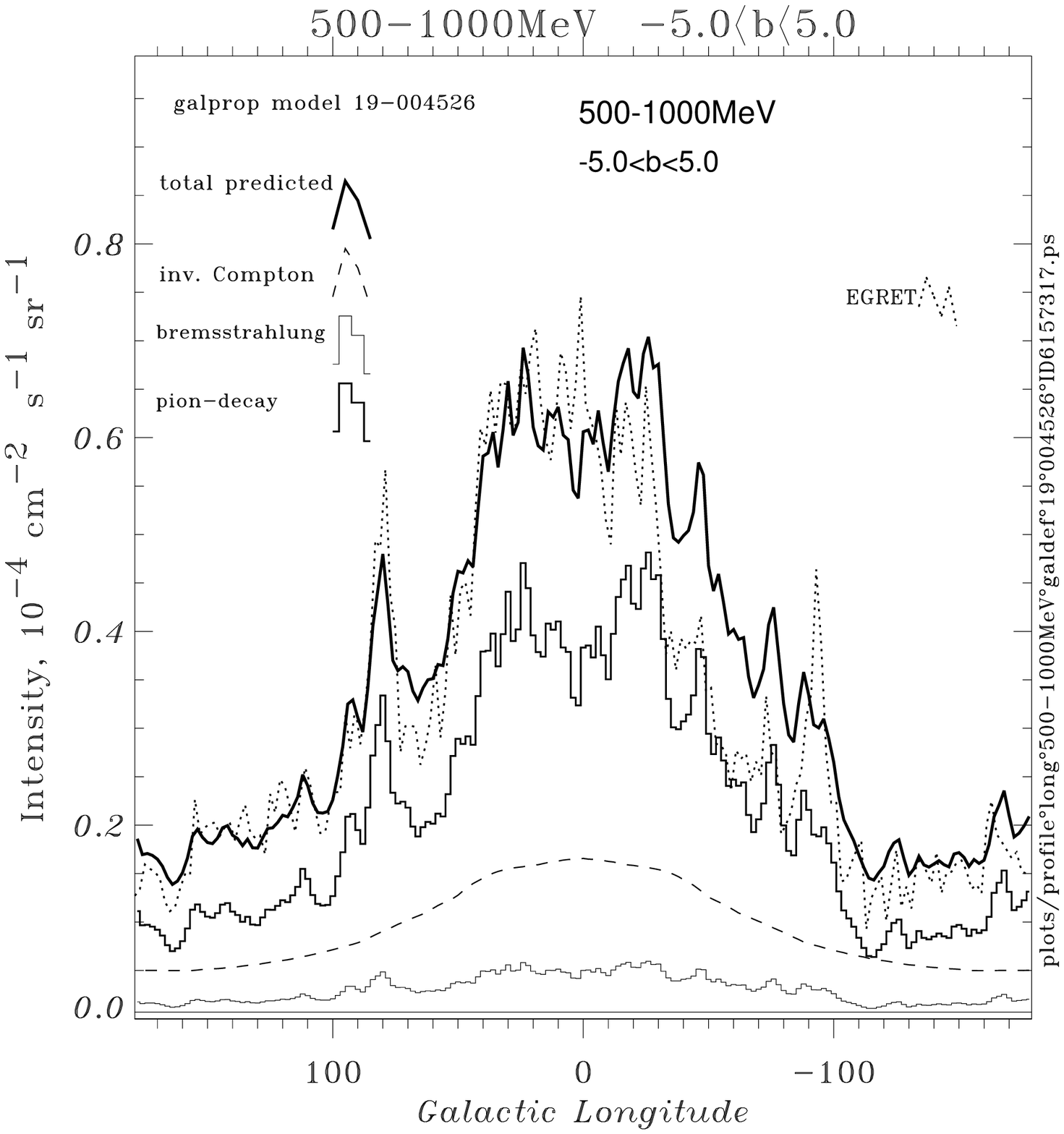,width=\fwc,clip=}
\psfig{file=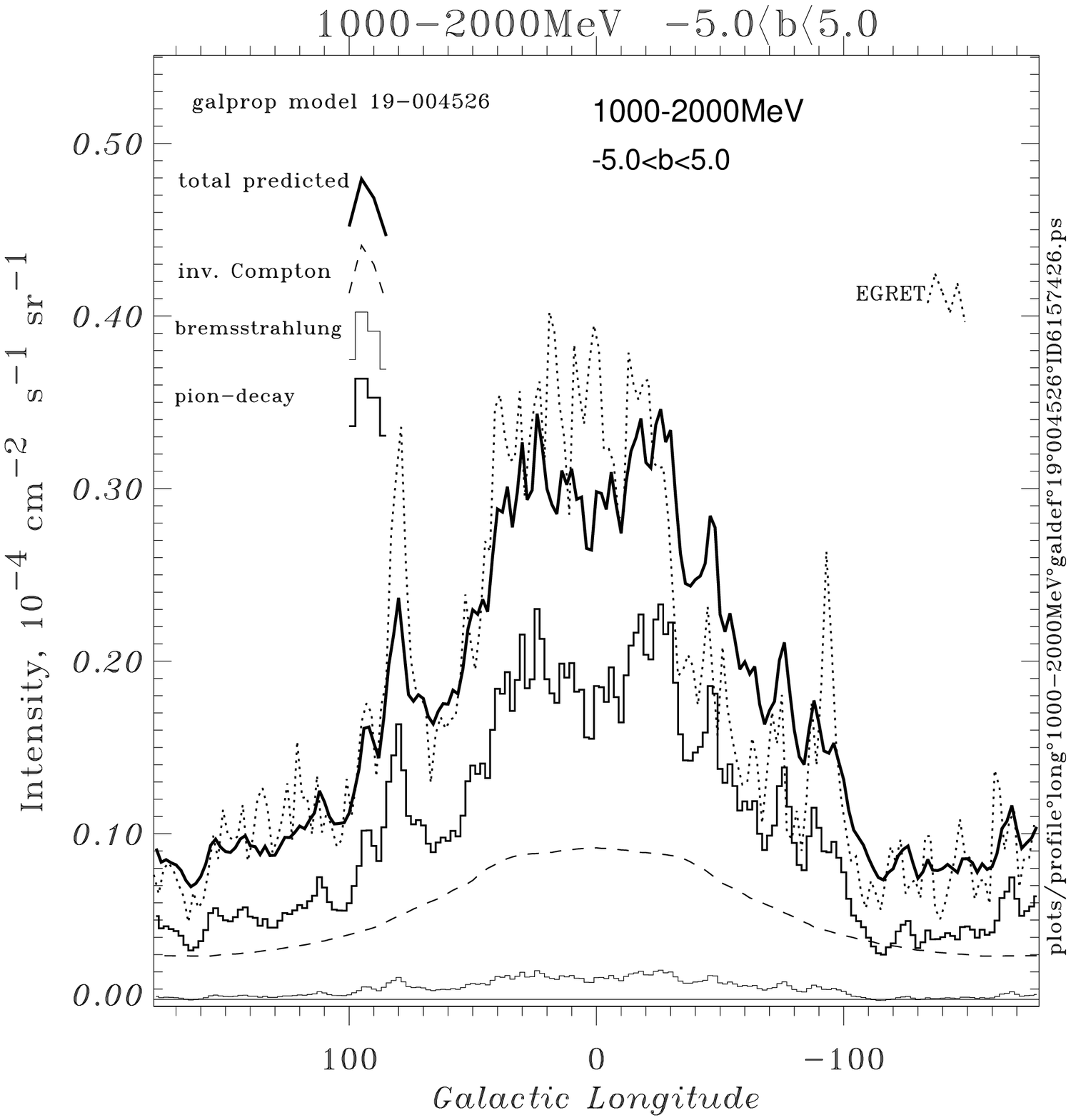,width=\fwc,clip=}
\psfig{file=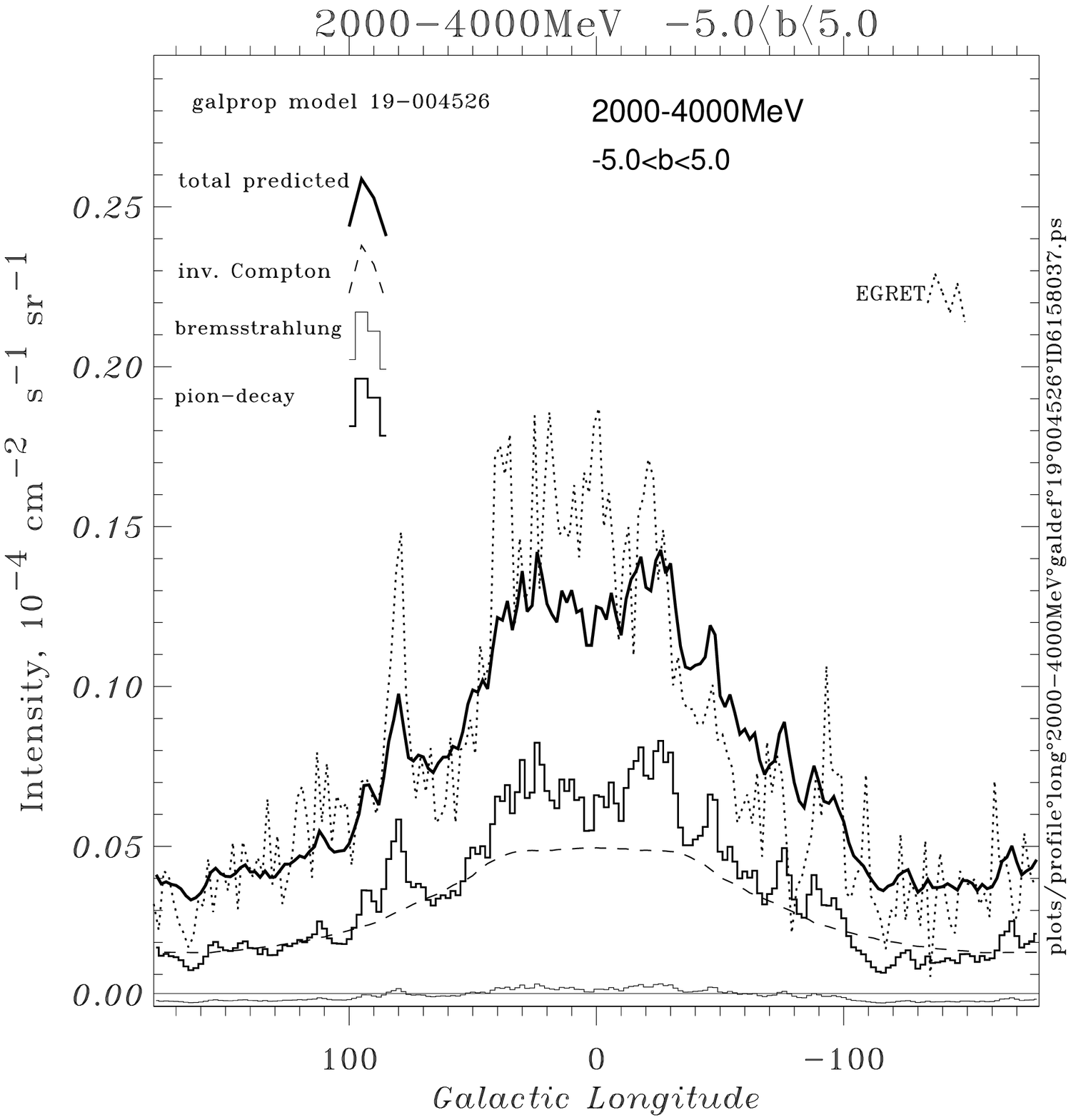,width=\fwc,clip=} }
\centerline{
\psfig{file=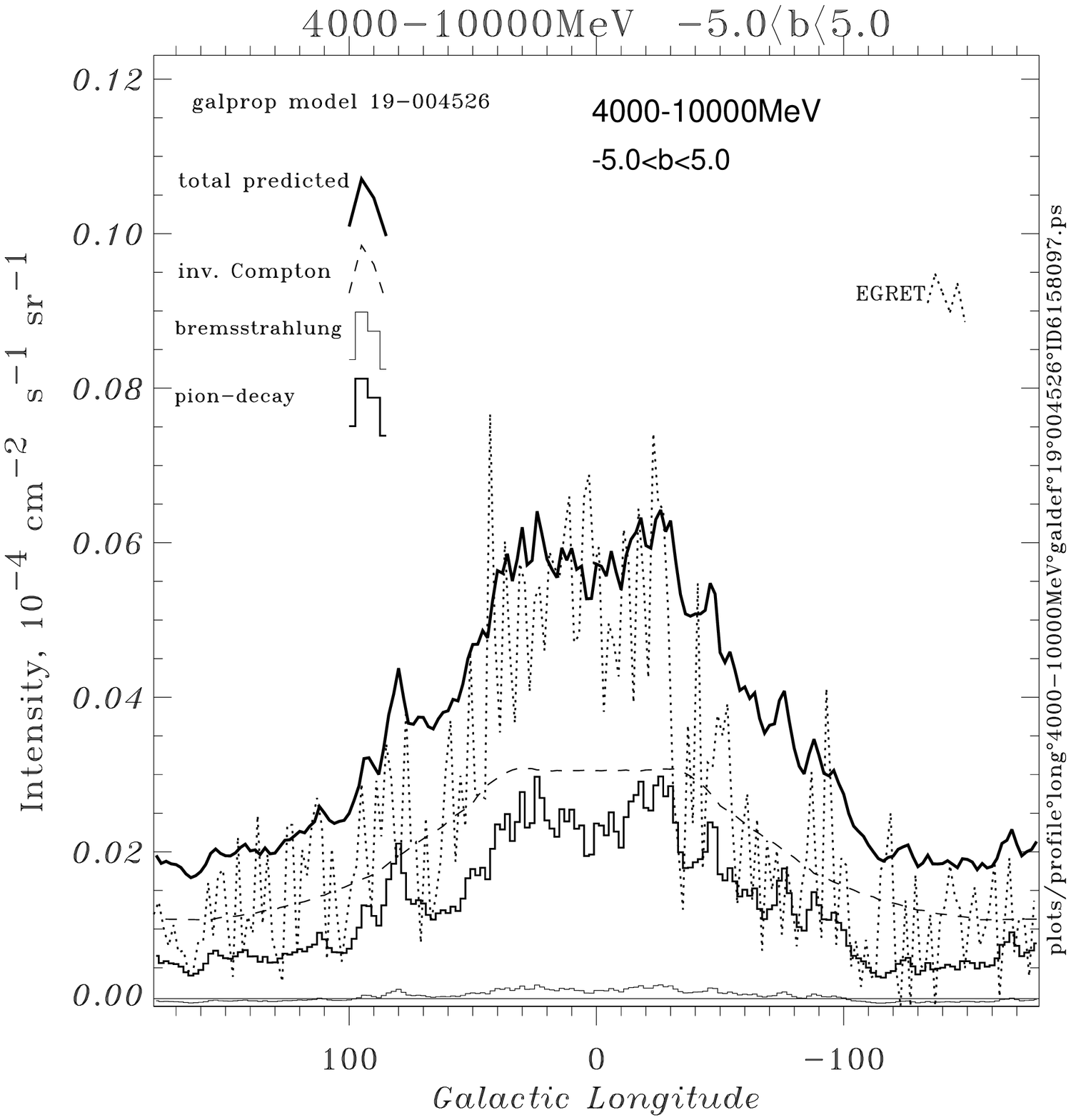,width=\fwc,clip=} }
\figcaption[fig15a.ps,fig15b.ps,fig15c.ps,fig15d.ps,fig15e.ps,fig15f.ps,%
fig15g.ps,fig15h.ps,fig15i.ps,fig15j.ps]{
Longitude distribution of \grays ($|b|\le 5^\circ$) for model
HEMN.  Coding of lines for components is the same as in
Fig.~\ref{Fig_gamma_latitude_profile}.  EGRET data: dotted line.
\label{Fig_gamma_longitude_profile}}
\end{figure*}

\begin{figure*}[tbh]%***************************************************** 16
\centerline{
\psfig{file=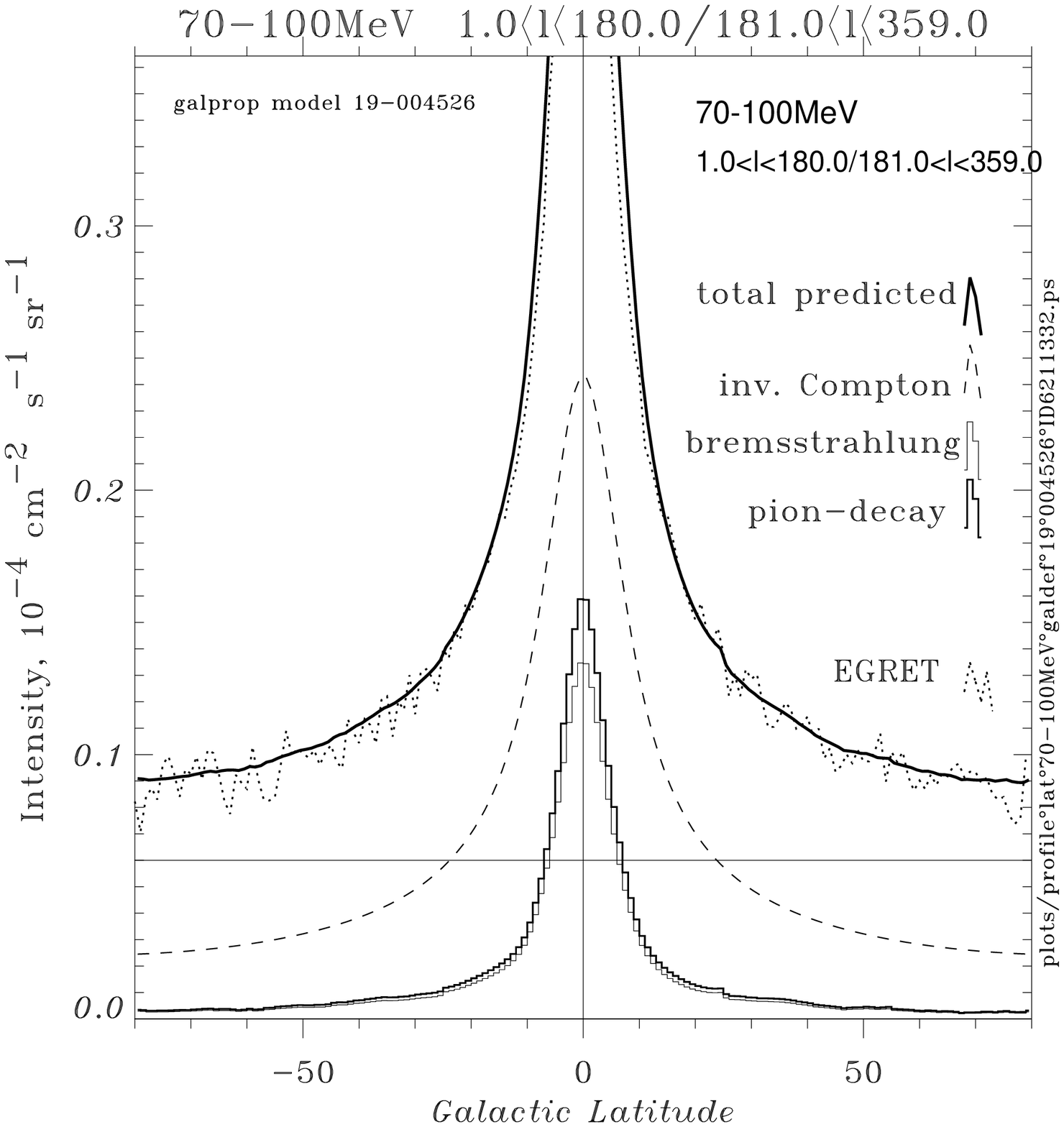,width=\fwb,clip=}
      \hspace{\hs}
\psfig{file=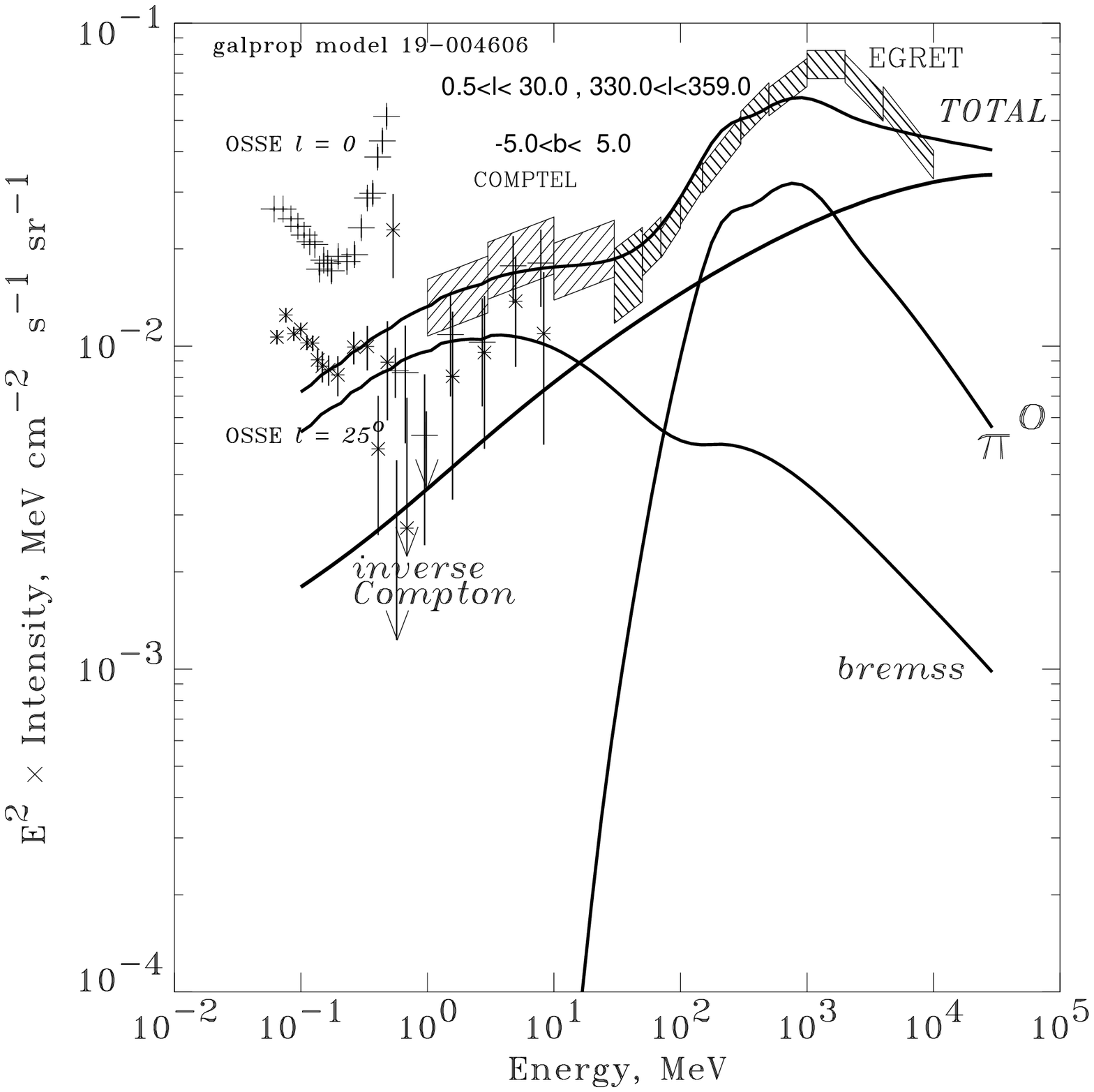,width=\fwb,clip=} } 
\parbox{89mm}{%
\figcaption[fig16.ps]{ High latitude
distribution (enlarged) of 70--100 MeV \grays from the EGRET
compared to model HEMN.  Lines are coded as in
Fig.~\ref{Fig_gamma_latitude_profile}.
\label{Fig_gamma_high_latitude_profile}}
}\hspace{7mm}
\parbox{89mm}{%
\figcaption[fig17.ps]{ 
Gamma-ray data as in
Fig.~\ref{Fig_gamma_spectrum_C} compared with SE model (electron
injection spectrum with upturn below 200 MeV).
\label{Fig_gamma_spectrum_SE}}
}
\end{figure*} 

\placefigure{Fig_gamma_spectrum_SE}

So far this is the most promising model (at least for \grays $>$30
MeV) and hence we consider it further by testing the angular
distribution of the emission.  The synchrotron predictions for this
model and comparison with data were presented in
Section~\ref{synch_emis}.  Figs.~\ref{Fig_gamma_latitude_profile} and
\ref{Fig_gamma_longitude_profile} show the HEMN model latitude and
longitude \gray distributions convolved with the EGRET point-spread
function, compared to EGRET Phase 1--4 data.  The separate components
are also shown. Since the isotropic component is here regarded as a
free parameter it was adjusted in each energy range to give agreement
with the high-latitude intensities in the latitude plots, and the same
value was used in the longitude plots.

In latitude the agreement is always quite good, in longitude the
maximum deviation is about 25\% in the 100--150 MeV, 150--300 MeV, and
4--10 MeV ranges, typically the agreement is better than 10\%.  We
believe, this is  satisfactory for a model which has not been
optimized for the spatial fit in each energy range, but anyway these
figures allow the reader to judge for himself.  In any case, it should
be remembered  that unresolved point-source components and
irregularities in the cosmic-ray source distribution are expected to
lead to deviations from our cylindrically symmetrical model at some level.

In this model the contributions to the spectrum of the inner Galaxy
from IC and $\pi^0$-decay are about equal at 100 MeV and 6 GeV,
$\pi^0$-decay dominates between these energies, and bremsstrahlung
produces $\la$10\% of the total.  The  comparison shows that a model
with large IC component can indeed reproduce the data.  A profile for
70--100 MeV enlarged to illustrate the high-latitude variation
(Fig.~\ref{Fig_gamma_high_latitude_profile}) shows that this model
also accounts very well for the observed emission; we regard this as
support for the large IC halo concept.

Turning to high energies,  consider the latitude and longitude profiles
for 4000--10000 MeV;  the agreement shows that the adoption of a hard
electron injection spectrum is a viable explanation for the $>$1 GeV
excess. The latitude distribution here is not as wide as at low
energies owing to the rapid energy losses of the electrons; both HN and
HEMN models reproduce the observed spectrum, and latitude and longitude
profiles almost equally well (\cite{MS98b}), and hence it is difficult
to discriminate between them on the basis of \grays alone. Independent
tests however argue against HN as described in Section~\ref{hn}.

It is interesting to note that in fitting EGRET data Strong \& Mattox
(1996) found that the IC component had a harder spectrum than expected
(see their Fig.~4), which was quite puzzling at that time.  Also the
study of Chen, Dwyer, \& Kaaret (1996) at high latitudes found a hard
IC component.  These results can now be understood in the context of
the HE or HEMN model; a renewed application of the fitting approach
with the new models would be worthwhile and is intended for the future.
All these results can be taken as adding support to the `hard electron
spectrum' interpretation of the \gray results, and for the idea that
the average interstellar electron spectrum is harder than that measured
in the heliosphere.

If this model is indeed correct it then implies that bremsstrahlung
plays a rather minor r\^ole at all energies, contrary to previous ideas,
with IC and $\pi^0$-decay accounting for $\sim$90\% of the diffuse
emission.

Although we have introduced rather arbitrary modifications to both
electron and nucleon spectra to better fit the \gray data, we note two
recent indications that add support to our approach from
independent studies.  Baring et al.\ (1999) recently presented models
for shock acceleration in SNR which produce very flat electron spectra
quite similar to what we require in the present case. Further study of
the possible link between these spectra is in progress.  A completely
independent line of evidence for a low-energy flattening of the proton
spectrum has recently been presented by Lemoine, Vangioni-Flam, \&
Cass\'e (1998), based on cosmic-ray produced light element abundances.
If the proton and He spectra do differ from that locally
measured, this could of course also apply other primaries, but an
investigation of this is beyond the scope of the present work, which
focusses on $\gamma$-rays.

\placefigure{Fig_gamma_spectrum_high_latitudes}

\begin{figure*}[tbh]%***************************************************** 18
\centerline{
\psfig{file=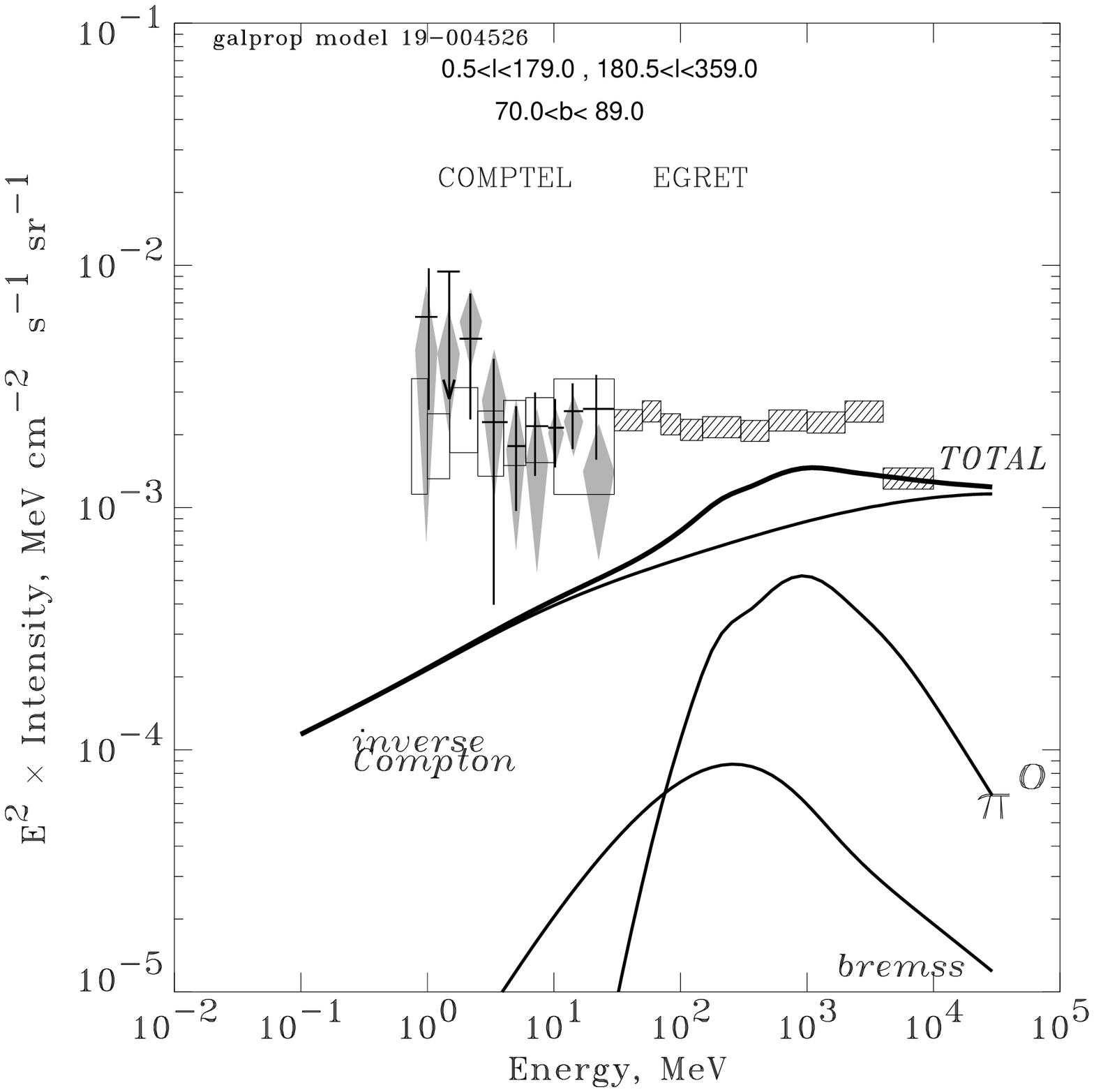,width=\fwb,clip=}
\psfig{file=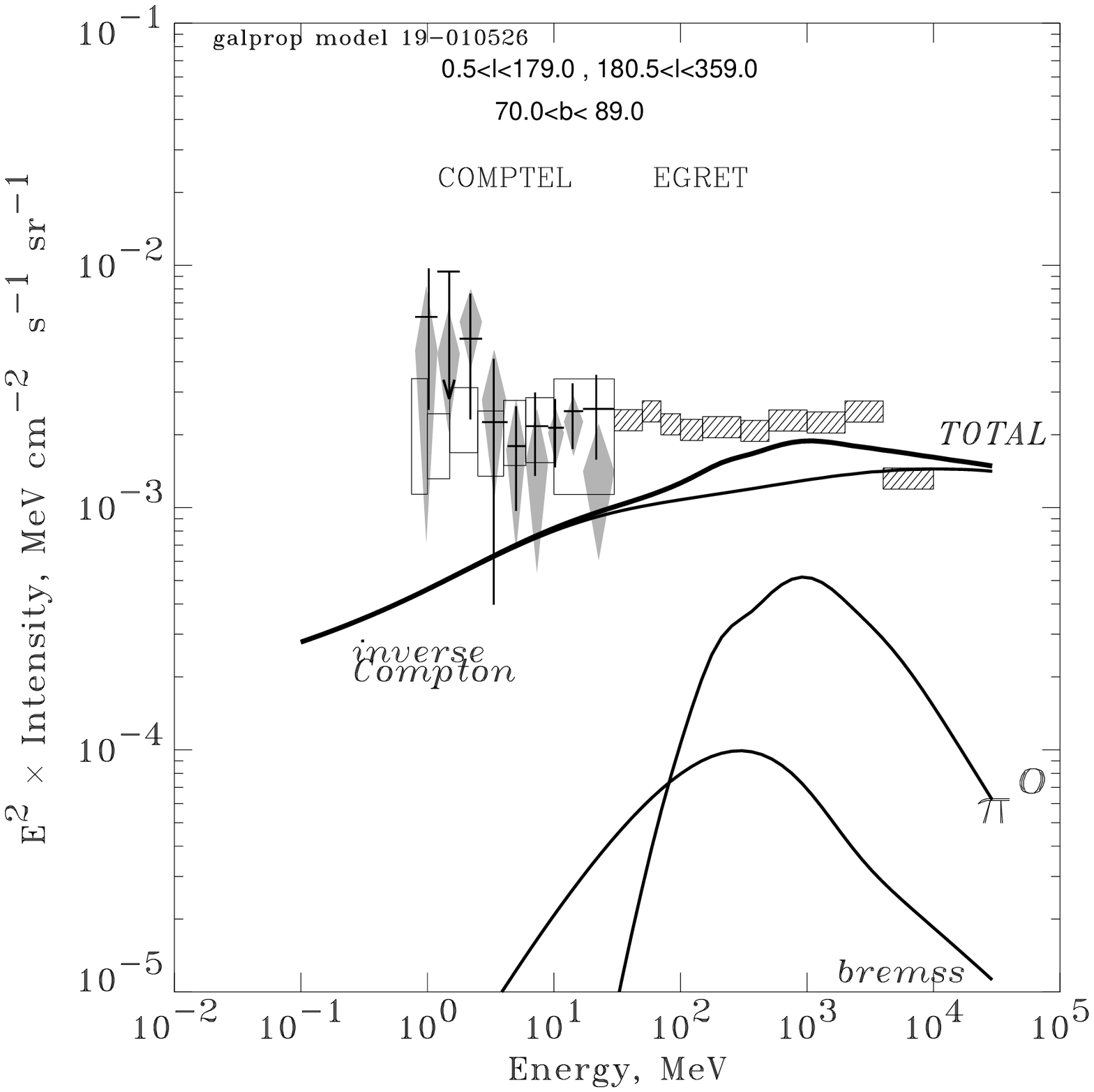,width=\fwb,clip=}
}
\figcaption[fig18a.ps,fig18b.ps]{
Energy spectrum of \grays from high Galactic latitudes ($|b|\ge
70^\circ$, all longitudes) for HEMN model (left) and HELH model
(right).  Shaded areas: EGRET total intensity from Cycle 1--4
data.  COMPTEL data: high-latitude total intensity (open boxes:
\protect\cite{Bloemen99}, diamonds: \protect\cite{Kappadath98},
crosses: \protect\cite{Weidenspointner99}).
\label{Fig_gamma_spectrum_high_latitudes}}
\end{figure*}

\subsection{A steeper spectrum of electrons at low energies (SE model)
or a population of MeV point sources in the Galactic plane ?}
\label{se}
%######################################################################
In order to reproduce the low-energy ($<$ 30 MeV) \gray emission via
diffuse processes it is necessary to invoke steepening of the electron
spectrum below about 200 MeV to compensate the increasing ionization
losses.  A steep slope continuing to higher energies would violate the
synchrotron constraints on the spectral index, as discussed in
Section~\ref{gray_data}.  For illustration we show a non-reacceleration
model with an injection index --3.2 below 200 MeV, and --1.8 above (SE
model).  This fits both the COMPTEL and EGRET data
(Fig.~\ref{Fig_gamma_spectrum_SE}) while remaining consistent with the
synchrotron constraints (Fig.~\ref{Fig_sync_index}).  The
synchrotron index increase occurs at frequencies $< 10$ MHz, below the
range where useful limits can be set (see \cite{Strong78}).
Although the behaviour of the electron diffusion coefficient
at energies below 100 MeV is quite uncertain (\cite{Bieber94}) the
propagated spectrum is here dominated by energy losses, 
which severely limit the electron range, so that this is
not critical for our model.

In this model 70\% of the emission is bremsstrahlung and 30\% IC at 1
MeV.  This is the only model in the present work which can reproduce
the entire \gray spectrum.  A possible mechanism for acceleration of
low-energy electrons has been proposed by Schlickeiser (1997).  However
the adoption of such a steep low-energy electron spectrum has problems
associated with the very large power input to the interstellar medium
(\cite{Skibo97}), and is {\it ad hoc} with no independent supporting
evidence. Moreover the OSSE-GINGA \gray spectrum is steeper than
$E^{-2}$ below 500 keV (\cite{Kinzer99}) which would require an even
steeper electron injection spectrum than adopted here.  It is more
natural to consider that the COMPTEL excess is just a continuation of
the same component producing the OSSE-GINGA spectrum.  Most probably
therefore the excess emission at low energies is produced by a
population of sources such as supernova remnants,  as has been
proposed  for the diffuse hard X-ray emission from  the plane observed
by RXTE (\cite{Valinia98}), or X-ray transients in their low state as
suggested for the OSSE diffuse hard X-rays (\cite{Lebrun99}).  The
contribution from point sources is then about 70\% at 1 MeV, the rest
being IC.  Above 10 MeV the diffuse emission dominates.  Yamasaki et
al.\ (1997) estimate a 20\% contribution from point sources to the hard
X-ray plane emission, the rest being attributed to young electrons in
SNR, but these are still localized \gray sources rather than truly
diffuse emission.

A model with a constant electron injection index --2.4 can also fit the
low-energy $\gamma$-rays, but conflicts with the synchrotron index and
fails to reproduce the high-energy $\gamma$-rays.  While this has been
a popular option in the past (e.g., \cite{Strong96}), it cannot any
longer be considered plausible.

We note that another possible origin for the low-energy $\gamma$-rays,
synchrotron radiation from $\sim$100 TeV electrons, has been suggested
by Porter \& Protheroe (1997).

\placefigure{Fig_gamma_high_latitude_profile_HELH}

\begin{figure*}[tbh]%***************************************************** 19
\centerline{
\psfig{file=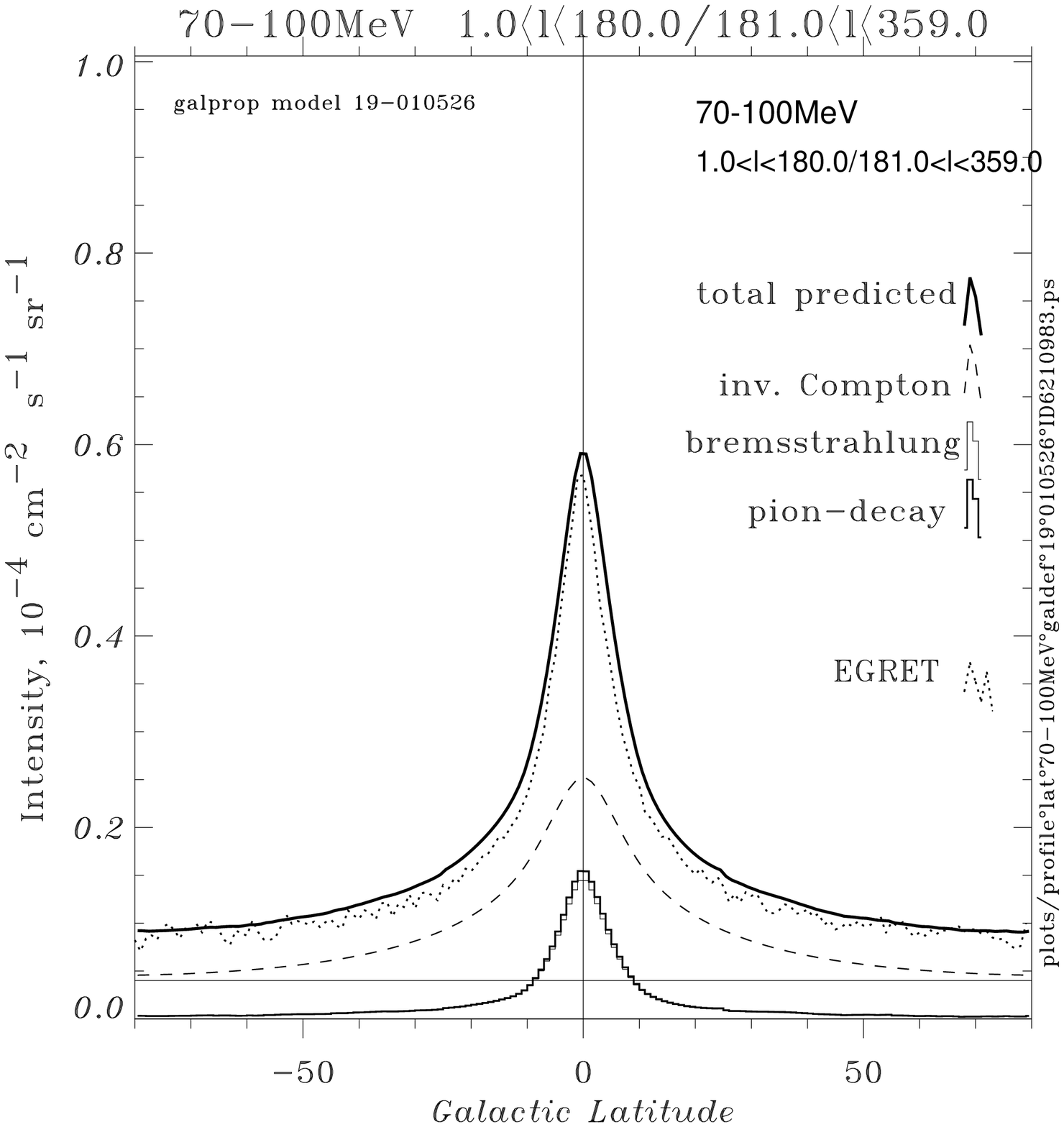,width=\fwb,clip=}
\psfig{file=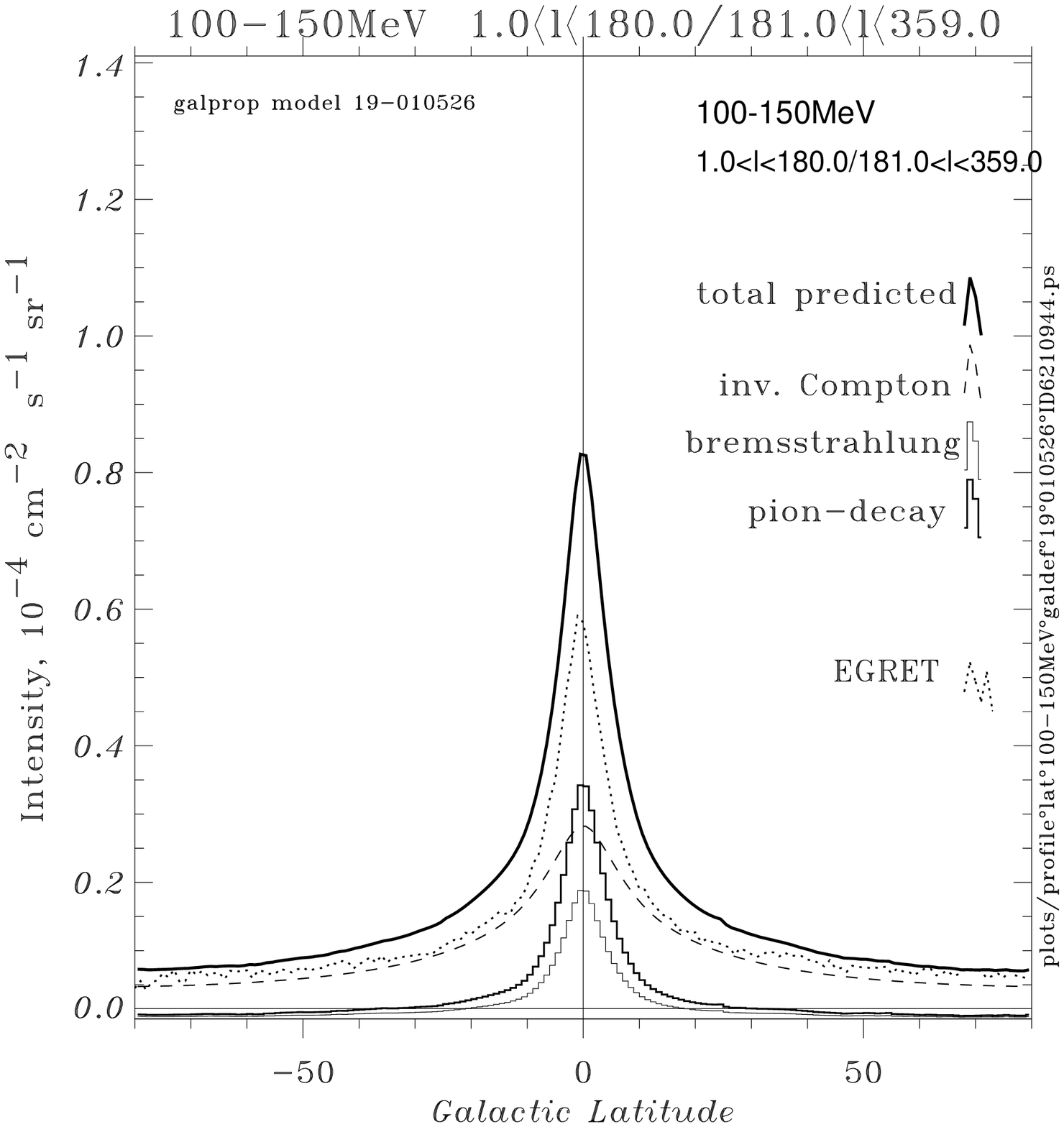,width=\fwb,clip=}
}
\figcaption[fig19a.ps,fig19b.ps]{
High latitude distribution of \grays for HELH model, 70--100 MeV
(left), and 100--150 MeV (right), compared to EGRET data.  Lines
are coded as in Fig.~\ref{Fig_gamma_latitude_profile}.
\label{Fig_gamma_high_latitude_profile_HELH}}
\end{figure*} 

\placefigure{Fig_gamma_luminosity}

\begin{figure*}[tbh]%***************************************************** 20
\centerline{
\psfig{file=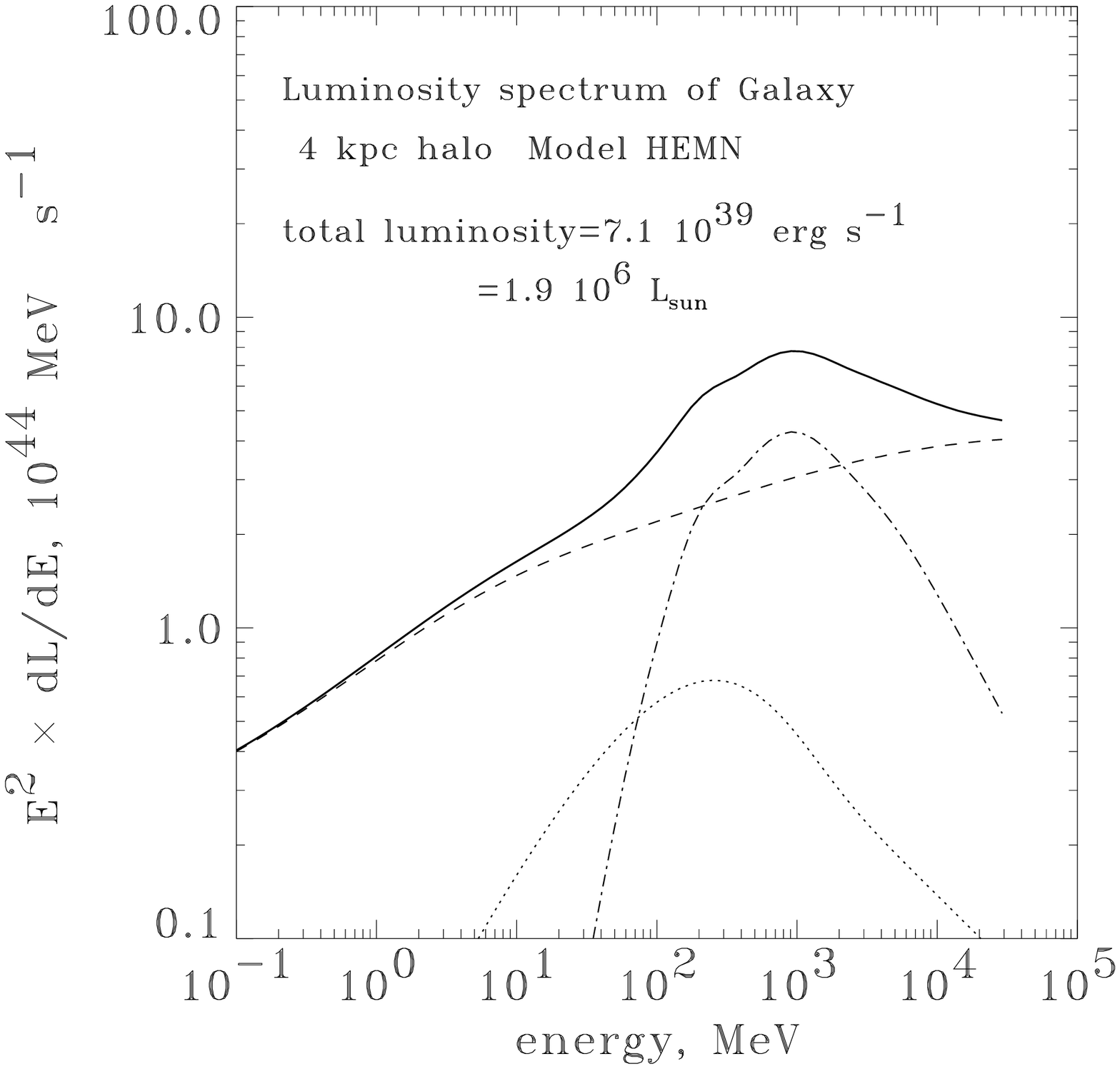,width=\fwb,clip=}
\psfig{file=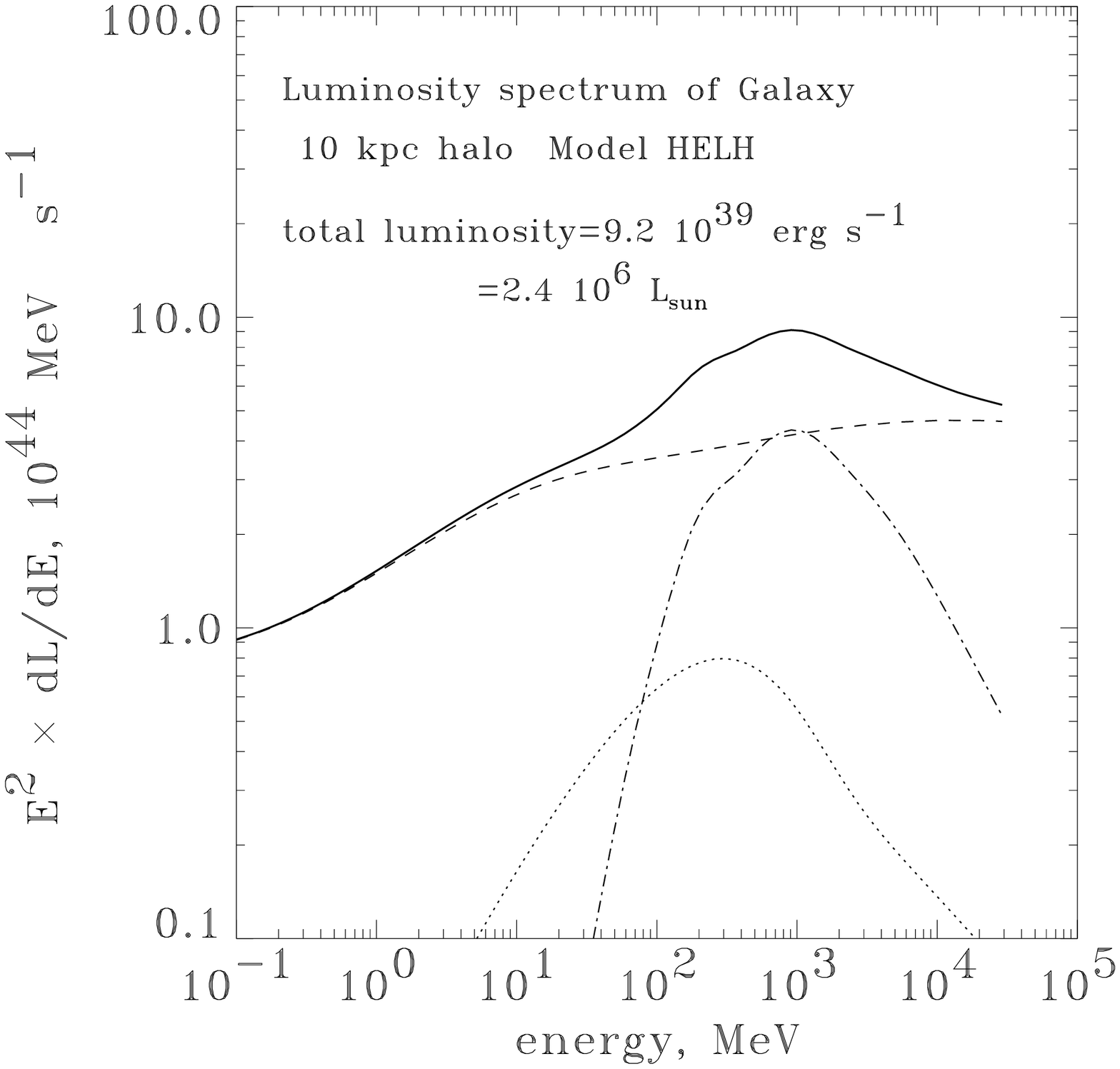,width=\fwb,clip=} }
\figcaption[fig20a.ps,fig20b.ps]{
Gamma-ray luminosity spectrum of diffuse emission from the {\it whole}
Galaxy using HEMN model (left), and HELH model (right).  Total
is shown as solid line.   Separate components: IC (dashes),
bremsstrahlung (dots), and $\pi^0$-decay (dash-dot).
\label{Fig_gamma_luminosity}}
\end{figure*}

\section{High latitude $\gamma$-rays and the size of the electron halo}
%######################################################################
Gamma rays provide a tracer of the electron halo via IC emission.  In
considering the HEMN model we showed that the high-latitude variation
of EGRET \grays is in good agreement with our large IC halo concept
(Fig.~\ref{Fig_gamma_high_latitude_profile}, Section~\ref{hard_e}).
Indication for a large \gray halo was also found by Dixon et al.\
(1998) from analysis of EGRET data.  Although we have used $z_h = 4$
kpc as the standard value in the present work we now test larger
values to derive limits. Studies of $^{10}$Be (\cite{SM98}) gave the
range $z_h = 4-12$ kpc for nucleons.  Webber \& Soutoul (1998) find
$z_h = 2-4$ kpc from $^{10}$Be and $^{26}$Al data.  Ptuskin \& Soutoul
(1998) find $z_h = 4.9^{+4}_{-2}$ kpc.  The high latitude \gray
intensity increases with halo size due to IC emission (though much
less than linearly due to electron energy losses), so that at least an
upper limit on the halo size can be obtained.
Fig.~\ref{Fig_gamma_spectrum_high_latitudes} shows the \gray spectrum
towards the Galactic poles for $z_h = 4$ kpc (HEMN model), and 10 kpc
(HELH model).  $z_h = 10$ kpc is possible although the latitude
profile for the 100--150 MeV range is then very broad and at the limit
of consistency with EGRET data
(Fig.~\ref{Fig_gamma_high_latitude_profile_HELH}).  Further the
isotropic component would have to approach zero above 300 MeV, so that
this halo size can be considered an upper limit.

If the halo size is 4--10  kpc as we argue, the contribution of
Galactic emission to the total at high latitudes is larger than
previously considered likely and has consequences for the derivation
of the diffuse extragalactic emission (e.g., \cite{Sreekumar98}).  An
evaluation of the impact of our models on estimates of the
extragalactic spectrum is beyond the scope of the present work.

\section{Luminosity spectrum of our Galaxy}
%######################################################################
So far some 90 extragalactic sources have been observed with the EGRET
telescope and several with COMPTEL (e.g., \cite{Hartman97}, 1999).
Most of these sources are blazars.  Such data usually serve as a basis
for estimates of the extragalactic \gray background radiation.
However, the number of normal galaxies far exceeds that of active
galaxies, it is therefore interesting to calculate the total diffuse
continuum emission of our Galaxy as an example.  Using cosmological
evolution scenarios, this can then be used as the basis for estimates
of the contribution from normal galaxies to the extragalactic
background.

The luminosity spectrum of the diffuse emission from the Galaxy is
shown in Fig.~\ref{Fig_gamma_luminosity}, based on models HEMN (4 kpc
halo) and HELH (10 kpc halo).  The total \gray diffuse luminosity of
the Galaxy above 1 MeV is $L_G= 7.1\times 10^{39}$ erg s$^{-1}$ for
the HEMN model and $9.2\times 10^{39}$ erg s$^{-1}$ for the HELH
model.  Above 100 MeV the values are $5.4\times 10^{39}$ erg s$^{-1}$
and $6.3\times 10^{39}$ erg s$^{-1}$, respectively. These values are
higher than previous estimates (e.g.\ \cite{Bloemen84}:
$[1.6-3.2]\times 10^{39}$ erg s$^{-1}$ above 100 MeV) due to our large
halo and the fact that our model incorporates the EGRET GeV excess.

\section{Conclusions}
%######################################################################
We have carried out a new study of the diffuse Galactic \gray
continuum radiation using a cosmic-ray  propagation model including
nucleons, electrons, antiprotons, positrons, and synchrotron emission.

We have shown that `conventional' models based on locally measured
cosmic-ray spectra are consistent with \gray measurements in the
30--500 MeV range, but outside this range excesses are apparent.  A
harder nucleon spectrum alone is considered but fitting to \grays
causes it to violate limits from positrons and antiprotons.  A harder
interstellar electron spectrum allows the \gray spectrum to be fitted
also above 1 GeV, and this can be further improved when combined with
a modified nucleon spectrum which still respects the limits imposed by
antiprotons and positrons. This is our preferred model, and it matches
the EGRET \gray longitude and latitude profiles reasonably in each
energy band.

Such a model produces only 25--50\% of the 1--30 MeV emission by
diffuse processes. The constraints provided by the synchrotron
spectral index do not allow all of the $<$30 MeV \gray emission to be
explained in terms of a steep electron spectrum unless this takes the
form of a sharp upturn below 200 MeV.  Therefore we prefer a source
population as the origin of the excess low-energy $\gamma$-rays, which
can then be seen as an extention of the hard X-ray continuum measured
by OSSE, GINGA and RXTE.  This is a quite natural scenario since it is
very likely that the hard X-rays are indeed from unresolved  sources,
and the switchover from source-dominated to diffuse-dominated has to
occur at some point; we propose here that it occurs at MeV energies.

The large electron/IC halo suggested here reproduces  well the
high-latitude variation of \gray emission, which can be taken as
support for the halo size for nucleons deduced from independent
studies of cosmic-ray composition.  Halo sizes in the range $z_h =
4-10$ kpc are favoured by both analyses.

Our models suggest that bremsstrahlung plays a rather minor r\^ole,
producing not more than $\sim$10\% of the Galactic emission at any
energy.

\acknowledgments 
A part of this work was performed while IVM held
a National Research Council/NASA GSFC Senior Research Associateship.

}
%\section*{APPENDIX}
%######################################################################
\appendix

\section{A.~~Spectrum of Electron Bremsstrahlung in the ISM} \label{bremss}
%######################################################################
In order to calculate the electron bremsstrahlung spectrum in the
interstellar medium, which includes neutral gas (hydrogen and Helium),
hydrogen-like and Helium-like ions as well as the fully ionized
medium, we use the works of Koch \& Motz (1959), Gould (1969), and
Blumenthal \& Gould (1970).  Our approach is similar to that used by
Sacher \& Sch\"onfelder (1984), but differs in some details.
Throughout this Section the units $\hbar=c=m_e=1$ are used.

The important parameter is the so-called screening factor defined as
\begin{equation}%=====================================================+
\label{A.1}
\delta=\frac{k}{2\gamma_0 \gamma},
\end{equation}%=======================================================+
where $k$ is the energy and momentum of the emitted photon, $\gamma_0$
and $\gamma$ are the initial and final Lorentz factor of the electron
in the collision.  If $\delta\to0$, the distance of the high energy
electron from the target atom is large compared to the atomic radius.
In this case screening of the nucleus by the bound electrons is
important.  Otherwise, for the low-energy electron only the
contribution of the nucleus is  important, while at high energies the
atomic electrons can be treated as unbound and can be taken into
account as free charges.

The cross section for electron-electron bremsstrahlung with one
electron initially at rest approaches, at high energies
($\gamma_0,\gamma,k\gg1$), the electron-proton bremsstrahlung cross
section with the proton initially at rest (\cite{Gould69}). Therefore
the contribution of atomic electrons at high energies can be accounted
for by a factor of $(Z^2+N)$ in  place of $Z^2$ in the formulas for
the unshielded charge, where $Z$ is the atomic number, and $N$ is the
number of the atomic electrons.  In the present paper we treat free
electrons in the ionized medium in the same way as protons, which is
an approximation, but it provides  reasonable accuracy for the range
3--200 MeV where the bremsstrahlung contribution into the diffuse
emission is most important.  In any case the contribution  from the
ionized medium is of minor importance in comparison with that of the
neutral gas.

\subsection{A.1.~~Low energies ($0.01\le E_{\rm kin}\le0.07$ MeV)}
%######################################################################

This is the case of nonrelativistic nonscreened bremsstrahlung,
$\Delta\equiv\delta/(2\alpha_f Z^{1/3})\gg1$. In the Born
approximation ($2\pi Z \alpha_f/\beta_0$, $2\pi Z \alpha_f/\beta\ll
1$) the production cross section is given by eq.~3BN(a) from Koch \&
Motz (1959)
\begin{equation}%=====================================================+
\label{A.2}
\frac{d\sigma}{dk}= f_E \frac{16}{3} \frac{Z^2 r_e^2 \alpha_f}{k
p_0^2} \ln\left(\frac{p_0+p}{p_0-p}\right),
\end{equation}%=======================================================+
where $\alpha_f$ is the fine structure constant, $p_0$ and $p$ are
initial and final momentum of the electron in the collision, $\beta_0$
and $\beta$ are initial and final velocity of the electron, and $f_E$
is the Elwert factor,
\begin{equation}%=====================================================+
\label{A.3}
f_E=\frac{\beta_0[1-\exp(-2\pi Z\alpha_f/\beta_0)]}
  {\beta[1-\exp(-2\pi Z\alpha_f/\beta)]},
\end{equation}%=======================================================+
which is a correction for the cross section eq.~(\ref{A.2}) at
nonrelativistic energies.

\subsection{A.2.~~Intermediate energies ($0.07\le E_{\rm kin}\le2$ MeV)}
%######################################################################

For the case of nonscreened bremsstrahlung ($\Delta\gg1$) the Born
approximation cross section is given by (\cite{KochMotz59}, eq.~3BN):
\begin{eqnarray}%=====================================================+
\label{A.4}
\frac{d\sigma}{dk}&=& \xi f_E Z^2 r_e^2 \alpha_f \frac{p}{k
p_0}\left\{\frac{4}{3}-2\gamma_0 \gamma \frac{p^2+p_0^2}{p_0^2
p^2}+\frac{\epsilon_0 \gamma}{p_0^3} +\frac{\epsilon
\gamma_0}{p^3}-\frac{\epsilon\epsilon_0}{p p_0} \right.  \nonumber\\
&& \left. +L\left[\frac{8}{3}\frac{\gamma_0 \gamma}{p_0 p}
+k^2\frac{\gamma_0^2 \gamma^2+p_0^2p^2}{p_0^3 p^3} +\frac{k}{2p_0
p}\left(\epsilon_0\frac{\gamma_0\gamma+p_0^2}{p_0^3}
-\epsilon\frac{\gamma_0\gamma+p^2}{p^3}
+2k\frac{\gamma_0\gamma}{p^2p_0^2}\right)\right] \right\},
\end{eqnarray}%=======================================================+
where
\begin{eqnarray}%=====================================================+
&&
\epsilon_0=\ln\left(\frac{\gamma_0+p_0}{\gamma_0-p_0}\right),\nonumber\\
&& \epsilon=\ln\left(\frac{\gamma+p}{\gamma-p}\right), \nonumber\\ &&
L=2\ln\left[\frac{\gamma_0\gamma+p_0p-1}{k}\right].  \nonumber\\
\end{eqnarray}%=======================================================+
The factor $\xi$ is given by
\begin{equation}%=====================================================+
\label{A.5}
\xi =  \left[1+\frac{N}{Z^2}\left(1-e^{(b-E_{\rm kin})/9b}\right)
   \right] \left(1-0.3\,e^{-k/c}\right)
\end{equation}%=======================================================+
where $b=0.07$ MeV, $c=0.33$ MeV, and the expression in square
brackets is a correction for the contribution of $N$ atomic electrons,
which is negligible at $E_{\rm kin}\sim0.1$ MeV, but becomes as large
as that of the protons at $E_{\rm kin}\sim2$ MeV (see also  general
comments at the beginning of Appendix~\ref{bremss}).  The second
factor, in round brackets, is a correction to obtain a smooth
connection between the approximations in the transition region near
0.1 MeV and is essential only for small $k$.

\subsection{A.3.~~High energies ($E_{\rm kin}\ge2$ MeV)}
%######################################################################

For the case of arbitrary screening we use eq.~3BS(b) from Koch \&
Motz (1959):
\begin{equation}%=====================================================+
\label{A.6}
\frac{d\sigma}{dk}=  r_e^2 \alpha_f\frac{1}{k}
\left[\left(1+\frac{\gamma^2}{\gamma_0^2}\right)\phi_1
-\frac{2}{3}\frac{\gamma}{\gamma_0}\phi_2\right].
\end{equation}%=======================================================+
If the scattering system is an unshielded charge, the functions are
$\phi_1=\phi_2=Z^2\phi_u$, where
\begin{equation}%=====================================================+
\label{A.7}
\phi_u =
   4\left[\ln\left(\frac{2\gamma_0\gamma}{k}\right)-\frac{1}{2}\right].
\end{equation}%=======================================================+
For the case where the scattering system is a nucleus with bound
electrons, the expressions for $\phi_1$ and $\phi_2$ are more
complicated and depend on the atomic form factor.

Corresponding expressions for one- and two-electron atoms ($N=1,2$)
have been given by Gould (1969). Rearranging these one can obtain
\begin{eqnarray}%=====================================================+ 
\label{A.8}
\phi_1(N) &=& (Z-N)^2\phi_u+ 8Z \left\{1 -\frac{N-1}{Z} +\int_\delta^1
dq\,\frac{R_N(q)}{q^3}(q-\delta)^2\right\}, \\ \phi_2(N) &=&
(Z-N)^2\phi_u+ 8Z \left\{\frac{5}{6}\left(1-\frac{N-1}{Z} \right)
%\right.\\&&\left.
+\int_\delta^1 dq\,\frac{R_N(q)}{q^4}
\left[q^3-6\delta^2q\ln(q/\delta)+3\delta^2q-4\delta^3\right]\right\},
\nonumber
\end{eqnarray}%=======================================================+
where
\begin{equation}%=====================================================+
\label{A.9}
   \begin{array}{ll} \displaystyle{R_1(q)=1-F_1(q)}, &
      \displaystyle{F_1(q)=\{1+q^2/[2\alpha_f Z]^2\}^{-2}};\\
      \displaystyle{R_2(q)=2[1-F_2(q)]-[1-F_2^2(q)]/Z}, &
      \displaystyle{F_2(q)=\{1+q^2/[2\alpha_f (Z-5/16)]^2\}^{-2}}.
      \end{array}
\end{equation}%=======================================================+
Equations~(\ref{A.8}), (\ref{A.9}) are valid for any $Z$, including
H$^-$ ions.  The formulas have been obtained under the assumption that
the two-electron wave function of He-like atoms can be approximated by
the product of one-electron functions in the form of Hylleraas or
Hartree.  For large $\Delta$ the expressions for $\phi_1$ and $\phi_2$
approach the unshielded value
$$ \phi_1=\phi_2\to(Z^2+N)\phi_u. $$

For the case of neutral He atoms Gould (1969) gives also numerical
values of $\phi_1$ and $\phi_2$ tabulated for the variable
$\{\delta/(2\alpha_f)\}=0...10$. The latter have been calculated for a
Hartree-Fock wave function, which are considered to be more accurate
than the Hylleraas function.  At low energies ($\Delta\gtrsim2$) both
functions provide the identical results.

At high energies where $k, \gamma_0, \gamma\gg1$, $\Delta\gtrsim4$,
eq.~(\ref{A.4}) with $\xi=1+N/Z^2$ and $f_E=1$ can be applied, where
electrons are treated as unbound in the same way as protons.

\subsection{A.4.~~Fano-Sauter Limit}
%######################################################################
The  formulas described above do not permit the evaluation of the
cross section at the high-frequency limit $k\to \gamma_0-1$.  The
cross section obtained in the Born-approximation becomes zero in this
limit, while the value is non-zero.  The corresponding expression has
been obtained by Fano in the Sauter approximation (\cite{KochMotz59}):
\begin{equation}%=====================================================+
\label{A.10}
\left(\frac{d\sigma}{dk}\right)_{FS}= 4\pi Z^3\alpha_f^2 r_e^2
\frac{\gamma_0\beta_0}{k(\gamma_0-1)^2}
\left\{\frac{3}4+\frac{\gamma_0(\gamma_0-2)}{\gamma_0+1}
\left[1-\frac{1}{2\beta_0 \gamma_0^2}
\ln\left(\frac{1+\beta_0}{1-\beta_0}\right) \right]\right\}.
\end{equation}%=======================================================+

\subsection{A.5.~~Heavier atoms}
%######################################################################
For the electron bremsstrahlung on neutral atoms heavier than He we
use the Schiff formula (\cite{KochMotz59}, eq.~3BN(e)):
\begin{eqnarray}%=====================================================+
\label{A.50}
\frac{d\sigma}{dk} &=& 2 Z^2 r_e^2 \alpha_f\frac{1}{k}
\left\{\left(1+\frac{\gamma^2}{\gamma_0^2}
-\frac{2}{3}\frac{\gamma}{\gamma_0}\right) \left(\ln
M(0)+1-\frac{2}{b}\arctan b\right)\right.\\ &&\left.
+\frac{\gamma}{\gamma_0}\left[\frac{2}{b^2}\ln(1+b^2)
+\frac{4(2-b^2)}{3b^3}\arctan b -\frac{8}{3b^2}+\frac{2}{9}\right]
\right\}, \nonumber
\end{eqnarray}%=======================================================+
where
$$
b=\frac{Z^{1/3}}{111\delta};  \qquad  M(0)=\frac{1}{\delta^2 (1+b^2)}.
$$

\section{B.~~Synchrotron radiation} \label{synchrotron}
%######################################################################
For  synchrotron emission we use the standard formula (see e.g.
\cite{Ginzburg79}).  After averaging over the  pitch angle for an
isotropic electron distribution, this gives the emissivity
$\epsilon(\nu,\gamma)$ of a single electron integrated over all
directions relative to the field in the form (\cite{Ghisellini88})
\begin{equation}%=====================================================+
\label{A.11}
\epsilon(\nu,\gamma) = 4\sqrt{3}\, \pi r_e m_e c\, \nu_B\,
x^2\left\{K_{4/3}(x) K_{1/3}(x) -{3\over 5}x\,
[K_{4/3}^2(x)-K_{1/3}^2(x)]\right\}
\end{equation}%=======================================================+
in units of (ergs s$^{-1}$ Hz$^{-1}$), where $\nu$ is the radiation
frequency, $\gamma$ is the electron Lorentz factor, $\nu_B=eB/(2\pi
m_e c)$, $B$ is the total magnetic field strength, $x\equiv
\nu/(3\gamma^2\nu_B)$, and $K_z$ is the modified Bessel function of
order $z$.

%\clearpage

\end{document}